\documentclass[12pt]{article}
\usepackage{latexsym}
\usepackage{amsmath}
\usepackage{amsfonts}
\usepackage{dsfont}
\usepackage{multibox}
\usepackage{verbatim}
\usepackage{hyperref}
\usepackage{tikz}
\usetikzlibrary{matrix,arrows}
\usepackage{subfigure}
\usepackage{amsmath,amssymb}
\usepackage{graphicx}
\usepackage{color}
\usepackage{booktabs}

\newcommand{\ra}[1]{\renewcommand{\arraystretch}{#1}}
 \csname
@addtoreset\endcsname{equation}{section}

\begin{document}

\def\be{\begin{equation}}
\def\ee{\end{equation}}
\def\bea{\begin{eqnarray}}
\def\eea{\end{eqnarray}}
\def\nn{\nonumber}
\def\n{n+\delta'}
\def\barn{\bar{n} + \bar{\delta}'}
\def\m{m+\delta''}
\def\barm{\bar{m}+\bar{\delta}''}

\begin{flushright}

\end{flushright}

\vspace{40pt}

\begin{center}

{\Large\sc T-duality twists and Asymmetric Orbifolds}

\vspace{50pt}

{\sc Hai Siong Tan}

\vspace{15pt}
{\sl\small
Institute for Theoretical Physics,
University of Amsterdam, \\
Science Park 904, Postbus 94485, 1090 GL, Amsterdam, The Netherlands
\\

\vspace{10pt}
{\it H.S.Tan@uva.nl}}

\vspace{70pt} {\sc\large Abstract}\end{center}

We study some aspects of asymmetric orbifolds of tori, with the orbifold group being some $\mathbb{Z}_N$ subgroup of the T-duality group and, in particular, provide a concrete understanding of certain phase factors that may accompany the T-duality operation on the stringy Hilbert space in toroidal compactification.  We discuss how these T-duality twist phase factors are related to the symmetry and locality properties of the closed string vertex operator algebra, and clarify the role that they enact in the modular covariance of the orbifold theory, mainly using asymmetric orbifolds of tori which are root lattices as working examples.

\newpage
\tableofcontents

\section{Introduction}
\label{sec:Intro}

Closed strings propagating on an orbifold \cite{DHVW1,DHVW2} present an important class of backgrounds in which stringy phenomena arise strikingly from the notion of twisted sectors. A geometric orbifold $\mathcal{O}$ is the quotient of a target space manifold $\mathcal{M}$ by the action of some discrete, metric-preserving group $G$, i.e. $\mathcal{O} = \mathcal{M}/G$.  Such an action typically leaves a number of points fixed, and at these fixed points, even though the manifold suffers from conical singularities, the orbifold string CFT is well-defined from the viewpoint of unitarity by simply including additional `twisted' sectors in which strings are closed up to the group action. Formally, the string functional integral is the sum over maps from the worldsheet into the orbifold, and in the neighborhood of the fixed points, the map is branched over the manifold's covering space \cite{Martinec,Horava}. Such a concept may be generalized to generic CFTs with discrete symmetries. Thus, given any CFT $\mathcal{C}$ admitting some discrete symmetry group $G$, it is natural to ask if it makes sense to construct $\mathcal{C}/G$. In the generic case, modding the theory out by $G$ can no longer be described geometrically as a closed string propagating on an orbifold, and we need to rely on more abstract principles in place of our geometrical intuition in developing the notion of twisted sectors. A fundamental question to be addressed is whether one could appropriately lift $G$ to be an automorphism of the operator algebra in a Hilbert space construction, and whether the twisted theory remains local.

In this paper, we shall study closed string orbifolds of tori, with the orbifold group being a $\mathbb{Z}_N$ subgroup of the T-duality group. Recall that for a closed bosonic string wrapped on a $d$-dimensional torus and coupled to a background Kalb-Ramond magnetic field, the T-duality group is $O(d,d;\mathbb{Z})$. This group contains, apart from the geometric $GL(d,\mathbb{Z})$ subgroup, orbifold elements which act differently on the string's left and right movers. The notion of such asymmetric orbifolds \cite{Narain1,Narain2}
is of course natural in the context of the heterotic string where it was first considered.  For a $d$-dimensional toroidal compactification of the heterotic string, on the even and self-dual lattice $\Gamma^{16+d, d} = \Gamma^{16,0} + d\Gamma^{1,1}$, we can, for example, consider modding out the Hilbert space by left-right asymmetric action on $d\Gamma^{1,1}$. 

In general, it is a difficult question to derive the sufficient conditions for a duality group to be a genuine automorphism of the operator algebra, and here we shall content ourselves mainly with the conditions related to the broad principles of locality and modular covariance. To put it in the context of T-duality twists of toroidal compactifications, schematically, let us denote the T-duality operation to be $\hat{g}$ and consider its action in the untwisted sector. When acting on a lowest-weight state labelled by the left and right momenta $\alpha_{L,R}$, a phase factor ambiguity arises which we can express as follows.
\be
\label{T}
\hat{g} | \alpha_L, \alpha_R \rangle = U(g, \alpha) | g(\alpha_L), g(\alpha_R) \rangle
\ee
where $g(\alpha)$ are $O(d,d;\mathbb{Z})$ transformations of the momenta zero modes, and $U(g,\alpha)$ is a phase factor which will play a role in the modular covariance of an orbifold of which twist is generated by $\hat{g}$. Our main purpose here is to develop a concrete understanding of this phase factor and compute it for some asymmetric orbifolds of the tori. By the state-operator correspondence, we shall see that this phase factor is related to the symmetry properties and mutual locality of the string's vertex operator algebra in the untwisted sector. It will inevitably make an appearance when we consider the modular covariance of the orbifold CFT since it might be non-trivial when evaluated on any residual zero modes' sublattices invariant under the orbifold twist.

The fundamental ideas invoked in our analysis are established ones. For example, it is well-known that the mutual locality of vertex operators requires the presence of certain one-cocycle translation operators acting on the zero-mode space. Already in the seminal papers \cite{Narain1,Narain2}, it was mentioned that it is important to include such cocycle factors attached to vertex operators to preserve their mutual locality. Physically, this can be  understood as the statement that the order of emission of gauge particles does not change the closed string amplitude whereas mathematically, this is a familiar notion in the formal study of vertex operator algebra \cite{Lepowsky}. As we shall discuss in detail, the presence of these factors leads one to consistency equations that determine the phase factor in \eqref{T} in order to preserve the $\mathbb{Z}_N$ symmetry of the operator algebra. They are then generically manifest in the genus one characters of the orbifold theory twisted by the same $\mathbb{Z}_N$ symmetry.

At least to our knowledge, the relationship between the cocycle factors and the phase factor in \eqref{T} has not been extensively explored as much as it deserves. Apart from the seminal papers \cite{Narain1,Narain2},
there is a delicate account of it in the theses of Hollowood \cite{Hollowood} and Myhill \cite{Myhill} in the general context of orbifolds, where their focus was on symmetric orbifolds of tori. Building upon their work, we will explore the role of the twist phase factor in the context of asymmetric orbifolds. A more recent inspiration comes from \cite{Simeon} where the phase factor in \eqref{T} was understood along these lines to motivate a modular-covariant form of the one-loop partition function of a single self-dual compact boson $(X)$ twisted by the only non-trivial element of $O(1,1;\mathbb{Z})$, i.e. $X_L \rightarrow X_L, X_R \rightarrow -X_R$. In the absence of these considerations, one may be misled to thinking that this background is anomalous (see \cite{Aoki} for a separate proposal for the Hilbert space construction). Yet as we shall discuss later, the inclusion of these phase factors does not always preserve one-loop modular covariance in the general case.

For the rest of the paper, we will present illustrative examples of how to derive and compute the phase factor in \eqref{T} in the context of $\mathbb{Z}_N$ asymmetric toroidal orbifolds of which twist lies in $O(d,d;\mathbb{Z})$. We will find that this problem is reduced to solving some constraint equations that descend from preserving the corresponding symmetry of the operator algebra diagonalized in the eigenspaces of the twist. These equations involve the cocycle factors which are needed to write down mutually local OPEs, and 
moreover can be neatly interpreted as the triviality condition of a 2-cocycle valued in the second cohomology of the Narain momenta lattice with $U(1)$ coefficients. In formulating the problem for the class of asymmetric orbifolds in which the twist in one chiral sector is trivial, we stumble upon a simple relationship between the moduli space for these backgrounds and automorphisms of the Lie algebra of which roots generate the toroidal lattice. In the simplest example of a chiral $\mathbb{Z}_2$ orbifold of a compact boson, we will also observe that this phase factor arises in a similar fashion at higher worldsheet genus. Other related results that are presented in this paper include a straightforward discussion of the modular covariance of shift orbifolds and comments on some asymmetric orbifold points of $CY_3$ compactification of the heterotic string.

The toroidal orbifolds that we consider in this paper are simple examples of closed strings whose boundary conditions are twisted by elements of the automorphism group of the string CFT's operator algebra. Orbifolding by T-duality offers a manageable class of non-geometric backgrounds that are more clearly understood from the viewpoint of orbifold constructions.  More broadly speaking, the automorphism group can be non-perturbative when we  twist boundary conditions by S-duality (see for example \cite{Ganor:Stwist,Ganor:2012mu, Ganor:2008,Ganor:2014}) or more generally U-duality \cite{Kumar,Ganor:Utwist,Ganor:2012ek}. Understanding these stringy monodrofolds \cite{Dabholkar} is not only interesting in its own right, but they may also have some implications for string cosmology \cite{Kachru,Maloney,Hassler:2014mla,Silverstein:2001xn}, string phenomenology \cite{Ibanez,Anastasopoulos:2009kj,Bianchi:2012} and modern duality-covariant frameworks of string theory such as the likes of `Double Field Theory' (see for example \cite{Hull}) and gauged supergravity theories \cite{Condeescu:Gsugra,Condeescu:Asym,Lust:2015}. The backgrounds that we are considering are sometimes called {\it T-folds} at their self-dual points, and they may furnish a stage upon which we can further address the notion of stringy non-geometry (see \cite{Bakas:2015gia} for an interesting recent work).

The plan of our paper is as follows. In Section 2, we present the basic outline of our approach, including a brief review of various basic conceptual ingredients, such as the notion of modular covariance of genus-one characters, level-matching, cocycles etc. that we need in shaping our narrative. In Section 3,  as a warm-up, we consider the simplest T-fold in detail (in this aspect, please also see \cite{Simeon} and \cite{Williams}). In Section 4, we explore asymmetric toroidal orbifolds, of which the orbifold group $G$ is a $\mathbb{Z}_N$ subgroup of $O(d,d;\mathbb{Z})$, and derive the constraint equations for the T-duality twist phase factor as the triviality condition of a certain 2-cocycle of the momentum lattice. Section 5 contains concrete two-dimensional and six-dimensional orbifold examples, some details of which are delegated to the Appendices. In Section 6, we generalize our observations for the simplest T-fold at higher worldsheet genus (which is a simple application of a related result presented in \cite{DVV} for $c=1$ CFTs on Riemann surfaces). Finally, we end with concluding remarks and a few suggestions for future work. 

{\bf Relations to previous work }: Orbifold twist phase factors were previously discussed in \cite{Hollowood, Myhill} in the general context of orbifolds and in \cite{Narain1, Narain2} for asymmetric orbifolds. Our work builds on related ideas mentioned in these seminal papers, and can be regarded as a more explicit exploration of these phase factors and their relation to modular covariance. In \cite{Narain2}, the origin of the twist phase factor is traced to a consistency condition that arises in defining the bosonic partition function of any asymmetric orbifold as the square root of that of a `parent' non-chiral boson theory. Holomorphic factorization of stringy instanton sum in the doubled theory requires the presence of a winding number-dependent phase factor which then leads to a non-trivial phase factor in the partition function of the asymmetric orbifold. Our starting point is different and it would be interesting to interpret the various results particularly in the setting of modern T-duality covariant frameworks like in \cite{Hull} where asymmetric twists can possibly be treated as symmetric ones. Another important and more modern inspiration for this work comes from \cite{Simeon} where the role of the twist phase factor in the simplest T-fold was discussed in detail.


\section{Generalities}
\label{sec:Gen}

In the following, we present the broad outline of our approach in our study of asymmetric orbifolds of tori. The key entity that lies at the heart of our discussion is the phase factor in \eqref{T} that accompanies the action of T-duality on the stringy states. This phase factor is related to three essential notions: (i)the mutual locality of vertex operators (ii)preservation of the T-duality symmetry in the closed string vertex operator algebra (iii)one-loop modular covariance of the partition traces. In this section, we will present the main points of this relationship, leaving more explicit examples, technical details and generalizations to subsequent sections. 

A consistency principle which we allude to in this paper is the modular covariance of one-loop partition traces, defined as 
\be
\label{Ztrace}
Z^g_h  (\tau) = \textrm{Tr}_h \left[  g\,q^{L_0}\bar{q}^{\bar{L}_0} \right],\,\,\,q \equiv e^{2\pi i \tau},
\ee
where $\tau$ is the complex structure of the Euclidean toroidal string worldsheet and $Z^g_h (\tau)$ denotes the partition function with the insertion of some twist element $g$ that belongs to the orbifold group and evaluated in the sector twisted by another element $h$. Under the mapping class group of the torus, the partition traces transform onto one another under a generic $SL(2,\mathbb{Z})$ element as follows (see Appendix~\ref{sec:OnMod} ).
\be
\label{traceSL2}
Z^g_h \left(  \frac{a\tau + b}{c\tau + d} \right)  = Z^{g^d h^{-b}}_{g^{-c}h^a} \left( \tau \right).
\ee
In this paper, we are mainly working with $\mathbb{Z}_N$ orbifolds, and the modular orbits can be organized straightforwardly. Now, level-matching conditions are typically taken by requiring an absence of global modular anomaly at one-loop. Take some partition trace $Z^g_h (\tau)$ and consider the subgroup $\Gamma_{(g,h)}$ of $SL(2,\mathbb{Z})$ that fixes the boundary condition, i.e. $\Gamma_{(g,h)}$ are the stability groups for the abelian group generated by $g$ and $h$. We then demand that the partition trace picks up no phase under $\Gamma_{(g,h)}$ (see for example \cite{Vafa} for an illuminating discussion). Of course, it is sufficient to check this for any single representative of each class of partition traces closed under modular transformations.\footnote{ To see this, start with a partition trace $Z^r_s (\tau) \equiv Z((r,s);\tau)$ that is mapped back to itself under $\Gamma$. Now under an $SL(2,\mathbb{Z})$ element $M$, the twist indices and stability group transform as $(r,s) \rightarrow M^{-1} (r,s)$ and $\Gamma \rightarrow M^{-1} \Gamma M$. But $ Z\left( \Gamma' \circ (r',s');\tau \right) = Z\left( \Gamma \circ (r,s); M(\tau) \right) = Z \left( (r,s);M(\tau) \right) = Z \left( (r',s') ;\tau \right)$. } We can use the twisted sectors without any twist insertion to be representatives of each closed orbit. Consider the trace $Z_{h_1}^0 (\tau)$ which transforms as 
$$
Z_{h_1}^0 \left( \frac{a \tau + b}{c \tau + d} \right) = Z^{h_1^{-b}}_{h_1^a} (\tau). 
$$ 
Then, by Bezout's lemma, it can be transformed onto another twisted sector $Z^0_{h_2}$, if there exists $a$ such that $ah_1 = h_2\,\,\textrm{mod}\,\,N$ and $\textrm{gcd}(a, \text{ord}(h_1)) =1$. For example, if $N = p \times q$, with ${p,q}$ being distinct primes, then there are three independent modular orbits of which representatives we can pick to be $Z^0_1$, $Z^0_p$ and $Z^0_q$. Level-matching translates into checking their invariances under $\tau \rightarrow \tau +N$, $\tau \rightarrow \tau + q$ and $\tau \rightarrow \tau + p$ respectively. Since the other partition traces are related to them by modular transformations, the level-matching conditions are equivalent to the modular covariance condition in \eqref{traceSL2}. 

In the context of toroidal compactification, we can let the twist to be some T-duality group element specified by an $O(d,d;\mathbb{Z})$ matrix acting on the space of winding and momenta zero modes equipped with some toroidal lattice and a constant Kalb-Ramond $B$-field, together with corresponding action on the left and right oscillators. Of course, the moduli have to be self-dual under the twist. Now, there is a q-number phase ambiguity when we lift T-duality symmetry to be a symmetry of the stringy Hilbert space, or equivalently by the state-operator correspondence, of the operator algebra. Recall that the untwisted Hilbert space $\mathcal{H}$ can be decomposed into the Fock space representation of the Heisenberg algebra of the oscillators $(\mathcal{F})$ and a zero mode space - the discrete space on which the left and right-moving momenta zero modes reside, i.e.
\be
\mathcal{H} = \mathcal{F} \otimes \Lambda_{L,R} = \sum_{p_L,p_R} \mathcal{H}_{p_L} \otimes \overline{\mathcal{H}}_{p_R} 
\ee
where we have also indicated a splitting of the Hilbert space into left and right sectors each labeled by momenta zero modes. In the twisted sector labelled by $h$ in the orbifold CFT,  we can imagine tensoring the orbifold twist $g$ with a phase factor $U_h (g,\alpha)$\footnote{In the untwisted sector, we shall denote the phase factor by $U(g,\alpha)$ (as in \eqref{T}).} where $\alpha$ are the momenta zero modes, then each partition trace gets modified as 
\be
\label{phaseDefin}
Z^g_h (\tau) \rightarrow \textrm{Tr}_h  \left[ U_h (g,\alpha) \theta_L q^{L_0} \theta_R q^{\bar{L}_0} \right],
\qquad g = (\theta_L, \theta_R),
\ee
where the twist element $g$ can be described by independent twists $\theta_{L,R}$ in the left and right sectors respectively. In this paper, we wish to understand these phases $U_h(g,\alpha)$ more carefully. As we shall discuss later, we will find that they are generally of the form $e^{i\Omega^{ij} \alpha_i \alpha_j}$ where $\Omega$ is some constant matrix. We shall invoke two general well-known principles for their construction as follows (see for example \cite{Hollowood} and \cite{Myhill}).

The first principle relates to writing down the vertex operators in \eqref{prelimvertex} correctly with the inclusion of appropriate cocycle factors which act on the zero mode space of the stringy Hilbert space $\Lambda_{L,R}$.  The Hilbert space decomposition is compatible with the quantization rule
\be
\label{quantizationasymmetric}
[x^0_L, p_L] = [x^0_R, p_R] = i,
\ee
where $x^0_{L,R}$ refer to the position zero modes in the left and right sectors of the theory. In the classical string theory, there is not much meaning in assigning `left' and `right' to the position zero modes, but in our study of asymmetric orbifolds, we will find that \eqref{quantizationasymmetric} is a fine assumption. We can now separately discuss the zero-mode space of each chiral sector. We assume a toroidal background equipped with some orbifold action, and define $\hat{C}(\alpha)$
to be the zero mode part of the string vertex operators refined with some possibly operator-valued prefactor that preserves mutual locality (see for example \cite{Polchinski} for a nice review). They are of the form 
\be
\label{cocycles}
\hat{C}(\alpha) = e^{i \delta(p_{L,R}, \alpha_{L,R})} e^{i \alpha_R x^{(0)}_R + i \alpha_L x^{(0)}_L},
\ee
where $\delta(p_{L,R}, \alpha_{L,R})$ is a non-unique function of the momenta which we shall describe explicitly in Section~\ref{sec:T-fold} and \ref{subsec:Cocycles} . 
We note that $\hat{C}(\alpha)$ furnishes a projective representation of the compactification lattice as follows, 
\be
\label{projectivecondition}
\hat{C}(\alpha) \hat{C}(\beta) = \epsilon(\alpha,\beta) \hat{C}( \alpha + \beta)
\ee
for some non-commutativity phase $\epsilon(\alpha,\beta)$ which plays the crucial role of removing the branch cut in the OPE of the otherwise unrefined vertex operators. Now the non-zero mode parts of each of two vertex operators (separated in their insertion points by $\delta z$) give rise to a factor $(\delta z)^{\frac{1}{2} \alpha_R \beta_R} (\delta \bar{z})^{\frac{1}{2} \alpha_L \beta_L}$ in their OPE. Thus, if we let
\be
\label{NCphase}
\epsilon(\alpha, \beta) = e^{\frac{i\pi }{2} \left(\alpha_R \beta_R - \alpha_L \beta_L \right)} \epsilon(\beta,\alpha),
\ee
we preserve the mutual locality of the vertex operators. From the associativity of the OPEs among the vertex operators, one can show that 
\be
\label{cocycleC1}
\epsilon(\alpha, \beta + \gamma) \epsilon(\beta, \gamma) = \epsilon(\alpha,\beta)\epsilon(\alpha + \beta,\gamma).
\ee
It turns out that we can interpret $\epsilon(\alpha,\beta)$ as an element of the two-cohomology of the Narain momenta lattice. One can define a lattice's $R$-cochain $c$ as a map from $R$ copies of the lattice to $U(1)$, and the coboundary operation acting on $c$ to be
\bea
\label{coboundary1}
\delta c \left( \alpha_1,...,\alpha_{R+1} \right) &=& 
\frac{c(\alpha_2,...,\alpha_{R+1})}{c(\alpha_1+\alpha_2, \alpha_3,...,\alpha_{R+1})}
\times 
\frac{c(\alpha_1, \alpha_2 + \alpha_3,...,\alpha_{R+1})}{c(\alpha_1,\alpha_2, \alpha_3+\alpha_4,...,\alpha_{R+1})} \times \ldots \cr
&&\times c(\alpha_1,...,\alpha_R)^{(-1)^{R+1}}.
\eea
Then one can see that \eqref{cocycleC1} is nothing but a 2-cocycle condition. Mutual locality demands \eqref{NCphase} which implies that $\epsilon (\alpha,\beta)$ cannot be a two-coboundary. 
( In \eqref{projectivecondition}, if $\hat{C}(\alpha)$ is a c-number function, then $\epsilon(\alpha,\beta)$ is indeed a two-coboundary.) There is however an equivalence condition that we should impose that will fix $\epsilon(\alpha,\beta)$ to be a class of the second cohomology group. In \eqref{projectivecondition}, there is a gauge degree of freedom preserving \eqref{NCphase} that corresponds to
\be
\label{phasetransformation1}
\hat{C}(\alpha) \rightarrow e^{i\delta(\alpha)} \hat{C}(\alpha), \qquad 
\epsilon (\alpha,\beta) \rightarrow e^{i \left(\delta(\alpha+\beta) - \delta(\alpha) -\delta(\beta) \right)} \epsilon (\alpha,\beta)
\ee
where $\delta(\alpha)$ is some scalar function of $\alpha$. Given some $\epsilon (\alpha,\beta)$, one can construct an equivalence class of it via \eqref{phasetransformation1} which preserves mutual locality. Later in Section~\ref{subsubsec:cocycleExpression}, we shall develop an explicit expression for it which turns out 
to be simply 
$$
\epsilon (\alpha,\beta) = e^{\frac{i\pi}{2} (n_\alpha \cdot m_\beta - n_\beta \cdot m_\alpha)},
$$
where $n_\alpha,m_\alpha$ are the momentum and winding numbers associated with momentum zero mode  $\alpha$. Secondly, we recall that as explained in \cite{Dijkgraaf}, there is an elegant way of interpreting the fusion algebra of a holomorphic CFT which enjoys a symmetry group $G$.  We begin with the untwisted sector, and let $G$ be some $\mathbb{Z}_N$ subgroup of the T-duality group. Under the action of $G$, the operator algebra in the untwisted sector decomposes into sectors filled with states transforming in the irreducible representations of $G$. For each of $N$ non-isomorphic representations of $\mathbb{Z}_N$, we can associate it with the following linear combination of vertex operators $V$ 
\be
\label{prelimvertex}
V_{[a]}= \sum_m e^{-\frac{2\pi i m a}{N}} \hat{g}^m \cdot V, \qquad \hat{g} \cdot V_{[a]} = e^{\frac{2\pi i a}{N}} V_{[a]}, \qquad a = 0,1,\ldots N-1,
\ee
where $\hat{g}$ is the T-duality twist operator that generates the $\mathbb{Z}_N$ action. As explained in \cite{Dijkgraaf} for a general $G$, the representation algebra should be identical to the fusion algebra
of the representations. Invoking this principle then, we see that this translates to
\be
\label{fusionrule}
V_{[a]} \times V_{[b]} \sim V_{[a+b]}.
\ee
This furnishes constraint conditions for $\hat{g}$, which in turn relate the cocycle factors and the phase factors in the untwisted sector which appear in \eqref{T} and thus, by the state-operator correspondence in \eqref{fusionrule}. As we shall explain in detail in Section \ref{subsec:Cocycles}, the constraints can be straightforwardly derived to read
\bea
\label{coboundaryconditions1}
\frac{\epsilon(g(\alpha),g(\beta))}{\epsilon(\alpha,\beta)} &=& \frac{U(g,\alpha+\beta)}{U(g,\alpha) U(g,\beta)}, \\
\label{coboundaryconditions2}
U(g^{p+1},\alpha) = U(g^p, \alpha) U(g, g^p(\alpha)), && \prod_{j=1}^N U(g, g^j(\alpha)) = 1.
\eea
Since $g$ is an automorphism of the Narain lattice, $\epsilon\left( g(\alpha), g(\beta) \right)$ is an element of $H^2( \Lambda, U(1))$ just like $\epsilon (\alpha,\beta)$. Diagonalizing the OPEs among the vertex operators leads us to consider the ratio between the two which, from \eqref{coboundaryconditions1} is clearly a two-coboundary. For the working examples that we consider in this paper, the twist is trivial in one chiral sector, and the ratio reduces to a trivial element of $H^2( \Lambda, \mathbb{Z}_2)$. Generally, for any orbifold twist, we can compute this ratio given a solution to $\epsilon (\alpha,\beta)$. Solving for the one-cochain phase factor $U(g,\alpha)$ is then equivalent to solving a cohomological problem. Formally, all this means that the consistency conditions we have derived can be understood as the triviality of the ratio $\epsilon \left(g(\alpha),g(\beta) \right) / \epsilon \left(\alpha,\beta \right)$, with the twist phase factors being one-cochains that have to satisfy \eqref{coboundaryconditions2}. Thus far, our considerations pertain to the untwisted sector. To obtain the appropriate form of twist phase factors in the twisted sector labelled by $h$, we can perform an $\mathcal{S}$ transformation on $Z^h_0$ in which the phase factors are evaluated on the residual sublattice invariant under the twist $h$, and thus are trivial elements of $H^1( \Lambda, U(1))$ satisfying the group composition law 
\be
\label{grouplaw}
U(g^{p_1 + p_2},\alpha_{inv.} ) = U(g^{p_1},\alpha_{inv.} ) U(g^{p_2},\alpha_{inv.} ),
\ee
where $\alpha_{inv.}$ refer to a momentum vector in the invariant sublattice. In particular, for asymmetric twists which are trivial in one chiral sector, this implies that for $\mathbb{Z}_{N}$ orbifolds where $N$ is odd, assuming that the consistency conditions \eqref{coboundaryconditions1}-\eqref{coboundaryconditions2} can be solved, there is no non-trivial twist phase factors appearing in $Z^h_0$.

Thus far, our discussion holds for a generic orbifold whether it is asymmetric or not.  For symmetric orbifolds, the $O(d,d;\mathbb{Z})$ element is some geometric $GL(d,\mathbb{Z})$ transformation of the toroidal basis. If the Kalb-Ramond B-field is zero, we find (see Section~\ref{subsubsec:fusionrule} for details) that the twist phase factor is trivial. Suppose now we turn on a B-field $B_0$ that commutes with the geometric twist as in
\be
\label{Bfieldsym}
\theta B_0 \theta^T - B_0 = 0,
\ee
then we still have the same $O(d,d;\mathbb{Z})$ twist. Further, let us perform a gauge transformation 
by shifting $B_0$ with an antisymmetric integral matrix $\delta B$ which does not obey \eqref{Bfieldsym}. This is equivalent to moving to another T-dual frame where the same twist $\theta$ is no longer a geometric one but still acts symmetrically. If we compute the twist phase factor, it turns out that it depends precisely on the LHS of \eqref{Bfieldsym}, with $B_0$ replaced by $\delta B$. Thus, even for symmetric orbifolds, this twist phase factor is not trivial for those which are non-geometric. These non-geometric orbifolds are simple to describe since they are related to the corresponding geometric ones by a suitable shift in the B-field, yet they furnish an explicit class of examples where the twist phase factors are non-trivial and which the non-geometry of the background is precisely understood as arising from a gauge transformation of the B-field in the original geometric orbifold. Generally, for asymmetric orbifolds, the twist phase factor is not trivial even in the case of vanishing B-field. 

In the subsequent sections, we shall solve for the phase factors explicitly for a class of asymmetric toroidal orbifolds in which the twist is trivial in the left-moving sector, and discuss how they appear in the partition traces by evaluating them on the residual sublattices in the partition traces $Z^g_0$. We find that while their inclusion does not completely guarantee level-matching as orbifold CFTs on their own, in all the cases that we consider, their presence ensures that 
\be
\label{levelMatching}
Z^0_h (\tau + N_h) = e^{i\delta} Z^0_h (\tau),
\ee
where $N_h$ is the order of the twist $h$ and $\delta$ is some real constant. In a broader sense, these phase factors arise as necessary conditions for T-duality to be an automorphism of the operator algebra yet they are not always sufficient for a consistent orbifold construction. Nonetheless, the phase factor $\delta$ should be taken into account together with other possible similar factors when we tensor the orbifold CFT with other CFTs like that of twisted fermions, shift orbifolds, etc. In Appendix \ref{sec:OnMod}, we provide a review of the modular transformation properties of chiral bosonic and fermionic blocks capturing the oscillators' degrees of freedom. They transform like in \eqref{levelMatching}, while the phase factors ensure the bosonic lattice sums also transform likewise.

On this note, we should also mention that in \cite{Freed:1987qk}, level-matching conditions are explained to be the vanishing of certain characteristic classes in the orbifold group cohomology. In \cite{Freed:1987qk}, the analysis pertains to symmetric orbifolds of the heterotic string and thus only chiral fermionic partition traces are taken into account, since the anomalous factors of the left- and right-bosonic chiral blocks should cancel each other and there is no residual bosonic lattice sum (apart from the internal 16-dimensional lattice). It would be interesting to furnish an equivariant geometric understanding of the modular covariance of asymmetric orbifolds by studying how the methods of \cite{Freed:1987qk} extend to twist phase factor-refined lattice sums. 

To summarize, one can decompose the stringy Hilbert space into eigenspaces of the T-duality symmetry operation, and demand that the operator algebra preserves the symmetry in such a basis. Preserving mutual locality of the vertex operators in the untwisted sector leads to non-trivial constraints among the two-cocycles and phase factors that accompany the T-duality operations. These constraints can be interpreted as solving for 
the ratio $\epsilon (\alpha, \beta)/ \epsilon (g(\alpha), g(\beta))$ to be a trivial class of $H^2 (\Lambda, U(1))$, subject to certain orbifold group action-dependent constraints for the twist phase factors (which are the one-cochains). When evaluated upon the invariant sublattices, the twist phase factors are trivial elements of $H^1 (\Lambda, U(1))$, and their appearances in the partition traces preserve the modular covariance of the orbifold CFT, up to a constant phase factor as in \eqref{levelMatching}. In the following sections, we will provide various illustrations mainly using asymmetric orbifolds of root lattices as working examples.

\section{The simplest T-fold}
\label{sec:T-fold}

As a warm-up, we first study a simple asymmetric orbifold $S^1/\mathbb{Z}_2$, where the $\mathbb{Z}_2$ acts as a chiral reflection on the right-movers. This twisted circle compactification is a T-fold because the twist  
\be
\label{Sreflection}
X_L \rightarrow X_L,\qquad X_R \rightarrow -X_R
\ee
is the only non-trivial element of $O(1,1;\mathbb{Z})$. For this twisted compactification, the circle radius must be frozen at the self-dual point, an important feature that distinguishes it from its symmetric counterpart where the $\mathbb{Z}_2$ acts as a geometric reflection. In the latter, after the orbifold identification, the $S^1$ becomes a finite interval of which end-points are fixed points of the $\mathbb{Z}_2$, leaving no restriction on the original radius modulus.

For the T-fold of \eqref{Sreflection}, the partition traces were proposed in \cite{Simeon} to read 
\bea
\label{Tfoldpartitiontraces}
Z^0_0 (\tau) &=& \frac{1}{|\eta(\tau)|^2} \sum_{n,w} q^{\frac{1}{4}(n+w)^2} \bar{q}^{\frac{1}{4}(n-w)^2} \cr
Z^1_0 (\tau) &=& \frac{1}{\eta(\tau)} \left[  \bar{q}^{-\frac{1}{24}} \prod_n (1+\bar{q}^n )^{-1} \right] \sum_m (-1)^{m^2} q^{m^2} \cr
Z^0_1 (\tau) &=& \frac{1}{\eta(\tau)} \left[  \bar{q}^{\frac{1}{48}} \prod_n (1-\bar{q}^{n-\frac{1}{2}} )^{-1} \right] \sum_m  q^{\frac{1}{4}(m-\frac{1}{2})^2} \cr
Z^1_1 (\tau) &=& \frac{1}{\eta(\tau)} \left[  \bar{q}^{\frac{1}{48}} \prod_n (1+\bar{q}^{n-\frac{1}{2}} )^{-1} \right] 
\sum_m  e^{\frac{\pi i}{2}(m^2 - m)}   q^{\frac{1}{4}(m-\frac{1}{2})^2}.
\eea
In the above form, these partition traces satisfy the modular covariance \eqref{Ztrace} and there are two features in \eqref{Tfoldpartitiontraces} which are absent in the symmetric orbifold case. The first is that the T-duality operator that is inserted in the sum does not merely switch the winding $(w)$ and momenta $(n)$ but has an additional $\mathbb{Z}_2$ phase factor. Secondly, the momenta in the twisted sector are different from what we would naively expect. Since T-duality switches $n \leftrightarrow w$, the left-moving momenta modes vanish and we are left with the right-moving ones but $P_R \neq n+w = 2n$, and are instead quantized as $m-\frac{1}{2}$ for some integer $m$.  We now proceed to understand the form of the partition traces in \eqref{Tfoldpartitiontraces} in detail. First, we shall elaborate on the cocycle factors and twist phase factors for the simplest T-fold. From \eqref{cocycles}, we shall take the cocycles to be of the form
\be
\label{cocyclesgen}
\hat{C}(\alpha ) =  e^{i\frac{\pi}{2}\zeta_L (\alpha) \hat{P}_L + i \frac{\pi}{2} \zeta_R (\alpha) \hat{P}_R} e^{i\alpha_L \hat{X}_L + i \alpha_R \hat{X}_R}.
\ee
Since they furnish a projective representation  (see eqn. \eqref{projectivecondition}) of the Narain momenta lattice, they are of the form 
\be
\zeta_L = m^R_L \alpha_R - m^L_L \alpha_L,\,\,\, \zeta_R = m^R_R \alpha_R - m^L_R \alpha_L,
\ee
and from \eqref{NCphase}, we obtain the condition
\be
m^R_L + m^L_R = \pm 1.
\ee
As mentioned earlier, there is a gauge degree of freedom corresponding to 
$$
\hat{C}(\alpha) \rightarrow e^{i\delta(\alpha)} \hat{C}(\alpha),\,\,\, \epsilon (\alpha,\beta) \rightarrow
e^{i\left[ \delta(\alpha) + \delta(\beta) - \delta(\alpha + \beta)  \right]} \epsilon(\alpha,\beta).
$$ 
The phase $\delta(\alpha)$ may be fixed by imposing the hermiticity condition $\hat{C}^\dagger (\alpha) = \hat{C}(-\alpha)$, and if so desired, we can compute $\delta(\alpha)$ to read
\be
\label{phaseSimplest}
\delta (\alpha) = \frac{\pi}{4} \left( m^L_L \alpha^2_L - m^R_R \alpha^2_R + (m^L_R - m^R_L) \alpha_L \alpha_R   \right).
\ee
We note that in \cite{Simeon}, the choice of $m^R_L = m^L_L = m^R_R = m^L_R = \frac{1}{2}$ was made
in \eqref{phaseSimplest}. Now, following \eqref{prelimvertex} and \eqref{fusionrule}, we now consider the OPE of two vertex operators labeled by $\mathbb{Z}_2$ indices $a,b$.
\bea
\label{OPEuntwisted}
V(\alpha,z_1)_{[a]} V(\beta,z_2)_{[b]} &=& \frac{1}{4} \left[ V(\alpha, z_1) + e^{-i\pi a} \hat{g} \circ V(\alpha, z_1) \right] \times \left[ V(\beta, z_2) + e^{-i\pi b} \hat{g} \circ V(\beta, z_2) \right] \cr
= \frac{1}{4} \Bigg[  V(\alpha, z_1) \times V(\beta, z_2) &+& e^{-i\pi(a+b)}\left( \hat{g} \circ V(\alpha,z_1) \right) \times \left( \hat{g} \circ V(\beta,z_2) \right) \cr
&+& e^{-i\pi a}\left( \hat{g} \circ V(\alpha,z_1) \right) \times
V(\beta,z_2) + e^{-i\pi b} V(\alpha,z_1) \times \left( \hat{g} \circ V(\beta,z_2) \right) \Bigg] \cr
= \frac{1}{4} (\delta z)^{\frac{1}{2} k^\alpha_R k^\beta_R} (\overline{\delta z})^{\frac{1}{2} k^\alpha_L k^\beta_L} &\Bigg[& 
\epsilon(\alpha,\beta) \hat{C}(\alpha + \beta) + e^{-i\pi (a+b)} \epsilon( t(\alpha), t(\beta)) t_\alpha t_\beta \hat{C}(t(\alpha+\beta)) \Bigg]\cr
+\frac{1}{4} (\delta z)^{\frac{1}{2} \alpha_R \beta_R} (\overline{\delta z})^{-\frac{1}{2} \alpha_L \beta_L} &\Bigg[& e^{-i \pi a} g_\alpha
\epsilon(g(\alpha),\beta) \hat{C}(g(\alpha) + \beta) + e^{-i\pi b} g_\beta \epsilon( \alpha, g(\beta)) \hat{C}(\alpha+g(\beta)) \Bigg], \nonumber \\
\eea
where $g(\alpha)$ are the twisted momenta, and $g_\alpha$ are the $U(1)$ phases that appear when $\hat{g}$ acts on the cocycles. We seek conditions on these phases for the fusion rule \eqref{fusionrule} to be preserved, and this should be done for each bracketed expression in the last line of \eqref{OPEuntwisted}. For the first, we have the constraint
\bea
&&\epsilon(\alpha,\beta) g_{\alpha+\beta} \hat{C}(g(\alpha + \beta))  + e^{-i\pi (a+b)} \epsilon(g(\alpha), g(\beta)) g_\alpha g_\beta g_{g(\alpha + \beta)} \hat{C} (\alpha + \beta) \cr
&=& \epsilon(g(\alpha), g(\beta)) g_\alpha g_\beta \hat{C}(g(\alpha + \beta)) + e^{-i\pi (a+b)} \epsilon(\alpha,\beta) \hat{C}(\alpha + \beta),
\eea
from which we obtain
\bea
\label{firstcond}
\epsilon(\alpha,\beta) g_{\alpha+\beta} &=& \epsilon(t(\alpha), g(\beta)) g_\alpha g_\beta \\
\label{secondcond}
\epsilon(\alpha,\beta) &=& \epsilon(g(\alpha),g(\beta)) g_\alpha g_\beta g_{g(\alpha + \beta)}.
\eea
Identical constraints come from the second bracketed term, and \eqref{firstcond}-\eqref{secondcond} yield
\be
g_{\alpha} g_{g(\alpha + \beta)} = 1.
\ee
For the $\mathbb{Z}_2$ orbifold that we are considering, the non-commutativity phase $\epsilon$ reads
\be
\epsilon(\alpha,\beta)= e^{\frac{i\pi}{2}(n_\beta w_\alpha - n_\alpha w_\beta) }.
\ee
Assuming that all the quantum numbers are integer-valued, we find that the most general solution to the phases $g_\alpha$ reads
\be
\label{simplestTfoldtwist}
g_\alpha = e^{i\pi n_\alpha w_\alpha}e^{i\pi c (n_\alpha + w_\alpha)}
\ee
where $c$ can be an arbitrary integer. In \cite{Simeon}, $c$ is set to be zero. This expression reproduces the form of the partition trace $Z^1_0$. Note that when $\hat{g}$ is inserted in the partition function, it forces all contributing states to have equal momenta and winding numbers, and thus $e^{i\pi c (n_\alpha + w_\alpha)}=1$.

Finally, let us comment on the twist phase factor in the twisted sector. 
We have seen that in the untwisted sector, preserving the symmetry of the operator algebra leads to a nice relationship between cocycle factors and the twist phase factors. 
In the twisted sector, there exists the notion of a twisted vertex operator ${}^\sigma V$ corresponding to the states in the twisted sector. Since two twisted vertex operators close onto an untwisted one, so one also needs the notion of an untwisted vertex operator ${}^uV$ in the twisted sector, with the OPE relation
\be
\label{symtwisted}
{}^uV \times {}^\sigma V \sim {}^\sigma V.
\ee
Like in the untwisted sector, one needs to ensure the mutual locality of the untwisted vertex operators with suitable inclusion of cocycles for ${}^u V$ too. This would be treated carefully in Section \ref{subsubsec:UntwistedVOA}. For the simplest T-fold, ${}^uV$ creates a state of vanishing left-moving momentum and right-moving momentum of an even integer say $2n$, whereas ${}^\sigma V$ creates one of vanishing left-moving momentum and right-moving momentum quantized as $\mathbb{Z} - \frac{1}{2}$. Earlier we have derived the twist phase factor \eqref{simplestTfoldtwist} in the untwisted sector which yields the twist eigenvalue of ${}^u V$ to be $e^{\pi in^2}$.
Demanding the twist eigenvalue of twisted state on the RHS of \eqref{symtwisted} to be identical to the product of $e^{\pi i n^2}$ and that of ${}^\sigma V$ on the LHS of \eqref{symtwisted} then leads to 
the form of the phase factor in the twisted sector (see last line of eqn. \eqref{Tfoldpartitiontraces}) as predicted by modular covariance of the partition traces. Now, let us proceed to apply these observations to general asymmetric toroidal orbifolds.

\section{Asymmetric Toroidal Orbifolds}
\label{sec:Asym}

We can write the left and right momentum zero modes in a $D$-dimensional toroidal background as
\be
P^R_i = n_i - \left( B_{ij} + G_{ij} \right) m^j,\qquad P^L_i = n_i - \left( B_{ij} - G_{ij} \right) m^j
\ee
where $n_i, m^j$ are integral momenta and winding numbers,
$\{G,B \}$ are the $D^2$ metric and B-field moduli respectively.
Under any $O(d,d;\mathbb{Z})$ element $\mathcal{O}$, the $O(d,d;\mathbb{R})/(O(d;\mathbb{R})\times O(d;\mathbb{R}))$ coset representative $\mathcal{G}$ and quantum numbers transform as
\be
\mathcal{G} \equiv \left( \begin{array}{cc} 
G-BG^{-1}B & BG^{-1}  \\ 
-G^{-1}B & G^{-1} \end{array} \right),\,\,\mathcal{G} \rightarrow \mathcal{O}^{-1} \mathcal{G} \mathcal{O}^{T^{-1}},\,\,\,
\left( \begin{array}{c} m \\ n \end{array} \right) = \mathcal{O}^T \left( \begin{array}{c} m\\ n \end{array} \right),\,\,\,
\mathcal{O}\equiv \left( \begin{array}{cc} 
a & b  \\ 
c & d \end{array} \right).
\ee
We now construct an orbifold of the toroidal background by adopting $\mathcal{O}$ as the twist. Further, we wish to restrict ourselves to cases where the twist is realized as independent linear transformations of the left and right-moving momenta, i.e.
\be
\label{leftrighttwists}
P^{R}_i (n,m) \rightarrow \left( \theta^{R} \right)^k_i P^{R}_k ,\,\,\,P^{L}_i (n,m) \rightarrow \left( \theta^{L} \right)^k_i P^{L}_k (n,m)
\ee
We find that for \eqref{leftrighttwists} to hold, the background moduli must be self-dual. Defining $E=G+B$ upon which T-duality is realized as a fractional linear transformation, the self-duality condition and twists read
\bea
\label{selfdual}
&&E = \frac{a E +b}{c E + d},\qquad  G = \theta G \theta^T\\
&&\theta^R = d^T - E c^T = (Ea^T - b^T)E^{-1},\,\,
\theta^L = d^T + E^T c^T = (E^T a^T + b^T)E^{T^{-1}}.\,\,
\eea
The self-duality condition is equivalent to keeping the conformal weights of $P^2_{L,R}$ invariant or simply the metric-preserving condition for both left and right sectors. For later purposes, we find it convenient to parametrize $\mathcal{O}$ in terms of $\theta$. Defining 
\be
\theta_\pm \equiv \theta_L \pm \theta_R,
\ee
the element $\mathcal{O}$ can be written as\footnote{One can check that our expression differs from equations 14-16 of \cite{Erler} purely due to a difference in the normalization of the moduli.}
\be
\label{Ointermsoftwists}
\mathcal{O}\equiv \frac{1}{2}\left( \begin{array}{cc} 
\theta^{-1}_L+ \theta^{-1}_R + B\theta^T_- G^{-1} & -(\theta_R^{-1}+\theta_L^{-1} + B\theta^T_- G^{-1})B+(G\theta^T_- + B\theta^T_+) \\ 
\theta^T_- G^{-1} & -\theta^T_- G^{-1}B + \theta^T_+ \end{array} \right),
\ee
from which we see that we obtain an asymmetric orbifold whenever we have a non-zero $c$ or $\theta_- = 2Gc^T$. 

\subsection{T-duals of geometric twists in $T^2$ compactification}
\label{subsec:T-duals}

$\mathbb{Z}_{3,4,6}$ asymmetric orbifolds can be realized in each of two T-dual frames of the respective symmetric $\mathbb{Z}_{3,4,6}$ orbifolds by twisting with their crystallographic symmetries. 
In each frame, both the left and right twists are rotations (they turn out to be inverses of each other), so there are no surviving zero modes in the twisted sectors. Let us first describe the well-understood symmetric orbifolds. In the lattice basis, the $\mathbb{Z}_N$ rotations are realized as $SL(2,\mathbb{Z})$ matrices acting on the complex structure $\tau$.\footnote{In this and the next subsections, we will sometimes let $\tau$ denote the complex structure of the target space torus, whereas for the rest of the paper, $\tau$ typically denotes that of the Euclidean worldsheet torus.}
Now let $\mathcal{O}_g$ denote the geometric $SL(2,\mathbb{Z})$ action. The generator of the orbifold group reads $\left( \begin{array}{cc} \mathcal{O}_g & 0 \\
0 & \mathcal{O}_g^{T^{-1}} \end{array} \right)$. The self-duality conditions lead to the following backgrounds ($\nu$ denotes the order, with the rotation angle being $2\pi/\nu$): 
\bea
&&(i)\tau = i, \mathcal{O}_g =
\left( \begin{array}{cc} 0 & -1 \\ 1 & 0 \end{array} \right),
E = \left( \begin{array}{cc}  V & B \\ -B & V \end{array} \right),
\,\,\,\nu = 4
\\
&&(ii)\tau = e^{\frac{i\pi}{3}}, \mathcal{O}_g =
\left( \begin{array}{cc} 0 & -1 \\ 1 & -1 \end{array} \right),
E = \frac{V}{\sqrt{3}} \left( \begin{array}{cc}  2 & 1+\frac{\sqrt{3}B}{V} \\ 1-\frac{\sqrt{3}B}{V}  & 2 \end{array} \right),
\,\,\, \nu = 3 \\
&&(ii)\tau = e^{\frac{i\pi}{3}}, \mathcal{O}_g =
\left( \begin{array}{cc} 1 & -1 \\ 1 & 0 \end{array} \right),
E = \frac{V}{\sqrt{3}} \left( \begin{array}{cc}  2 & 1+\frac{\sqrt{3}B}{V} \\ 1-\frac{\sqrt{3}B}{V}  & 2 \end{array} \right),
\,\,\, \nu = 6 
\eea
In these orbifolds, the Kahler modulus $\rho = B + iV$ is left unfixed. Let us now perform T-dualities along each of the two cycles of $T^2$. For some $\mathcal{O}_g = \left( \begin{array}{cc} \alpha & \beta \\ \chi & \delta \end{array} \right)$, the orbifold twist now possibly develops an asymmetric component. Letting $\mathcal{O}^{(1),(2)}$ denote the generator after T-dualizing along the $X^{1,2}$ directions,
\be
\label{OGeoT}
\mathcal{O}^{(1)} = \left( \begin{array}{cc}  \delta \mathds{1}_2 & -i\chi \sigma_2 \\ i\beta \sigma_2  & \alpha \mathds{1}_2 \end{array} \right), \qquad
\mathcal{O}^{(2)} = \left( \begin{array}{cc}  \alpha \mathds{1}_2 & i\beta \sigma_2 \\ -i\chi \sigma_2  & \delta \mathds{1}_2 \end{array} \right).
\ee 
The background moduli are the T-duals of the former, and are rather simple to describe as follows.
\bea
&&\textrm{T-duality along $X^1$:    } \tau \leftrightarrow -\bar{\rho} \\
&&\textrm{T-duality along $X^2$:    } \tau \rightarrow \frac{\rho}{|\rho|^2},\, \rho \rightarrow \frac{\tau}{|\tau|^2}:
\eea
It is useful to write \eqref{OGeoT} in terms of the left and right twists $\theta^{L,R}$ via \eqref{Ointermsoftwists}. In the T-dual frame of each of the cases (i)-(iii), we find that the left and right twists ($\theta^{L,R}$) are inverses of each other, and are $SL(2,\mathbb{R})$ transformations in general. They are thus asymmetric orbifolds. The twists in both frames are related by 
$
\theta_{T_1} \theta^T_{T_2} = 1.
$
Below, we display $\theta^R = (\theta^L)^{-1}$ and their eigenvalues $\lambda$ for each case. Note that the moduli parameters are the original ones before we perform the respective T-dualities.
\bea
&&\text{(i)}\tau = i, \theta^R = \frac{1}{V}
\left( \begin{array}{cc} -B & -1 \\ B^2+V^2 & B \end{array} \right), \lambda = \pm i
\\
&&\text{(ii)}\tau = e^{\frac{i\pi}{3}}, (\theta^R)^{-1} =
\left( \begin{array}{cc} -\frac{1}{2}+\frac{\sqrt{3}B}{2V} & -\frac{\sqrt{3}(B^2+V^2)}{2V} \\ \frac{\sqrt{3}}{2V} & -\frac{1}{2}-\frac{\sqrt{3}B}{2V} \end{array} \right), \lambda = e^{\pm \frac{2\pi i}{3}}
\\
&&\text{(iii)}\tau = e^{\frac{i\pi}{3}}, \theta^R =
\left( \begin{array}{cc} \frac{1}{2} -\frac{\sqrt{3}B}{2V} & -\frac{\sqrt{3}}{2V} \\ \frac{\sqrt{3}(B^2+V^2)}{2V}  & \frac{1}{2}+\frac{\sqrt{3}B}{2V} \end{array} \right), \lambda = e^{\pm \frac{\pi i}{3}}
\eea
Finally we note that if we T-dualize along both toroidal directions, we obtain a symmetric orbifold with the inverse identification. Since both left and right-movers are rotated (in opposite directions), there are no surviving zero modes, and the background moduli do not appear in the expression of the one-loop partition function. 

\subsection{Constructing chiral asymmetric orbifolds}
\label{subsec:Constructing}

In the following, we shall construct asymmetric orbifolds which are not T-duals of geometric ones. As a start, we restrict ourselves to those in which the orbifold actions are of the form $\theta_{L,R} \in \mathbb{Z}_N, \theta_{R,L} = \mathds{1}_2$. We shall henceforth refer to this special class of orbifolds as `chiral' asymmetric orbifolds in this paper.

Letting either the left or right action to be trivial, the respective $O(d,d,\mathbb{Z})$ element then reads 
\bea
\label{leftO}
&&(\textrm{i}) \mathcal{O}_L = \left( \begin{array}{cc} \mathds{1}_2 + Ec & EcE^T \\ c & \mathds{1}_2 + cE^T \end{array} \right), \qquad \theta_L = \mathds{1}_2 + 2Gc^T, \theta_R = \mathds{1}_2, \\
\label{rightO}
&&(\textrm{ii}) \mathcal{O}_R = \left( \begin{array}{cc} \mathds{1}_2 + E^Tc & -E^TcE \\ -c & \mathds{1}_2 + cE \end{array} \right), \qquad \theta_R = \mathds{1}_2 + 2Gc^T, \theta_L = \mathds{1}_2. 
\eea
Given a metric-preserving twist $\theta$, there is no further restriction on the B-field apart from the requirement that the matrix elements of $\mathcal{O}$ are integer-valued. Starting from any $\mathcal{O}_{L,R}$ above, we can construct a symmetric orbifold by taking the product 
\be
\label{symTwist}
\mathcal{O}_{sym.} = \mathcal{O}_L \mathcal{O}_R = \left( \begin{array}{cc} \mathds{1}_2 + 2Gc & 
-E^T c E + E c E^T \\ 0 & \mathds{1}_2 + 2cG \end{array} \right) = \left( \begin{array}{cc} \theta^{-1}
& B\theta^T - \theta^{-1}B \\ 0 & \theta^T \end{array} \right).
\ee
This element correponds to a symmetric orbifold with twist $\theta = \mathds{1}_2 + 2Gc^T, \theta^{-1} = \mathds{1}_2 + 2Gc$. From \eqref{symTwist}, we see that any asymmetric T-fold of the above form must descend from a symmetric orbifold, and that an element of the symmetric orbifold group does not necessarily belong to the $GL(d;\mathbb{Z})$ subgroup since we can perform a integral shift of the $B$ field. The converse is however not true. From a generic symmetric orbifold, one cannot always take the `square root' to obtain an asymmetric one with the original twist acting on the right or left moving momenta. 

In the following, we will describe some two-dimensional examples followed by a more systematic description of appropriate moduli for higher-dimensional tori. Let us work with $\tau = i$ and $\tau = e^{\frac{i\pi}{3} }$. For these moduli, we can only find the following asymmetric orbifolds of the form above (where the orbifold generator in one chiral sector is trivial). We display the twists and their corresponding $O(2,2;\mathbb{Z})$ elements in Tables 1 and 2 below.

\begin{table}[h]\centering \ra{1.1} 
\begin{tabular}{c|c|c|c}
\toprule 
$\theta_R $ & $\left( \begin{array}{cc} 1 & 0 \\ 0 & -1 \end{array} \right)$  & $\left( \begin{array}{cc} -1 & 0 \\ 0 & 1 \end{array} \right)$ & $\left( \begin{array}{cc} -1 & 0 \\ 0 & -1 \end{array} \right)$
\\
$\mathcal{O}_R$ & 
$\left( \begin{array}{cccc} 
1 & 0 & 0 & 0 \\ 
0 & 0 & 0 & 1\\
0 & 0 & 1 &0\\
0 & 1 & 0 & 0\end{array} \right)$ &
$\left( \begin{array}{cccc} 
0 & 0 & 1 & 0 \\ 
0 & 1 & 0 & 0\\
1 & 0 & 0 &0\\
0 & 0 & 0 & 1\end{array} \right)$ &
$\left( \begin{array}{cccc} 
0 & 0 & 1 & 0 \\ 
0 & 0 & 0 & 1\\
1 & 0 & 0 &0\\
0 & 1 & 0 & 0\end{array} \right)$ 
\\
\bottomrule
\end{tabular} 
\caption{$\tau=\rho=i$. Note that contrary to the symmetric case, there is no $\mathbb{Z}_4$ element. Together with the identity, the three $\theta_R$ form the group $O(1,1;\mathbb{Z})$ or equivalently the Weyl group of $A_1 \times A_1$.} 
\end{table}

\begin{table}[h]\centering \ra{1.1} 
\begin{tabular}{c|c|c|c|c}
\toprule 
$\theta_R $ & $\left( \begin{array}{cc} -1 & 0 \\ -1 & 1 \end{array} \right)$  & $\left( \begin{array}{cc} 0 & 1 \\ 1 & 0 \end{array} \right)$ & $\left( \begin{array}{cc} 1 & -1 \\ 0 & -1 \end{array} \right)$ & $\left( \begin{array}{cc} 0 & -1 \\ 1 & -1 \end{array} \right)$
\\
$\mathcal{O}_R$ & 
$\left( \begin{array}{cccc} 
1 & 0 & 0 & 0 \\ 
0 & 0 & 0 & 1\\
0 & 0 & 1 &0\\
0 & 1 & 0 & 0\end{array} \right)$ &
$\left( \begin{array}{cccc} 
0 & 1 & 1 & 0 \\ 
0 & 1 & 0 & 0\\
1 & -1 & 0 & 0\\
-1 & 1 & 1 & 1\end{array} \right)$ &
$\left( \begin{array}{cccc} 
0 & 0 & 1 & 1 \\ 
-1 & 1 & 1 & 1\\
1 & 0 & 0 & -1\\
0 & 0 & 0 & 1\end{array} \right)$ &
$\left( \begin{array}{cccc} 
0 & 1 & 1 & 0 \\ 
-1 & 1 & 1 & 1\\
1 & -1 & 0 & 0\\
0 & 1 & 0 & 0\end{array} \right)$ 
\\
\bottomrule
\end{tabular} 
\caption{$\tau=\rho=e^{i\pi/3}$. Note that contrary to the symmetric case, there is no $\mathbb{Z}_6$ element. Each element generates a $\mathbb{Z}_2$ action except for the last which generates a $\mathbb{Z}_3$ action. We have omitted the inverse of the last entry. Together with the identity, they yield the discrete group $S_3$ - the Weyl group of $A_2$.} 
\end{table}

For each case in Tables 1 and 2, there is a corresponding orbifold in which the same twist defines a non-trivial $\theta_L$ instead of $\theta_R$. Worldsheet parity symmetry yields the corresponding commuting $\mathcal{O}_L$ element which one can read off from \eqref{leftO}. Thus, we can use them as building blocks to generate independent $\mathbb{Z}_N$ actions on the left and right-moving sectors. We can of course consider their $\mathcal{O}(2,2;\mathbb{Z})$ orbits. Let $\mathcal{T}$ be the T-duality element which defines the dual frame. In our conventions, the new orbifold element $\tilde{\mathcal{O}}$ now reads
\be
\tilde{\mathcal{O}} = \mathcal{T}^{-1} \mathcal{O} \mathcal{T},\qquad
\tilde{E} = \mathcal{T}^{-1} \circ E,\,\,\,\, \left( \begin{array}{c} \tilde{m} \\ \tilde{n} \end{array} \right) = 
\mathcal{T}^T \left( \begin{array}{c} \tilde{m} \\ \tilde{n} \end{array} \right).
\ee
It is crucial to note that this does not exhaust all the possibilities of orbifold actions which are asymmetric. We have encountered the class of T-duals of the geometric rotational orbifolds, which in particular contains asymmetric $\mathbb{Z}_4$ twist for $\tau = i$ and $\mathbb{Z}_6$ twist for $\tau = e^{\frac{i\pi}{3}}$. The left- and right-actions are however not independent. Another class of asymmetric orbifolds can be constructed in which there is a $\mathbb{Z}^{(\tau)}_N \times \mathbb{Z}^{(\rho)}_M$ action arising from the $SL(2,Z)_\tau \times SL(2,Z)_\rho$ subgroup. Factorized T-dualities exchange the complex structure and the Kahler modulus, and from the T-duals of the geometric orbifolds, it is easy to deduce that in terms of their right- and left twists, 
$\mathbb{Z}^{(\tau)}_N$ is generated by $\theta_L = \theta_R$ while $\mathbb{Z}^{(\rho)}_M$ is generated by $\theta_L = \theta^{-1}_R$. 

Now, the $T^2$ orbifolds we have considered in Tables 1 and 2 can be equivalently described as
orbifolds of the root lattices $A_1 \times A_1$ and $A_2$ by their Weyl groups. For their geometric orbifold counterparts, there are more possible orbifolds, since automorphisms of these root lattices do not just comprise of Weyl reflections but also outer automorphisms which are geometrically realized as discrete symmetries of the respective Lie algebras' Dynkin diagrams. For example, the $\mathbb{Z}_4$ and $\mathbb{Z}_6$ orbifold elements cannot be embedded in $O(2,2;\mathbb{Z})$ as chiral asymmetric twists, but they do act legitimately in symmetric orbifolds of the same tori of which they are associated with the $\mathbb{Z}_2$ outer automorphism group elements of the respective root lattices. This simple fact prompts a broader question, namely, for a generic $d$-dimensional chiral asymmetric orbifold of the torus of the form $\Lambda_R/\mathcal{G}$, where $\Lambda_R$ is a simple Lie algebra's root lattice and $\mathcal{G}$ some symmetry group, is the set of embeddable $\mathcal{G}$ always equivalent to its Weyl group? 

Let us first focus on root lattices of simply laced algebras of which rank is equal to the torus dimensionality.
From \eqref{leftO}, it is clear that we require both $c = \frac{1}{2} (\theta^T_L - \mathds{1}) G^{-1}$ and $E=G+B$ to be integral. We can endow the torus with a metric 
$$G_{ij} = k (\alpha_i, \alpha_j),$$
where $\alpha_i$ are the simple roots and $k$ is some suitable constant which we shall fix shortly. Recall that the Weyl group is a Coxeter group generated by Weyl reflections about the hyperplanes orthogonal to each simple root $\alpha_m$. In the language of \eqref{leftO}, they are realized as asymmetric twists $\theta^{(m)}$ of the form 
\be
\theta^{(m)}_{ij} = \delta_{ij} - C_{im} \delta_{mj}
\ee
where $m$ is not summed over, and $C_{im} \equiv 2(\alpha_i, \alpha_m)/(\alpha_m, \alpha_m)$ is the Cartan matrix. Normalizing all the roots' lengths to be two, we then have
\be
c^j_i = -\frac{1}{2k}\delta_{m}^j \delta_{mi}.
\ee
After also taking into account the integrality of $E$, it is clear that we should set $k=1/2$ which give us the self-dual moduli
\be
\label{simplylacedmoduli}
G_{ij} = \frac{1}{2} C_{ij},\,\,\, B_{ij} =  G_{ij}\,\,\forall\,i>j.
\ee
Thus, all Weyl reflections can be embedded in $O(d,d;\mathbb{Z})$ as asymmetric twists for the special moduli \eqref{simplylacedmoduli}. What about the outer automorphism groups? For $A_1 \times A_1$ and $A_2$, we have seen that they cannot be be embedded in the T-duality group as chiral twists. There are not many of them and we can quickly check their relevance.


\begin{figure}[h]

\begin{tikzpicture}
\begin{scope}
[roots/.style={circle,draw=black!100,fill=white!20,thick,inner sep=0pt,minimum size=5mm}, bend angle=25, pre/.style={<->,shorten <=1pt, shorten >=1pt, semithick}]

\node at (0,0) [roots] (a1) {};
\node (dot) at (3.4,0) {$\ldots \, \ldots\, \ldots\, \ldots$};
\node at (7,0) [roots] (an) {}
edge [pre, bend left] (a1);
\node at (5.7,0) [roots] (an1) {}
edge [-] (dot)
edge [-] (an);
\node at (1.3,0) [roots] (a2) {}
edge [-] (a1)
edge [-] (dot)
edge [pre, bend right] (an1);
\node (captionA) at (3.5,-1.6) {(a)$A_N$};
\end{scope}

\begin{scope}
[roots/.style={circle,draw=black!100,fill=white!20,thick,inner sep=0pt,minimum size=5mm}, bend angle=35, pre/.style={<->,shorten <=1pt, shorten >=1pt, semithick}]

\node at (10,0) [roots] (e61) {};
\node at (11.3,0) [roots] (e62) {}
edge [-] (e61);
\node at (12.6,0) [roots] (e63) {}
edge [-] (e62);
\node at (12.6,1.3) [roots] (e64) {}
edge [-] (e63);
\node at (13.9,0) [roots] (e65) {}
edge [-] (e63)
edge [pre, bend left] (e62);
\node at (15.2,0) [roots] (e66) {}
edge [-] (e65)
edge [pre, bend left] (e61);
\node (captionB) at (12.7,-1.6) {(b)$E_6$};
\end{scope}

\begin{scope}
[roots/.style={circle,draw=black!100,fill=white!20,thick,inner sep=0pt,minimum size=5mm}, bend angle=35, pre/.style={<->,shorten <=1pt, shorten >=1pt, semithick}]

\node at (0,-4) [roots] (d1) {};
\node (ddot) at (3.4,-4) {$\ldots \, \ldots\, \ldots\, \ldots$};
\node at (7,-4) [roots] (dn) {};
\node at (7.7,-2.8) [roots] (dw1) {}
edge [-] (dn);
\node at (7.7,-5.2) [roots] (dw2) {}
edge [-] (dn)
edge [pre, bend right] (dw1);
\node at (5.7,-4) [roots] (dn1) {}
edge [-] (ddot)
edge [-] (dn);
\node at (1.3,-4) [roots] (d2) {}
edge [-] (d1)
edge [-] (ddot);
\node (captionC) at (3.7,-5.9) {(c)$D_N,\,\,N > 4$};

\node at (13,-4) [roots] (d4n) {};
\node at (13.7,-2.8) [roots] (d4w1) {}
edge [-] (d4n);
\node at (13.7,-5.2) [roots] (d4w2) {}
edge [-] (d4n)
edge [pre, bend right] (d4w1);
\node at (11.6,-4) [roots] (d4n1) {}
edge [-] (d4n)
edge [pre, bend right] (d4w2)
edge [pre, bend left] (d4w1);
\node (captionC) at (13,-5.9) {(d)$D_4$};
\end{scope}

\end{tikzpicture}

\caption{Outer automorphisms corresponding to discrete symmetries of the Dynkin diagrams. The twist groups are all $\mathbb{Z}/\mathbb{Z}_2$ apart from the triality of $D_4$ in which case the symmetry group is $S_3$.}
\label{fig:outertwists}
\end{figure}
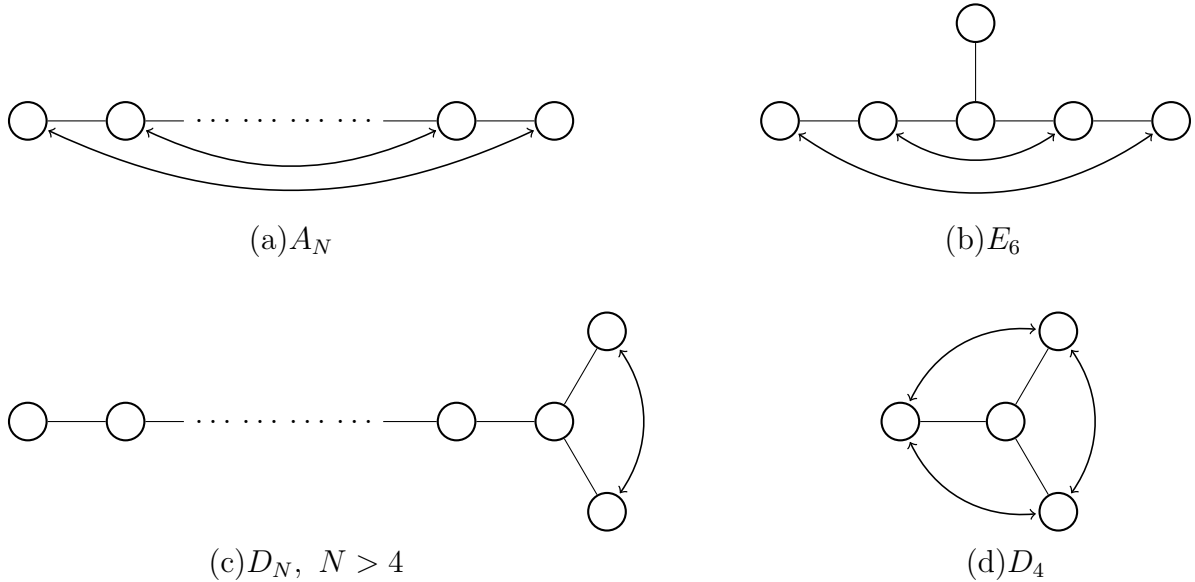

We display these twists in Figure 1 where each outer automorphism descends from a permutation symmetry of the nodes of the respective Dynkin diagrams. Although they are realizable as geometric twists, we checked that all of them unfortunately yield non-integral $c$ in \eqref{leftO}, and thus they cannot be embedded in $O(d,d;\mathbb{Z})$ as chiral twists. With a tad more work, we can extend what we have learnt to the non-simply laced algebras too. The crucial point is to invoke the fact that as root lattices, we have the equivalence
\be
\label{latticeequivalence}
G_2 \sim A_2, \qquad B_N \sim A^N_1, \qquad C_N \sim D_N, \qquad F_4 \sim D_4.
\ee
Each equivalence can be represented by a integral linear transformation matrix $\mathcal{F}$ that maps the simple roots between members of each pair in \eqref{latticeequivalence}. Using an explicit description of the roots, we can derive these transformation matrices straightforwardly and relate between the non-simply laced variables (tilded) and the simply laced ones as follows 
\be
\label{map}
\tilde{G} = \mathcal{F} G \mathcal{F}^T, \qquad \tilde{B} = \mathcal{F} B \mathcal{F}^T,\qquad
\tilde{\theta} = \mathcal{F} \theta \mathcal{F}^{-1}.
\ee
Since the outer automorphism groups of the non-simply laced algebras are trivial by inspection, the Weyl $(\mathcal{W})$ and outer automorphism groups $(\Gamma)$ of the simply laced algebras should map to the Weyl group of the non-simply laced ones $(\tilde{\mathcal{W}})$. Formally, this is captured in an exact sequence
\be
1 \rightarrow \tilde{\mathcal{W}} \rightarrow \mathcal{W} \rightarrow \Gamma \rightarrow 1.
\ee
Thus, for the root lattices of the non-simply laced algebras, not all members of the Weyl group yield a chiral asymmetric orbifold, but only those which related to the simply laced elements by \eqref{map}. For example, for $C_N$, the Weyl group is $S_N \ltimes (\mathbb{Z}_2)^N$, but allowed chiral asymmetric orbifolds elements derive from a smaller group $S_N \ltimes (\mathbb{Z}_2)^{N-1}$ since we have to mod out by $\Gamma = \mathbb{Z}_2$. Similarly, the chiral twists for $F_4$ and $B_N$ yield orbifold groups that lie
in $S_4 \ltimes (\mathbb{Z}_2)^3$ and $(\mathbb{Z}_2)^N$ respectively. 

Also, we wish to point out that there is a subalgebra within the operator algebra of these asymmetric orbifold theories that is isomorphic to Kac-Moody algebras (associated with the loop extension of the finite-dimensional Lie algebras of which roots generate the toroidal lattice). This basically descends from the fact that, without orbifolding, toroidal backgrounds which are root lattices of simply-laced Lie algebras admit such enhanced symmetries, and thus are equivalent to WZW theories based on the same Lie algebras. For the chiral asymmetric orbifolds discussed above, it turns out that our choice of the metric and B-field are compatible with the emergence of these enhanced affine symmetries (see also \cite{Elitzur,Schellekens} for related results). 

Let us briefly review the well-known fact that affine algebras admit vertex operator representations. Recall that the affine currents $J^a$ of conformal dimension one satisfy the OPE
\be
\label{KM1}
J^a (z) J^b (w) \sim \frac{kG^{ab}}{(z-w)^2} + \frac{if^{ab}_c J^c (w)  }{z-w}
\ee
where $k$ is the Kac-Moody level, $G$ is an appropriate Killing form and $f^{abc}$ are the structure constants of the associated finite-dimensional Lie algebra. It turns out one can realize \eqref{KM1} for simply-laced algebras with the theory of a free chiral boson $\Phi(z)$ with suitably normalized zero-mode momenta. Now, let us denote $H^k$ to be the generators of the maximal torus, and $E^\alpha$ to be the raising/lowering operators
associated with some root $\alpha$. To realize \eqref{KM1}, we let the identity to be the extension operator, and 
\be
 H^k (z) = i \partial \Phi^k (z), \,\,\,E^\alpha (z) = V(\alpha, z) = e^{i\alpha_k \Phi^k (z)}, \,\,|\alpha|^2 = 2. 
\ee
And thus we see that the zero mode momenta are selected to lie in the root space of the associated finite simple Lie algebra. This constraint sets up an operator subalgebra that reads 
\bea
\label{toroidalKM1}
H^k (z) E^\alpha (w) &\sim& \frac{\alpha^k E^\alpha (w)}{z-w}, \\
\label{toroidalKM2}
E^\alpha (z) E^\beta (w) &\sim& (z-w)^{(\alpha,\beta)} E^{\alpha + \beta} (w) + (z-w)^{1+(\alpha,\beta)} \alpha_k H^k (w) E^{\alpha + \beta} (w). 
\eea
For a simply-laced algebra, singular terms arise when we have $(\alpha, \beta) = -1$ or $(\alpha,\beta) = -2$ (and thus $\alpha +\beta = 0$). In the former,  the simple pole picks up a negative sign when we exchange
$\alpha \leftrightarrow \beta, z\leftrightarrow w$. Mutual locality thus requires the presence of cocycle factors, of which insertion implies that \eqref{toroidalKM1} and \eqref{toroidalKM2} are equivalent to \eqref{KM1} in the Cartan-Weyl basis. In toroidal compactifications with non-zero Kalb-Ramond B-field, such enhanced affine symmetries can then arise whenever there exists some set of left and/or right momenta which act as roots of some simply-laced algebra, i.e. $|p_{L,R}|^2 =2$. 

In the context of chiral asymmetric orbifolds of toroidal root lattices of the $ADE$ series\footnote{For non-simply-laced algebras, roots of other lengths require addition of free fermions whereas for other levels, it is known that one needs free parafermions.}, chiral twists which kill off all modes in either left or right sector imply that the surviving zero modes read $\alpha^k_{L,R} = \pm 2G^{k}_jm^j$. Since we adopt the metric to be half the Cartan matrix, we arrive at precisely the correct normalization for \eqref{KM1} to be realized.

Finally, we comment on a class of asymmetric orbifolds considered in some papers \cite{Condeescu:Asym,Condeescu:Gsugra}, which are not T-folds. Recall that there is an $O(d;\mathbb{R}) \times O(d;\mathbb{R})$ subgroup in $O(d,d;\mathbb{R})$ which preserves the spectrum. These transformations are symmetries of the theory in the sense that they are tranformations acting on the $O(d,d)$ coset metric $\mathcal{G}$ which preserves the Hamiltonian, but they are not automorphism of the Hilbert space, or in this case, there is no relabelling of winding and momenta numbers which are consistent with quantization. The only elements for which the transformation is an automorphism belong to the subgroup $O(d;\mathbb{Z}) \times O(d;\mathbb{Z})$ as embedded in the T-duality group. In the lattice basis, defining the metric $G$ in terms of the vielbein $G=ee^T$, we have $\theta = eRe^{-1}$, where $R \in O(2,\mathbb{R})$.Explicitly, in our chart 
$$
e= \frac{1}{\sqrt{\tau_2}} \left( \begin{array}{cc} 1 & 0 \\ \tau_1 & \tau_2 \end{array} \right), R= \left( \begin{array}{cc} \cos \theta & \sin \theta \\ -\sin \theta & \cos \theta \end{array} \right), \theta_R = \frac{1}{\tau_2} \left( \begin{array}{cc} -\tau_1 \sin \theta + \tau_2 \cos \theta & \sin \theta \\ -\sin \theta |\tau|^2 & \tau_1 \sin \theta + \tau_2 \cos \theta \end{array} \right).
$$
Thus, we could have an asymmetric $\mathbb{Z}_4$ acting chirally for $\tau=i$ (or $\mathbb{Z}_6$ for $\tau = e^{\frac{i\pi}{3}}$) but these orbifold elements are not contained in $O(2,2,\mathbb{Z})$.

\subsection{2-cocycles and asymmetric twists in toroidal orbifolds}
\label{subsec:Cocycles}

In the simplest T-fold, we have seen that for twisted sectors in which there are surviving zero modes, the mutual locality of the vertex operators leads us to the correct construction of a phase-factor refined orbifold twist. The modular orbit of $Z^1_0 (\tau)$ yields the other two partition traces $Z^1_1 (\tau)$ and $Z^0_1 (\tau)$, revealing how the twisted zero modes are quantized and the twist phase factor in the twisted sector. The former should be compatible with level-matching conditions, and the latter compatible with a fusion rule that furnishes a representation of the orbifold group. Below, we shall explore the universality of these relations for a general asymmetric toroidal orbifold. 

\subsubsection{An expression for the 2-cocycle}
\label{subsubsec:cocycleExpression}

We now present a simple expression for the cocycle. From \eqref{cocyclesgen},
\be
\epsilon (\alpha, \beta) = e^{i\left( \delta(\alpha) + \delta(\beta) - \delta(\alpha+\beta) \right)}
e^{-\frac{i\pi}{2} \left( \alpha_L \cdot \zeta_L (\beta) + \alpha_R \cdot \zeta_R (\beta)      \right)        },
\ee
and thus, to preserve mutual locality of the vertex operators, we need
\be
\beta_L \cdot \zeta_L(\alpha) + \beta_R \cdot \zeta_R (\alpha) - \alpha_L \cdot \zeta_L (\beta) - \alpha_R \cdot \zeta_R (\beta) = \alpha_R \cdot \beta_R - \alpha_L \cdot \beta_L.
\ee
Now we adopt the following ansatz for $\theta_{L,R}$. Restoring the indices,
\be
\zeta^m_L (\alpha) = {}^RY^{mn} \alpha_{Rn} - {}^LY^{mn} \alpha_{Ln},\,\,\, 
\zeta^m_R (\alpha) = {}^RZ^{mn} \alpha_{Rn} - {}^LZ^{mn} \alpha_{Ln}.
\ee
The condition that we are seeking for can thus be written as
\be
\label{localitycondition}
\beta_R \cdot \left( {}^RZ - {}^RZ^T \right) \cdot \alpha_R + \beta_L \cdot \left( {}^LY^T - {}^LY \right) \cdot \alpha_L + \beta_L \cdot \left( {}^RY + {}^LZ^T \right) \cdot \alpha_R -
\beta_R \cdot \left( {}^LZ + {}^RY^T \right) \cdot \alpha_L= \alpha_R \cdot \beta_R - \alpha_L \cdot \beta_L.
\ee
Guided by the simplest T-fold, we find the following solution for $\zeta_{L,R}(\alpha)$ and $\delta(\alpha)$.
\bea
\label{cocycle1}
\zeta_L (\alpha) &=& \frac{1}{2} \left( G^{-1} - G^{-1}BG^{-1} \right) (\alpha_R - \alpha_L), \\
\label{cocycle2}
\zeta_R (\alpha) &=& \frac{1}{2} \left( G^{-1} + G^{-1}BG^{-1} \right) (\alpha_R - \alpha_L), \\
\label{cocycle3}
\delta(\alpha)&=& -\frac{\pi}{4} \left( \alpha_L \cdot \zeta_L(\alpha) + \alpha_R \cdot \zeta_R (\alpha) \right).
\eea
The phase \eqref{cocycle3} was derived by imposing the hermiticity condition for the cocycles, i.e. $\hat{C}^\dagger (\alpha) = \hat{C} (-\alpha)$ which yields $\delta (-\alpha) + \delta (\alpha) = -\frac{\pi}{2} \left( k_L \zeta_L + k_R \zeta_R   \right)$. From \eqref{cocycle1} - \eqref{cocycle3}, we can compute
the non-commutativity phase to read
\be
\epsilon (\alpha, \beta) = e^{\frac{i\pi}{4} \left( \beta_{Lm} G^{mn} \alpha_{Rn} - \beta_{Rm} G^{mn} \alpha_{Ln} - (\alpha_{Rm} - \alpha_{Lm}) B^{mn} (\beta_{Rn} - \beta_{Ln})  \right) }
=e^{\frac{i\pi}{2} (n_\alpha m_\beta - n_\beta m_\alpha)}.
\ee
On the other hand,
we note that the $\hat{C}(\alpha)$ can be expressed in terms of $\epsilon (\alpha,\beta)$ by defining the momentum states created by $\hat{C}(\alpha)$ acting on the vacuum, i.e.
$$
|\alpha \rangle = \hat{C}(\alpha) |0 \rangle,
$$
upon which it is easy to see that 
\be
\label{cocyclesandepsilon}
\hat{C} (\alpha) = \sum_\beta \epsilon (\alpha, \beta) |\alpha + \beta \rangle \langle \beta |.
\ee
As mentioned earlier in Section \ref{sec:Gen},
from the associativity of the OPEs among the vertex operators, one can show that 
\be
\label{cocycleC}
\epsilon(\alpha, \beta + \gamma) \epsilon(\beta, \gamma) = \epsilon(\alpha,\beta)\epsilon(\alpha + \beta,\gamma),
\ee
from which we can interpret $\epsilon(\alpha,\beta)$ as a 2-cocycle of the Narain momenta lattice $\Lambda$. Further, there is an equivalence condition that we should impose that will imply that $\epsilon(\alpha,\beta)$ is a class of the second cohomology group. In \eqref{cocyclesandepsilon} (or \eqref{projectivecondition}), there is a gauge degree of freedom preserving \eqref{NCphase} that corresponds to
\be
\label{phasetransformation}
\hat{C}(\alpha) \rightarrow e^{i\delta(\alpha)} \hat{C}(\alpha), \qquad 
\epsilon (\alpha,\beta) \rightarrow e^{i \left(\delta(\alpha+\beta) - \delta(\alpha) -\delta(\beta) \right)} \epsilon (\alpha,\beta)
\ee
where $\delta(\alpha)$ is some scalar function of $\alpha$. For example, for the chiral asymmetric orbifolds considered earlier in Section \ref{subsec:Constructing} where there are no surviving right-moving momentum zero modes, upon evaluation on the invariant sublattice ($\Lambda^*$), $\alpha_R = 0, \alpha_{Li} = 2G_{ij} m^j = C_{ij}m^j$ and the 2-cocycle reduces to
\be
\epsilon (\alpha, \beta) \vert_{\alpha,\beta \in \Lambda^*.}= e^{-i\pi B_{ki} m^k_\alpha m^i_\beta}.
\ee
We can perform a gauge transformation with 
$$
\delta (\alpha) = -\frac{\pi}{4} \sum_{k\neq i} C_{ki} m^k_\alpha m^i_\alpha
$$
which takes us to 
\be
\label{VOAcycle}
\tilde{\epsilon}(\alpha,\beta) = e^{i\pi \sum_{k>i} C_{ki} m^k_\alpha m^i_\beta }.
\ee
This particular form of two-cycle has appeared more frequently in the literature of vertex operator algebra (see for example \cite{GSW}).

\subsubsection{Twist operators from 2-cocycles and a fusion rule}
\label{subsubsec:fusionrule}

Let $\hat{g}$ denote a $\mathbb{Z}_N$ twist, and consider the following linear combination of vertex operators
\be
V(\alpha, z_1)_{[a]} = \sum_{m=1}^N e^{-\frac{2\pi i ma}{N}}\hat{g}^m \circ V(\alpha, z_1)
\ee
which has a $\hat{g}$-eigenvalue of $e^{\frac{2\pi i a}{N}}$. The OPE between two of them which reads 
\be
\label{OPEtwist}
V(\alpha, z_1)_{[a]} \times V(\beta, z_2)_{[b]}  = \sum_{m,k} e^{ -\frac{2\pi i}{N}(ma+kb)} 
\left( \hat{g}^m \circ V(\alpha, z_1) \times \hat{g}^k \circ V(\beta,z_2) \right)
\ee
should give us operators which have eigenvalues $e^{\frac{2\pi i}{N}(a+b)}$.  As we saw in the case of the simplest T-fold, we shall refine the T-duality twist by tensoring it with a $U(1)$ phase factor which we shall call $U(g,\alpha)$, i.e. 
\be
\label{phasefactorTwist}
\hat{g} \circ V(\alpha, z_1) = U(g, \alpha) V( g(\alpha), z_1).
\ee
The phase factor $U(g,\alpha)$ is nothing but the generalization of \eqref{simplestTfoldtwist} for the simplest T-fold. With \eqref{phasefactorTwist}, the RHS of \eqref{OPEtwist} now reads
\bea
\label{RHSOPEtwist}
&&\sum_{\delta_ \equiv l-k}
(\delta z)^{\frac{1}{2} \alpha^T_R G^{-1} g^{l-k}_R \cdot \beta_R }
(\delta \bar{z})^{\frac{1}{2} \alpha^T_L G^{-1} g^{l-k}_L \cdot \beta_L } \times \cr
&&\Bigg[
\sum_{k=0}^{N-1} e^{-\frac{2\pi i( k(a+b) + b\delta_-)}{N}} U( g^k, \alpha) U(g^{\delta_- + k},\beta ) \epsilon \left(g^k(\alpha), g^l (\beta) \right) 
\hat{C} \left( g^k (\alpha) + g^l (\beta)   \right)
\Bigg] 
\eea
where we have used the fact that 
\bea
&&(g^m \cdot \alpha )^T  G^{-1}  (g^k \cdot \beta) = \alpha^T  G^{-1}  (g^{k-m} \cdot \beta) \cr
&&\hat{C}(g^{1+p}(\alpha) ) \times \hat{C}(g^{k+p}(\beta)) = U(g^{1+p},\alpha) U(g^{k+p},\beta) \epsilon \left( g^{1+p}(\alpha), g^{k+p}(\beta) \right) \hat{C}\left(g^{1+p}(\alpha) + g^{k+p}(\beta) \right), \nonumber \\
\eea
since the twist $g$ is metric preserving and thus satisfies $G^{-1} = g^\dagger G^{-1} g^{-1}$. 
Acting on \eqref{RHSOPEtwist} with twist $g$, we obtain 
\bea
\label{RHSOPEtwist2}
&&\sum_{\delta_ \equiv l-k}
(\delta z)^{\frac{1}{2} \alpha^T_R G^{-1} g^{l-k}_R \cdot \beta_R }
(\delta \bar{z})^{\frac{1}{2} \alpha^T_L G^{-1} g^{l-k}_L \cdot \beta_L }  \Bigg[
\sum_{k=0}^{N-1} e^{-\frac{2\pi i( (k-1)(a+b) + b\delta_-)}{N}} \times \cr
&&U( g^{k-1}, \alpha) U(g^{\delta_- + k-1},\beta ) \epsilon \left(g^{k-1}(\alpha), g^{l-1} (\beta) \right)  U\left(  g, g^{k-1}(\alpha) + g^{l-1}(\beta) \right) 
\hat{C} \left( g^k (\alpha) + g^l (\beta)   \right)
\Bigg]. \nonumber \\
\eea
Comparing \eqref{RHSOPEtwist} and \eqref{RHSOPEtwist2}, we obtain
\bea
\label{conditionGtwist}
&&U(g^{1+p},\alpha) U(g^{k+p},\beta) \epsilon \left(g^{1+p}(\alpha), g^{k+p}(\beta) \right) \cr
&=&U(g^p, \alpha) U(g^{k+p-1},\beta) U(g, g^p(\alpha) + g^{k+p-1}(\beta)) \epsilon \left(g^p(\alpha), g^{k+p-1}(\beta) \right).
\eea
We can recast \eqref{conditionGtwist} in more illuminating forms.
Let's judiciously take $k=l=N-1$ which yields 
\be
\label{conditionGtwist2}
\frac{U(g,\alpha+\beta)}{U(g,\alpha) U(g,\beta)} = \frac{\epsilon(g(\alpha),g(\beta))}{\epsilon(\alpha,\beta)}
\ee
where we have invoked the boundary condition $U(1,\alpha)=1$. In \eqref{conditionGtwist}, replacing $\alpha \rightarrow g^{p}(\alpha), \beta \rightarrow g^{k+p-1} (\beta)$, we can remove the appearance of the 2 cocycles and obtain 
\be
\label{conditionGtwist3}
\frac{U( g^{p+1},\alpha)}{U(g, g^p(\alpha)) U(g^p, \alpha)} = \frac{U( g^{p+1},\beta)}{U(g, g^p(\beta)) U(g^p, \beta)}. 
\ee
Each side of \eqref{conditionGtwist3} can be taken to be unity, and thus we arrive at the relations 
\be
\label{conditionGtwist4}
U(g^{p+1},\alpha) = U(g^p, \alpha) U(g, g^p(\alpha)),\qquad \prod_{j=1}^N U(g, g^j(\alpha)) = 1.
\ee
As mentioned earlier in Section \ref{sec:Gen}, these constraints can be interpreted as solving for
the ratio $\epsilon (\alpha, \beta)/ \epsilon (g(\alpha), g(\beta))$ to be a trivial class of $H^2 (\Lambda, U(1))$, subject to certain orbifold group action-dependent constraints for the twist phase factors (which are the one-cochains). Further, when evaluated upon the invariant sublattices, the twist phase factors are trivial elements of $H^1 (\Lambda, U(1))$. Shortly in Section~\ref{sec:Examples}, we will compute the twist phase factor from \eqref{conditionGtwist2} and \eqref{conditionGtwist4} for some two-dimensional and six-dimensional examples of chiral asymmetric orbifolds by solving the triviality condition. 

For symmetric orbifolds, in the absence of a B-field, the orbifold twist is always a geometric one (see eqn.\eqref{symTwist}), and in this case, the metric-preserving relation $G = \theta G \theta^T$ suffices to show that the RHS of \eqref{conditionGtwist2} reduces to unity. Let us now turn on a B-field, in which case we find that the RHS of \eqref{conditionGtwist2} reads
\be
\label{Bphasefactor}
e^{-\frac{i\pi}{4} (\alpha_R - \alpha_L)^m \left( \theta^{-1} B \theta^\dagger - B \right)_{mn}  (\beta_R - \beta_L)^n   }
\ee
Therefore, as mentioned earlier in Section~\ref{sec:Gen}, for B-fields which do not satisfy \eqref{Bfieldsym}, the twist phase factor is non-trivial even for symmetric orbifolds. As \eqref{symTwist} reveals, this condition turns out to be the defining one for the symmetric orbifold twist to be geometric. 
Such a non-geometric background can nevertheless be regarded as the T-dual of a geometric bakground (with the same metric and any B-field $B_0$ satisfying \eqref{Bfieldsym}) with the T-duality element 
\be
\mathcal{T} = \left(   \begin{array}{cc}  \mathds{1} & \delta B \\ 0 & \mathds{1} \end{array} \right),
\ee
from which it is easy to see that the non-geometric nature arises simply from a gauge transformation of the B-field via the shift $B_0 \rightarrow B_0 - \delta B$ for some suitable $\delta B$ that appears in the twist phase factor from \eqref{Bphasefactor}.

Thus we have seen that the twist phase factor is non-trivial even for symmetric orbifolds and precisely for those of them which are not geometric. In these cases, the effect crucially depends on an appropriate B-field to be turned on. In the general case where the twist can be asymmetric, even in the absence of the B-field, the twist phase factor can be non-trivial.

\subsubsection{Untwisted vertex operators in twisted sectors}
\label{subsubsec:UntwistedVOA}

In the twisted sectors, we should also preserve the mutual locality of untwisted vertex operators invariant under the twist which we shall denote by ${}^uV$. States in the twisted sectors can be created by acting on a twisted state with these vertex operators. In the following, we shall derive the equation to be satisfied by the cocycles for ${}^uV$ when we demand mutual locality, and point out how the twist operators in the untwisted and twisted sectors should relate to one another to preserve the symmetry of the operator algebra. 

For definiteness, we shall restrict ourselves to chiral asymmetric orbifolds with a $\mathbb{Z}_N$ twist 
acting only on the right-movers, i.e. $g = \left( 1, \theta \right)$,
so $\theta$ denotes the twist that defines the string's boundary conditions in the twisted sector, and we let $\tilde{\alpha}_L, \tilde{\alpha}_R$ denote the momenta zero modes that lie in the invariant Narain sublattice. The untwisted vertex operators in the twisted sector can then be written as 
\be
^uV(z,\bar{z}) = \hat{C}(\tilde{\alpha} ) e^{i\tilde{\alpha}_L {}^uX_L (z)} e^{i\tilde{\alpha}_R {}^uX_R (\bar{z})}  
\ee
where $\hat{C}(\tilde{\alpha})$ is a cocycle operator, and
\bea
{}^uX^i_L &=& x^i_L - \frac{i}{2} G^{ij} p_{Lj} \text{ln}z + \frac{i}{\sqrt{2}} \sum_{r\in \mathbb{Z}} \frac{a^i_r}{rz^r}, \cr
{}^uX^i_R &=& x^i_R - \frac{i}{2} G^{ij} p_{Rj} \text{ln}\bar{z} + \frac{i}{\sqrt{2N}} \sum_{r\in \mathbb{Z}/N} \frac{\bar{C}^i_r}{r \bar{z}^r}, \qquad 
\bar{C}^i_r = \frac{1}{N} \sum_{s=0}^{N-1} e^{2\pi i rs} \left( \theta^s \right)^i_k \bar{a}^k_{Nr}.
\eea
We can compute the VEV of the twisted oscillators of which commutator reads
\bea
\left[ \bar{C}^i_r, \bar{C}^j_s \right] &=& \frac{Nr}{N^2} \delta_{N(r+s),0} \sum_{\mu,\nu = 0}^{N-1} 
e^{2\pi i (r\nu + s\mu)} G^{mk} \left( \theta^\nu \right)^i_m \left( \theta^\mu \right)^j_k \cr
&=& \sum_{\mu = 0}^{N-1} r e^{-2\pi i r \mu}G^{ik} \left( \theta^\mu \right)^j_k \delta_{r+s,0},
\eea
leading to the expectation value 
\be
-\langle \tilde{\alpha}_R \,{}^uX_{osc.} (\bar{z}) \tilde{\beta}_R \,{}^uX_{osc.} (\bar{w}) \rangle
= \frac{1}{2} \sum_{\mu = 0}^{N-1} \text{ln} \left( 1 - e^{-\frac{2\pi i \mu}{N}} \left( \frac{\bar{w}}{\bar{z}} \right)^{1/N} \right)^{\tilde{\alpha}_{Ri} \theta^{ij} \tilde{\beta}_{Rj}}.
\ee
Hence in the OPE of two $^u V$, the right oscillators contribute a factor of 
\be
\bar{z}^{-\frac{1}{2N} \tilde{\alpha}_{Ri} \sum_{\mu =0}^{N-1} \left(\theta^\mu \right)^{ij} \tilde{\beta}_{Rj}}
\times 
\prod^{N-1}_{\mu=0} \left(  \bar{z}^{1/N} - \bar{w}^{1/N} e^{-\frac{2\pi i \mu}{N}} \right)^{\frac{1}{2}
\tilde{\alpha}_{Ri} \left( \theta^\mu \right)^{ik} \tilde{\beta}_{Rk} }.
\ee
The prefactor is cancelled by its inverse which arises due to the other terms in the OPE.
Upon exchanging $\bar{z} \leftrightarrow \bar{w}$ and $\tilde{\alpha} \leftrightarrow \tilde{\beta}$, 
and taking into account the left-moving degrees of freedom, the anomalous factor that spoils the mutual locality (in the absence of the cocycle factors) reads
\be
\label{anomalousfactor}
\textrm{exp} \left[ \frac{\pi i}{N} \left( \tilde{\beta}_{Ri} \sum_\mu \mu \left( \theta^{-\mu} \right)^{ij} \tilde{\alpha}_{Rj} \right) - \frac{\pi i}{2} \left( \tilde{\beta}_{Ri} \sum_\mu \left( \theta^\mu \right)^{ij} \tilde{\alpha}_{Rj}   \right) \right].
\ee
As a consistency check, we note that if the twist is the identity, then independent of $N$, this factor reduces to $e^{-\frac{i\pi}{2} \tilde{\beta}_R \cdot \tilde{\alpha}_R}$ which is the appropriate expression in the untwisted sector. Similarly for the untwisted left-movers, we have $e^{-\frac{i\pi}{2} \tilde{\beta}_L \cdot \tilde{\alpha}_L}$. Thus, the overall factor which needs to be balanced by the cocycles reads
\be
\label{overallbranchcut}
\textrm{exp} \left[ \frac{\pi i}{N} \left( \tilde{\beta}_{Ri} \sum_\mu \mu \left( \theta^{-\mu} \right)^{ij} \tilde{\alpha}_{Rj} \right) - \frac{\pi i}{2} \left( \tilde{\beta}_{Ri} \sum_\mu \left( \theta^\mu \right)^{ij} \tilde{\alpha}_{Rj} - \tilde{\beta}_{Li} G^{ij} \tilde{\alpha}_{Lj}  \right) \right].
\ee
We now turn to the cocycles $\hat{C}(\tilde{\alpha} )$ which we write as 
\be
\label{cocycleexpression}
\hat{C} \left(\tilde{\alpha} \right) = e^{i\phi_L (\tilde{\alpha}) \hat{p}_L + i\phi_R ( \tilde{\alpha}  ) \hat{p}_R },
\ee
which in turn leads to the following expression for the 2-cocycle map 
\be
\epsilon (\tilde{\alpha}, \tilde{\beta} ) = e^{-i\tilde{\alpha}_L \cdot \phi_L (\tilde{\beta}) - i\tilde{\alpha}_R \cdot \phi_R (\tilde{\beta}) }
\ee
and the mutual locality condition is then obtained by equating $\epsilon (\tilde{\alpha}, \tilde{\beta} )/ \epsilon (\tilde{\beta}, \tilde{\alpha} )$ to \eqref{anomalousfactor}. In the case where there are no surviving zero modes in the right sector, i.e. $\tilde{\alpha}_R = 0$, and invoking the ansatz $\phi^n_{L} (\tilde{\alpha}) = \phi^{nm}_{L} \tilde{\alpha}_{Lm}$ where $\phi^{nm}_L$ is some constant matrix, the mutual locality condition yields the anti-symmetric part of $\phi_L$. 

Earlier, we have seen that for the untwisted sector, the cocycle factors are essential in deriving the twist phase factor. In the twisted sector, the situation is somewhat different. The twisted states are already eigenstates of the twist, and so are the invariant untwisted vertex operators. There is a consistency condition for the twist phase factor, unrelated to the cocycles, which can be simply expressed as
\be
\label{twistconstraint}
U_h \left( g,  \tilde{k} + k^{twisted} \right) = U\left( g,  \tilde{k}\right) U_h \left( g,  k^{twisted} \right) 
\ee
where $k^{twisted}$ is some momentum vector in the lattice of the twisted sector $h$, of which the (untwisted) lattice invariant under $h$ is a sublattice.

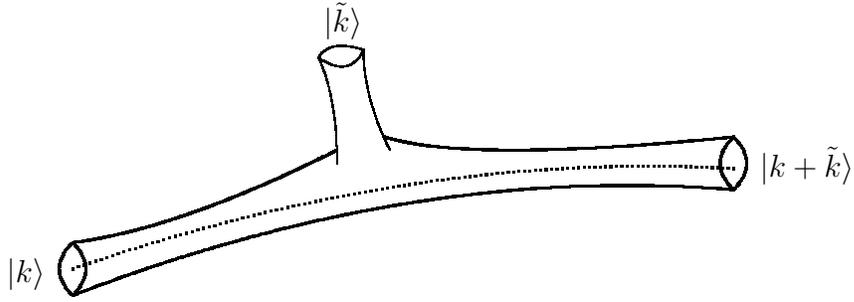
\begin{figure}[h]
\centering
\begin{picture}(350,130)(-60,-50)
\thicklines
\qbezier(0,-20)(40,-15)(100,15)
\thinlines
\qbezier(100,10)(100,35)(93,50)
\qbezier(110,53)(110,35)(120,15)
\qbezier(93,50)(105,42)(110,53)
\qbezier(93,50)(97,57)(110,53)
\thicklines
\qbezier(118,20)(150,10)(250,20)
\qbezier(0,-40)(130,10)(250,0)
\qbezier(0,-40)(10,-30)(0,-20)
\qbezier(0,-40)(-10,-30)(0,-20)
\qbezier(250,0)(260,10)(250,20)
\qbezier(250,0)(240,10)(250,20)
\thicklines
\qbezier[120](0,-30)(130,15)(250,8)
\put(95,60){$|\tilde{k} \rangle$}
\put(-25,-35){$| k \rangle$}
\put(260,5){$|k+\tilde{k} \rangle$}
\end{picture}
\caption{A schematic diagram depicting the absorption/emission of an untwisted state $|\tilde{k} \rangle$ that is associated with an invariant untwisted vertex operator ${}^uV$ from a twisted state with momentum $|k\rangle$. Note that the dotted line represents the cut that signifies twisted boundary conditions in the twisted sector.}
\label{fig:untwisted}
\end{figure}

As already mentioned in \cite{Simeon}, the relation in \eqref{twistconstraint} reflects the preservation of the symmetry of the operator algebra among the untwisted vertex operators and twisted ones (see Figure~\ref{fig:untwisted}). From a practical point of view, the other twist phase factors can be derived by performing Dehn twists on the partition trace $Z^0_1$. For all the consistent orbifold examples that we consider in this work, we find that \eqref{twistconstraint} is nicely satisfied.

\subsection{Some general points on modular covariance of chiral orbifolds}
\label{subsec:generalpoints}

In this section, we briefly present some general observations on the modular covariance of chiral asymmetric orbifolds. First consider the case where there is no surviving instanton sum in the right-moving sector. Then, in the untwisted sector with the insertion of the right twist with eigenvalue $e^{\pm \frac{2\pi i k}{N}}$, in the absence of twist phase factors, the surviving instanton sum can be expressed as 
\be
Z^1_{0, inst.} = \sum_{m \in \mathbb{Z}} e^{  \pi (2i \tau) m^i G_{ij} m^j         }
\ee
which after an $\mathcal{S}$ transformation yields
\be
\label{twistedsectorchiral}
Z^0_{1, inst.} = \frac{-i\tau}{\sqrt{\textrm{Det}(2G)}} \sum_{k \in \mathbb{Z}} 
q^{ \frac{1}{4}   k_i G^{ij} k_j    },
\ee
thus allowing one to read off the multiplicity of the twisted sector $Z^0_1$ as $\sqrt{4 \sin^2 \frac{\pi k}{N}}/\sqrt{\textrm{Det}(2G) }$ where the numerator 
arises from the bosonic oscillators. This is a special case of the well-known formula for the theta function of a $D$-dimensional lattice $\Lambda$ that reads
\be
\label{latticeinversion}
\Theta_{\Lambda^*} (\tau) = \sqrt{\textrm{Det}\,\,\Lambda} \left( \frac{i}{\tau} \right)^{D/2} \Theta_{\Lambda} \left(  -\frac{1}{\tau} \right),
\ee
where $\Lambda^*$ is the dual lattice. It is important to note that the zero mode quantization rule in the twisted sector can be read off. Given some non-trivial $U(g,\alpha)$ in the untwisted sector of the form $e^{2\pi i m\phi}$ in $Z^1_{0,inst.}$ for some constant $\phi$, the twisted left momentum zero modes read
$$ 
p_L = \frac{1}{2} (k + \phi), \qquad k \in \mathbb{Z}.
$$
Now for any chiral twist which kills off all right-moving momentum zero modes, on the invariant left sublattice ($n=E m$), the condition \eqref{conditionGtwist4} then translates to simply
\be
\label{specialgrouplaw}
U(\theta^k, \tilde{\alpha_L}) = U(\theta,\tilde{\alpha_L})^k,
\ee
where $\tilde{\alpha_L}$ refer to the residual left-moving momenta modes. 
Since $U(\theta,\tilde{\alpha_L})$ is $\mathbb{Z}_2$ valued, this means that for all $\theta$ of odd orders, there
is no non-trivial twist phase factor appearing in the partition trace $Z^1_0$.

With \eqref{specialgrouplaw},
we can just focus on $U(\theta,\tilde{\alpha_L})$. Generally, the non-diagonal elements of $Q$ in $U(g,\alpha)$ are uniquely fixed by \eqref{conditionGtwist2}, while $\eqref{conditionGtwist4}$ yields contraints on the diagonal elements depending on the lattice and choice of twist. As we shall see in explicit examples later, these constraints turn out to be among those which preserve the level-matching condition. From $Z^1_0$, performing a lattice Poisson resummation, we can obtain the lattice sum in the twisted sectors. These are theta functions of the dual lattices (eg. weight lattices if the tori are Lie root lattices) possibly weighted due to twist phase factors, and which should transform under a Dehn twist as
\be
\label{instantonDehntwist}
Z^{inst.}_\theta (\bar{\tau} + N_\theta) = e^{ i \delta_\theta} Z^{inst.}_\theta (\bar{\tau})
\ee
for some constant $\delta_\theta$, and where we have denoted $N_\theta$ to be the order of twist $\theta$. As we shall observe in some examples in Section~\ref{sec:Examples}, not all lattice theta functions transform like in \eqref{instantonDehntwist}.  The inclusion of an appropriate twist phase factor is crucial for \eqref{instantonDehntwist} to be true. Similarly for cases where there are some residual zero modes in the right-moving sector, we need to find the invariant sublattice and the twist phase factor, before performing a Poisson resummation and an $\mathcal{S}$ transformation to read off the twisted partition traces.

\subsection{Shift Orbifolds} 
\label{subsec:Shift}

Before moving onto explicit examples in which we compute the twist phase factors $U_h (g,\alpha)$, 
in this section, we wish to consider a class of orbifolds defined by twisting toroidal theories by translations. They are 
$\mathbb{Z}_N \times \mathbb{Z}_M$ orbifolds, in which we have independent shifts of momentum and winding numbers. These orbifolds may furnish the role of the base of a freely acting orbifold background constructed by fibering a rotational orbifold over it. First, let us recall some basic facts about the orbifold of a compact boson $X$ defined by geometric $\mathbb{Z}_N$ shifts
of the form 
\be
\label{symmetricOrbifold}
\hat{s}: \, X \rightarrow X + \frac{1}{N}.
\ee
With regards to the left- and right-movers, the translation operator $\hat{s}$ acts symmetrically on both. We can describe the shifted boundary conditions by the characteristics $\delta = (\delta', \delta'')$ where $\delta', \delta'' = 0,1/N,2/N,\ldots, (N-1)/N$. Corresponding to $h = \hat{s}^{N\delta'}$ and $g = \hat{s}^{N\delta''}$, the twisted boundary conditions of $X$ read 
$
X(\sigma_1 + 1, \sigma_2) = X(\sigma_1, \sigma_2) + \delta'' \,\, (\textrm{mod}\,\,1), \,\,\,
X(\sigma_1, \sigma_2 + 1) = X(\sigma_1, \sigma_2) + \delta' \,\, (\textrm{mod}\,\,1).
$
The classical zero modes have winding numbers along each worldsheet direction which we denote by $m,n$. Explicitly, we write
$
X^{cl.}_{m,n} (\sigma) = \sigma_1 (m + \delta'') +  \sigma_2 (n + \delta').
$
The string path integral can be split up into a product of a quantum part capturing degrees of freedom of $X-X^{cl.}$ coming from all the oscillators modes, and the classical zero modes of which contributions read
\be
\sum_{m,n} e^{-\frac{\pi R^2}{\tau_2}  \lvert \tau(n + \delta') - (m + \delta'') \rvert^2    },
\ee
where we have restored its radius $R$. Performing a Poisson resummation in $m \rightarrow w$, we have
\be
\label{momentumshift}
{Z^g}_h (\tau) = \textrm{Tr}_h \left( g q^{\frac{1}{4} p^2_L} \bar{q}^{\frac{1}{4} p^2_R} \right)=
\frac{1}{|\eta|^2} \sum_{n,w}  e^{-2\pi i \delta'' w}
q^{\frac{1}{4} \left( \frac{w}{R} + R(n + \delta')  \right)^2} 
\bar{q}^{\frac{1}{4} \left( \frac{w}{R} - R(n + \delta')  \right)^2}.
\ee
In \eqref{momentumshift}, we see that in the basis labelled by momentum number $w$ and winding number 
$n$, $g = \hat{s}^{N\delta''}= e^{-i(p_L+p_R)\delta X}=e^{-2 \pi i \delta'' w}$. More generally, one can set up an orbifold by independent shifts in $X$ and its T-dual $\tilde{X}$, where the shifts are $
X \rightarrow X + \frac{1}{N},\tilde{X} \rightarrow \tilde{X} + \frac{1}{M}
$
Introducing two other shift parameters $\bar{\delta}, \bar{\delta}''$, altogether we have
$
\delta', \delta'' = 0,1/N, \ldots, (N-1)/N\,\,\,\, \textrm{and}\,\,\,
\bar{\delta}', \bar{\delta}'' = 0,1/M, \ldots, (M-1)/M.\,\,\,\,
$
In each twisted sector labeled by $\delta', \bar{\delta}'$, the instanton part of the partition function which corresponds to summing over all classical backgrounds with different winding modes reads 
\be
\label{orbifoldZ}
Z^{(\delta'', \bar{\delta}'')}_{(\delta',\bar{\delta}')} (\tau) = \sum_{w,n} e^{-2\pi i \delta''(w+\bar{\delta}') -2\pi i \bar{\delta}''(n+\delta')  } q^{\frac{1}{4} \left( \frac{w + \bar{\delta}'}{R} +   R(n + \delta')  \right)^2} 
\bar{q}^{\frac{1}{4} \left( \frac{w + \bar{\delta}'}{R} -   R(n + \delta') \right)^2}.
\ee
This is thus an asymmetric translational orbifold. The above considerations generalize straightforwardly for asymmetric shift orbifolds of tori, of which the one-loop partition traces read
\be
\label{orbifoldTorusZ}
Z^{(\delta'', \bar{\delta}'')}_{(\delta',\bar{\delta}')} (\tau)= \sum_{\vec{m},\vec{n}} e^{-2\pi i \delta''_{k}  (m^k+\bar{\delta}'^k) -2\pi i \bar{\delta}''^k (n_k+\delta'_k)  } q^{\frac{1}{4} P^2_L} \bar{q}^{\frac{1}{4} P^2_R}
\ee
where the left- and right-momenta now depend on shifted modes, i.e. $P_L = (n + \delta') + E^T (m+\bar{\delta'}), P_R= (n + \delta') - E (m+\bar{\delta'}).$. Under a Dehn twist $\tau \rightarrow \tau + 1$, 
\bea
\label{DehntwistShift}
Z^{(\delta'', \bar{\delta}'')}_{(\delta',\bar{\delta}')} (\tau +1) &=&
\sum_{\vec{m},\vec{n}}   e^{2\pi i  (m^k+\bar{\delta}'^k)(n_k+\delta'_k)}
e^{-2\pi i \delta''_{k}  (m^k+\bar{\delta}'^k) -2\pi i \bar{\delta}''^k (n_k+\delta'_k)  } q^{\frac{1}{4} P^2_L} \bar{q}^{\frac{1}{4} P^2_R} \cr
&=& e^{-2\pi i \bar{\delta'} \delta'} Z^{(\delta''-\delta', \bar{\delta}'' -\bar{\delta}')}_{(\delta',\bar{\delta}')} (\tau). 
\eea
As for its behaviour under the action of $\mathcal{S}$, we first perform a Poisson resummation in the momenta modes to rewrite the trace as
\be
Z^{(\delta'', \bar{\delta}'')}_{(\delta',\bar{\delta}')} (\tau)= \sum_{\vec{m},\vec{w}} e^{-2\pi i \bar{\delta}'' \delta'}
e^{2\pi i \left( \delta'_k w'^k - \delta''_k m'^k +w'^k B_{ik} m'^i  \right)} e^{-\frac{\pi}{\tau_2} 
\left( m'^k m'^l G_{kl} + |\tau|^2 w'^k w'^l G_{kl} + 2\tau_1 m'^k w'^l G_{kl}          \right)  }
\ee
where we have defined $m' = m+ \bar{\delta}'$ and $w' \equiv w + \bar{\delta}''$. Then it is straightforward to check that under $\tau \rightarrow -1/\tau$, we have
\be
Z^{(\delta'', \bar{\delta}'')}_{(\delta',\bar{\delta}')} \left(-\frac{1}{\tau} \right)=
e^{-2\pi i (\delta'' \bar{\delta}' + \delta' \bar{\delta}'')} Z^{(-\delta', -\bar{\delta}')}_{(\delta'',\bar{\delta}'')} (\tau).
\ee
Just like the chiral blocks, the shift orbifolds' partition traces may suffer from global anomaly in the sense that the twist labels may not furnish a faithful representation of $\mathbb{Z}_N$. In this case, 
\be
\label{shiftanomaly}
Z^{(\delta''-N \delta', \bar{\delta}'' - N\bar{\delta}' )}_{(\delta',\bar{\delta}')} (\tau) = e^{4\pi i N \delta'_k \bar{\delta}'^k}
Z^{(\delta'', \bar{\delta}''  )}_{(\delta',\bar{\delta}')} (\tau) .
\ee
Now, a purely momenta or winding translational orbifold has no anomaly. However, one can imagine fibering an anomalous $\mathbb{Z}_N$ rotational orbifold (such as the $\mathbb{Z}_3$ chiral orbifold of the $A_2$ root lattice as we shall explain shortly) over a shift orbifold with conjugate anomalous phase factor such that we have level-matching, i.e. $Z^0_1 (\tau + N) = Z^0_1 (\tau)$. To this end, consider the case where there is a $\mathbb{Z}_N$ action generated by a $1/N$ shift in both momenta and winding, parametrized as 
$ \delta_k = \frac{\nu_k}{N}, \bar{\delta}_k = \frac{\bar{\nu}_k}{N}$,
then from \eqref{shiftanomaly}
\be
Z^{(\delta'', \bar{\delta}''  )}_{(\delta',\bar{\delta}')} (\tau +N) = e^{\frac{2\pi i}{N}\nu_k \bar{\nu}^k }
Z^{(\delta'', \bar{\delta}''  )}_{(\delta',\bar{\delta}')} (\tau) .
\ee
By engineering $\{\nu,\bar{\nu}\}$, these asymmetric shift orbifolds can thus act as suitable bases for rotational orbifolds of which twisted partition traces satisfy level-matching up to a constant anomalous phase factor.

\section{On some chiral asymmetric orbifolds of $T^2$ and $T^6$}
\label{sec:Examples}

In this Section, we shall compute $U(g,\alpha)$ for some two and six-dimensional examples. 
Denoting $N_\alpha = (n_\alpha, m_\alpha)$, we let
$U(g,\alpha)$ take the form
\be
U(g,\alpha) = e^{i N^T_\alpha Q N_\alpha }
\ee
where $Q$ is some constant matrix. Consider now the cases where the twist is trivial on the left-movers. From \eqref{conditionGtwist2}, we have 
\be
Q+Q^T = \frac{\pi}{2} \left|
\left( \begin{array}{cc} c - c^T - 2cBc^T & (c+c^T)E+2cBc^TE
\\ 2E^T c B c^T - E^T(c+c^T)
& E^T (c^T-c)E - 2E^T cBc^TE
\end{array} \right)
\right|
\ee
where the equality sign is defined modulo $2\pi$. Note that we take the absolute value of the RHS to symmetrize the expression which is an antisymmetric matrix with every element being an integer multiple of $\pi$. We are also interested in the lattice direction invariant under the twist since they yield the residual string instanton sums in the partition traces. In the case where $c$ is invertible, this has the simple solution
\be
\label{asymZero}
n=Em.
\ee
In other cases, one has to solve for the sublattice which is invariant under the asymmetric twist. Now from \eqref{conditionGtwist4}, we have other constraints to be imposed on $Q$ and we find that generically, they do not fix $Q$ uniquely. But as we shall see shortly, we find that the constraints are compatible with that of modular covariance which fixes those parameters unconstrained by \eqref{conditionGtwist4}.
In the following, Sections \ref{subsec:ChiralZ2} and \ref{subsec:ChiralZ3} deal with the computations of the phase factors and partition traces in Tables 1 and 2, while Section \ref{subsec:Heterotic} examines asymmetric orbifold points of smooth $CY_3$ compactifications of the heterotic string.

\subsection{Chiral $\mathbb{Z}_2$ orbifolds}
\label{subsec:ChiralZ2}

We shall first comment on the orbifolds in Table 1 which can be understood as direct products of the ordinary self-dual circle theory and the simplest T-fold. For example, for (i), the twist acts trivially on an $S^1$ and reflects the right-movers of the second circle. Hence, it is a product of the circle theory and the simplest T-fold, and similarly so for (ii), while (iii) is nothing but the product of two simplest T-folds. Earlier, we mentioned that for chiral asymmetric orbifolds of tori with Lie root lattices, outer automorphisms are not embedded in the duality group. In this semi-simple case, there is yet an outer-automorphism defined by exchanging the two $A_1$-theories though neither can it be realized as an asymmetric twist. 

In Table 2 we have asymmetric orbifolds of the $A_2$ torus with the twist being an element of the Weyl group of $A_2$.
Building on our previous discussions, it is easy to see that
\be
Z^1_0 (\tau) = \frac{1}{\eta^2(\tau)  \eta(\bar{\tau})} \sqrt{\frac{2 \eta(\bar{\tau}) }{\theta_2 (\bar{\tau})}} Z^1_{0,inst.}(\tau).
\ee
The oscillators' contributions can be checked to be invariant under $\tau \rightarrow \tau + 1$, so it remains to see if the instanton sum is itself invariant. One can faithfully check that this is true for the cases (iv)-(vi). If we perform $\tau \rightarrow -\frac{1}{\tau}$, this takes us to the twisted sector for which we are obliged to ensure that the zero modes level-match by $L_0 - \bar{L}_0 = \frac{1}{2}\,(\textrm{mod}\,\,1)$. The partition trace reads
\be
\label{partitionTrace01}
Z^0_1 (\tau)=  \frac{1}{|\tau|\sqrt{-i\tau}\eta(\bar{\tau}) \eta^2(\tau)} \sqrt{ \frac{2\eta(\bar{\tau})}{\theta_4(\bar{\tau})}}Z^1_{0,inst.} \left( -\frac{1}{\tau} \right). 
\ee
For the $\mathbb{Z}_2$ orbifolds in Table 2, the invariant Narain sublattices $(\Lambda^*)$ are all three-dimensional ones. The main quantity of interest is $Z^1_{0,inst.}$, which we shall write as
\be
\label{instantonsumUntwisted}
Z^1_{0,inst.} (\tau) = \sum_{\tilde{\alpha}_L, \tilde{\alpha}_R \in \Lambda^*} U(\theta,\tilde{\alpha})
e^{\frac{\pi i \tau}{2} \tilde{\alpha}^2_L}e^{\frac{\pi i \bar{\tau} }{2} \tilde{\alpha}^2_R} = \sqrt{\textrm{Det}\,\Upsilon} \sum_{\vec{v}}
e^{-\pi (\vec{v}-\vec{\delta})_i \Upsilon^{ij} (\vec{v}-\vec{\delta})_j}
\ee
where $\vec{v}$ are integers, $\vec{\delta}$ are half-integer shifts (each $\delta_i \in \{0,\frac{1}{2} \}$ ) that are present in $U(\theta, \tilde{\alpha})$, and $\Upsilon$ is a $3\times 3$ matrix that depends on $\tau$, obtained after Poisson-resumming all quantum numbers (second equality). As noted earlier, \eqref{conditionGtwist2} does not fix the phase factor completely, and in particular we have the freedom to specify diagonal constants in the $Q$'s ( i.e. the exponential arguments of $U(\theta, \alpha)$ ). They are a set of four even integers which are further constrained by \eqref{conditionGtwist4}. When evaluated on the invariant sublattice, they yield $\mathbb{Z}_2$-valued phase factors of the form
\be
U(\theta, \tilde{\alpha}) = e^{2\pi i\left( \delta_l l + \delta_m m + \delta_n n \right)},
\ee
where the $\delta_i$'s are the ones that appear in \eqref{instantonsumUntwisted}. After Poisson resumming and performing $\tau \rightarrow -1/\tau$, we obtain the zero modes in the twisted sector which are shifted in each direction whenever the twist phase factor is non-trivial in $Z^1_0$. Upon taking $\tau \rightarrow \tau +1$, we obtain the twist phase factor in the twisted sector which reads simply as
\be
\label{twistOptwist}
U_\theta (\theta,\alpha) = e^{i\frac{\pi}{2}(\alpha^2_L - \alpha^2_R)}.
\ee
Modular covariance translates into the sufficient condition $Z^0_1(\tau+2) = Z^0_1(\tau)$. 
We find that this gives the same constraint on the $\delta$'s as \eqref{conditionGtwist4} and one more constraint which is precisly that imposed by \eqref{twistconstraint}. Below, we display the zero modes, conditions for the twist phase factors and the 3 by 3 matrix associated with the invariant sublattice (see eqn. \eqref{instantonsumUntwisted}) 
for (iv) - (vi) in Table 2. Please note that we have denoted the zero modes in the twisted sector by primed quantities, and the quantum numbers are shifted accordingly by the $\delta$'s, e.g. $m' \equiv m +\delta_m$. 
\\[2ex]
\noindent
(iv)For this orbifold, $\theta_R = \left(  \begin{array}{cc} -1 & 0 \\ -1 & 1 \end{array} \right)$, 
\bea
&&\tilde{\alpha}_L = (2m+l, n+m+l),\,\,\, \tilde{\alpha}_R = (0, n-l), \cr \cr
&& \alpha'_L = (-m', -l'-n'),\,\,\, \alpha'_R = \left(0, \frac{1}{2}(m'+n'-2l') \right), \,\,\, \delta_m - \delta_n = \frac{1}{2},\,\,\, \delta_l \delta_n = 0,\cr \cr
&&\Upsilon = \frac{1}{3|\tau|^2} \left(  
\begin{array}{ccc} 4\tau_2 & -2\tau_2 & 3i\tau_1 + \tau_2 \\
-2\tau_2 & \frac{3i}{2}\tau_1 + \frac{5}{2}\tau_2 & -\frac{3i}{2}\tau_1 - \frac{1}{2}\tau_2 \\
3i\tau_1 + \tau_2 & -\frac{3i}{2}\tau_1 - \frac{1}{2}\tau_2 & \frac{3i}{2}\tau_1 + \frac{5}{2}\tau_2
\end{array}
\right).
\eea
\\[1ex]
\noindent
(v)For this orbifold, $\theta_R = \left(  \begin{array}{cc} 0 & 1 \\ 1 & 0 \end{array} \right)$,
\bea
&&\tilde{\alpha}_L = (2m+n, n+m+l),\,\,\, \tilde{\alpha}_R = (n-l, n-l), \cr \cr
&& \alpha'_L = (-m', -l'-n'),\,\,\, \alpha'_R = \left(\frac{1}{2}(-m'-l'+2n'), \frac{1}{2}(-m'-l'+2n') \right), \,\,\, \delta_l - \delta_m = \frac{1}{2},\,\,\, \delta_l \delta_n = 0,\cr \cr
&&\Upsilon = \frac{1}{3|\tau|^2} \left(  
\begin{array}{ccc} \frac{1}{2}\left( 3i\tau_1 + 5\tau_2  \right) & -\frac{1}{2}\left( 3i\tau_1 + \tau_2  \right) & 3i\tau_1 + \tau_2 \\
-\frac{1}{2}\left( 3i\tau_1 + \tau_2  \right) &  \frac{1}{2}\left( 3i\tau_1 + 5\tau_2  \right)        & -2\tau_2 \\
3i\tau_1 + \tau_2 & -2\tau_2 & 4\tau_2
\end{array}
\right).
\eea
\\[1ex]
\noindent
(vi)For this orbifold, $\theta_R = \left(  \begin{array}{cc} 1 & -1 \\ 0 & -1 \end{array} \right)$,
\bea
&&\tilde{\alpha}_L = (m+n, m+2l),\,\,\, \tilde{\alpha}_R = (n-m-l,0), \cr \cr
&& \alpha'_L = (-m'-n', -l'-n'),\,\,\, \alpha'_R = \left(\frac{1}{2}(l'-2m'+2n'),0 \right), \,\,\, \delta_l = \frac{1}{2},\,\,\, \delta_m \delta_n = 0,\cr \cr
&&\Upsilon = \frac{1}{3|\tau|^2} \left(  
\begin{array}{ccc} \frac{1}{2}\left( 3i\tau_1 + 5\tau_2    \right) & -2\tau_2 & 2\tau_2 \\
-2\tau_2 & 4\tau_2 & 3i\tau_1 - \tau_2 \\
2\tau_2 & 3i\tau_1 - \tau_2 & 4\tau_2
\end{array}
\right).
\eea
The spectra of the orbifold CFTs (iv)-(vi) are identical, and thus these theories are dual to one another. A quick way to see this is simply to compare the partition trace $Z^1_0$, and to realize that each of them is related to the other two by a relabeling of the various winding and momenta numbers. The important simple point to be taken from these calculations is that the initial choice of a $U(\theta,\alpha)$ which we determine using \eqref{conditionGtwist2} and \eqref{conditionGtwist4} is compatible with level-matching. Modular transformations of the partition traces generate consistent expressions for the twisted zero modes and the twist operator in the twisted sector. Another important relation we observed is that the twist phase factors in the untwisted and twisted sectors derived by the above procedure also satisfy \eqref{twistconstraint}.

\subsection{A Chiral $\mathbb{Z}_3$ orbifold}
\label{subsec:ChiralZ3}

As for the last case in Table 2, where the asymmetric twist is $\mathbb{Z}_3$, we find that \eqref{conditionGtwist2} and \eqref{conditionGtwist4} lead to a trivial $U(\theta, \tilde{\alpha})$. Among the orbifolds in Table 2, this is the only case where any non-trivial duality phase factor will violate the mutual locality condition of the vertex operators in the untwisted sector. For this case, \eqref{conditionGtwist2} gives us
\be
Q= \frac{\pi}{2}\left( \begin{array}{cccc} a & 0 & 1 & 0 \\ 0 & b & 0 & 1 \\ 1 & 0 & c & 1 \\ 0 & 1 & 1 & d \end{array} \right),\,\,\,
U(\theta, \tilde{p}) = \textrm{exp} \Bigg[ \frac{i\pi}{2} m^T  \left( \begin{array}{cc} a+c+2 & 2+a \\
2+a & a+b+d+2 \end{array} \right) \Bigg]
\ee
where $\{ a,b,c,d\}$ are even integers left unfixed by \eqref{conditionGtwist2}. On the other hand, \eqref{conditionGtwist4} yields
\be
a+b+d = 2, \qquad a+c =2.
\ee
Thus, the residual instanton sum reads 
\be
\label{z3instanton}
Z^1_{0,inst.} = \sum_{m^1, m^2} q^{m^i G_{ij} m^j} .
\ee
Although the twist phase factor is trivial in the untwisted sector, it is not so in the twisted sectors. Let $h$ denote the twist corresponding to that of the last entry of table 2, and let $U_{h_1} (h_2, \vec{m})$ denote the factor refining the instanton sum in $Z^{h_2}_{h_1}$. We find
\bea
\label{twistoperatorsZ3}
U_h (h^2, \vec{m}) &=& U_{h^2} (h, \vec{m}) = e^{\frac{i\pi}{2} m_i G^{ij} m_j} \cr
U_h (h, \vec{m}) &=& U_{h^2} (h^2, \vec{m}) = e^{-\frac{i\pi}{2} m_i G^{ij} m_j}.
\eea
It can be checked that they satisfy \eqref{twistconstraint}, thus preserving the symmetry of the operator algebra among untwisted and twisted vertex operators. For a $\mathbb{Z}_3$ action, the modular covariance relations describe the following closed $\mathcal{S}$ and $\mathcal{T}$ orbits. 
\bea
\label{modularorbits}
&&\mathcal{S}\,\,\,\textrm{orbits}: Z^1_0 \rightarrow Z^0_1 \rightarrow Z^2_0 \rightarrow Z^0_2 \rightarrow Z^1_0,\qquad
Z^1_1 \rightarrow Z^2_1 \rightarrow Z^2_2 \rightarrow Z^1_2 \rightarrow  Z^1_1\cr
&&\mathcal{T}\,\,\,\textrm{orbits}: 
Z^0_1 \rightarrow Z^2_1 \rightarrow Z^1_1 \rightarrow Z^0_1,\qquad
Z^0_2 \rightarrow Z^1_2 \rightarrow Z^2_2 \rightarrow Z^0_2, \,\,\, Z^1_0, Z^2_0 \,\,\textrm{self-dual}. 
\eea
On $Z^0_1$, we need to check that the level-matching condition 
holds. This is equivalent to checking that it is invariant under $\tau \rightarrow \tau+3$. 
We find that the instanton sum \eqref{z3instanton} satisfies the identity
\be
\label{phiId}
Z^1_{0,inst.} \left(  -\frac{1}{\tau} \right) = \frac{i \tau}{\sqrt{3}}  Z^1_{0,inst.} \left(  \frac{\tau}{3} \right)
\ee
which implies invariance under $\tau \rightarrow \tau+3$ of the dual lattice sum since $Z^1_{0,inst.} (\tau)$
is invariant under $\tau \rightarrow \tau  + 1$ (with the factor of $\tau$ being cancelled by an identical factor in $\eta(-1/\tau)$). Generally, the dual lattice in the twisted sector contains the invariant sublattice that appears in $Z^1_0$ as a sublattice. In the twisted sectors, the right-moving zero modes read simply as 
\be
\label{quantizationtwisted}
\tilde{\alpha}_R = (m_1,m_2).
\ee
Thus far, the triviality of the twist phase factor appears to be compatible with modular covariance. Yet
for the overall modular covariance, one needs to take into account the oscillators' degrees of freedom. 
Then, the partition traces of this asymmetric $\mathbb{Z}_3$ orbifold read
\bea
&&Z^1_0 = Z^2_0 = \bar{q}^{-\frac{1}{12}} \prod_{m=1}^\infty \left( 1 - \bar{q}^m e^{\frac{2\pi i}{3}}   \right)^{-1}
\left( 1 - \bar{q}^m e^{\frac{-2\pi i}{3}}   \right)^{-1} \frac{1}{\eta^2 (\tau)}\sum_{m_i} q^{\frac{3}{4}m_i G^{ij}m_j} \\
&&Z^0_1 = Z^0_2 = \bar{q}^{\frac{1}{36}} \prod_{m=1}^\infty \left( 1 - \bar{q}^{m-\frac{1}{3}}   \right)^{-1}
\left( 1 - \bar{q}^{m-\frac{2}{3}}    \right)^{-1} \frac{1}{\eta^2 (\tau)}\sum_{m_i} q^{\frac{1}{4}m_i G^{ij}m_j}\\
&&Z^1_1 = Z^2_ 2 = e^{\frac{2\pi i}{9}}  
\bar{q}^{\frac{1}{36}} \prod_{m=1}^\infty \left( 1 - \bar{q}^{m-\frac{1}{3}} e^{-\frac{2\pi i}{3}}   \right)^{-1}
\left( 1 - \bar{q}^{m-\frac{2}{3}} e^{\frac{2\pi i}{3}}   \right)^{-1} \frac{1}{\eta^2 (\tau)} \cr
&& \qquad \qquad \qquad \qquad \qquad \times \sum_{m_i} e^{-\frac{\pi i}{2} m_i G^{ij} m_j }  
q^{\frac{1}{4}m_i G^{ij}m_j} \\
&&Z^2_1 = Z^1_ 2 = e^{\frac{4\pi i}{9}}  
\bar{q}^{\frac{1}{36}} \prod_{m=1}^\infty \left( 1 - \bar{q}^{m-\frac{1}{3}} e^{-\frac{4\pi i}{3}}   \right)^{-1}
\left( 1 - \bar{q}^{m-\frac{2}{3}} e^{-\frac{2\pi i}{3}}   \right)^{-1} \frac{1}{\eta^2 (\tau)} \cr
&& \qquad \qquad \qquad \qquad \qquad \times \sum_{m_i} e^{-\pi i m_i G^{ij} m_j }  
q^{\frac{1}{4}m_i G^{ij}m_j}.
\eea
We note that the constant phases of $e^{\frac{2\pi i}{9}}$ and $e^{\frac{4\pi i }{9}}$ can be absorbed into the twist operator, but this implies that these operators only realize the orbifold group projectively in the twisted sector. The related conclusion is that this theory is anomalous by itself as one can check that level-matching fails up to a constant phase factor $e^{2\pi i/3}$. Expanding the spectrum in $q, \bar{q}$, the degeneracies are non-integral. These problems disappear when we take the product of three identical copies of this orbifold CFT which yields a consistent six-dimensional background that can be interpreted as an asymmetric orbifold points of a smooth Calabi-Yau compactification. Also, as pointed out earlier, another consistent orbifold can also be obtained by fibering this anomalous orbifold over an asymmetric $\mathbb{Z}_3$ orbifold in which the twist is defined by a $1/3$ shift in both momenta and winding in each of the lattice directions. Finally, let us comment briefly that although the cocycles for the invariant untwisted vertex operators in the twisted sector do not directly affect the partition trace, one can solve the mutual locality condition to determine part of \eqref{cocycleexpression}. In this case, this requires the anti-symmetric part of $\phi_L$ to be $\pi/3$.

\subsection{Asymmetric orbifold points of Calabi-Yau compactifications of the heterotic string}
\label{subsec:Heterotic}

We now briefly comment on the modular covariance condition for a family of orbifolds of the
heterotic string which can be naturally regarded as asymmetric orbifold points of smooth Calabi-Yau compactifications of which geometric orbifold points are known and tabulated in Table \ref{table:CY}.
\begin{table}[h]
\centering
\begin{tabular}{c c c c}
\hline\hline
$$& $\mathbb{Z}_N$ & Twist vector $\vec{\nu}$ & Toroidal lattice  \\ [0.5ex]
\hline
(i) & $\mathbb{Z}_3$ &   $\frac{1}{3}(1,1,-2)$   & $SU(3)^3, E_6$ \\
(ii) & $\mathbb{Z}_4$   &     $\frac{1}{4}(1,1,-2)$   & $SU(4)^2$ \\ 
(iii) & $\mathbb{Z}_6$ &  $\frac{1}{6}(1,1,-2)$    &$SU(3)\times G^2_2$ \\ 
(iv) & $\mathbb{Z}_7$ & $\frac{1}{7}(1,2,-3)$  &$SU(7)$\\ 
(v) & $\mathbb{Z}_8$ &   $ \frac{1}{8}(1,2,-3)$  & $SO(5)\times SO(9)$       \\ 
(vi) & $\mathbb{Z}_{12}$&   $\frac{1}{12}(1,4,-5) $      &$E_6$\\ [1ex]
\hline
\end{tabular}
\caption{Some $T^6/\mathbb{Z}_N$ orbifolds which are symmetric orbifold points of $CY_3$ compactification of the heterotic string preserving $\mathcal{N}=1$ supersymmetry in four dimensions.}
\label{table:CY}
\end{table}
For these geometric orbifolds, the lattices are the root lattices of some suitable semi-simple Lie algebras with the twist being an element of its Weyl or outer-automorphism group. Those in Table 3 (see for example \cite{Takahashi:thesis,Ibanez:textbook,Blumenhagen} for excellent reviews) preserve $4D\,\, \mathcal{N}=1$ supersymmetry that descends from requiring the $\mathbb{Z}_N$ holonomy group to lie in $SU(3)$. Let $e^{\pm 2\pi i\nu_i}$ be the eigenvalues of the twists acting on the complex coordinates $z_i, i=1,2,3$ parametrizing $T^6$, then this condition simply translates to $\sum \nu_i =0 \,\, \textrm{mod}\, 1$. 
In the context of orbifolding the heterotic string, we should specify a simultaneous $\mathbb{Z}_N$ translation on the 16 internal compact left-moving bosons $X^I$ to preserve one-loop modular covariance. We can write the quotient structure of this class of orbifolds as 
\be
\label{hetorbifold1}
\left[ \Lambda^6/\mathcal{O}_{geo.} \right] \otimes \left[ \Lambda^{16}_L / \mathcal{O}_{shift} \right],
\ee
where $\Lambda^6$ is the $T^6$ lattice, $\mathcal{O}_{geo.}$ also acts on the right-moving worldsheet fermions, and we have excluded the possibility of including Wilson lines for simplicity. One can consider their asymmetric counterparts in a similar fashion that we have done so in the previous sections. Instead of restricting ourselves to \eqref{hetorbifold1}, we shall consider heterotic orbifolds of the form
\be
\label{hetorbifold}
\left[ \Gamma_{6,6}/O(6,6;\mathbb{Z}) \right] \otimes \left[ \Lambda^{16}_L / \left(\mathcal{O}_{shift} \right) \right],
\ee
where $\Gamma_{6,6}$ refer to the toroidal stringy Hilbert space,
and we have picked our orbifold group to lie in the $O(6,6;\mathbb{Z})$ of the T-duality group $O(22,6;\mathbb{Z})$ augmented with a set of suitable translations on the internal compact bosons. In particular, we shall focus explicitly on orbifolds in which $\theta_R$ is identical to the geometric twist, $\theta_L$ is trivial.
and there is a suitable shift in the 16-dimensional lattice. We need to turn on a suitable B-field, as explained in Section~\ref{subsec:Constructing}, and the above-mentioned twist can then be embedded if it purely consists of Weyl reflections. These are the cases (i), (ii), (iv), (vi), i.e. products of simply-laced algebras. 

We can straightforwardly check if the partition traces $Z^0_h$ develop an anomaly under $\tau \rightarrow \tau + N$. For these chiral asymmetric orbifolds, the twisted right-moving bosons contribute a factor of $e^{\frac{i\pi}{N} \sum_i k_i (1-k_i) }$ while the twisted fermions contribute a factor of $e^{\frac{i\pi}{N} \sum_i k_i^2 }e^{\frac{2\pi i N}{3}}$ (see Appendix~\ref{sec:OnMod} for a derivation). Since the twists sum to zero, the residual instanton sum must then be invariant under $\tau \rightarrow \tau+N$ (or at least up to some phase factor which can be cancelled by an appropriate shift in the internal 16-dim. lattice). 

In Appendix~\ref{sec:asymHet}, we compute the twist phase factors which characterize each asymmetric orbifold. We find that when the twist is restricted only to the right-moving sector, the asymmetric orbifolds  corresponding to (i), (ii), (iv), (vi) are modular covariant. It would be interesting to make our analysis of asymmetric orbifolds of the heterotic string more systematic in view of the twist phase factors as new ingredients in our understanding of modular invariance, by for example, including Wilson lines and twisting by other discrete subgroups of $O(22,6;\mathbb{Z})$.

\section{The twist phase factor at higher-genus for the simplest T-fold}
\label{sec:Genera}

We begin by reviewing higher-genera characters of a toroidal bosonic string background (see for example \cite{Verlinde} and \cite{Giveon} for an excellent review). 
Let $g$ denote the worldsheet genus. Then the complete string partition function reads 
\be
\label{higherG}
Z_g = \int dh\,dX \textrm{exp}\left[ - \frac{1}{4\pi} \int_{\mathcal{S}} d\tau d\sigma
\sqrt{h} h^{mn} G_{ij} \partial_m X^i \partial_n X^j + \epsilon^{mn} B_{ij} \partial_m X^i \partial_n X^j  \right]
\ee
where we integrate over all worldsheet metrics $h$ that are compatible with Riemann surface $\mathcal{S}$ of genus $g$. Let us define the canonical homology cycles $(a_\alpha, b_\alpha)$ of $\mathcal{S}$ as follows. Define $\omega_\alpha, \alpha=1,2,\ldots g$ be the holomorphic one-forms that span 
$H_{1} (\mathcal{S}, \mathbb{Z}) = \mathbb{Z}^{2g}$, and the $g\times g$ period matrix $\Omega = \Omega_1 + i\Omega_2$ be
\be
\label{periodmatrix}
\Omega_{\alpha \beta} = \int_{a_\alpha} \omega_\beta,\qquad \int_{b_\alpha} \omega_\beta = \delta_{\alpha \beta}, \qquad \int_{a_\alpha} dX^i = 2\pi n^i_\alpha, \qquad \int_{b_\alpha} dX^i = 2\pi m^i_\alpha
\ee
where we have defined $n^i_\alpha, m^i_\alpha$ to be the $d\times g-$dimensional winding and momentum vector modes. The complete partition function in \eqref{higherG} can be written as an integral over all the $3g-3$ modular parameters. In the following, we shall first consider the classical instanton sector of the partition function. The winding numbers along each cycle are now $d\times g$-dimensional vectors. The partition function reads 
\bea
Z_{cl.}(G,B,\tau) &=& \sum_{n,m}
\textrm{exp} \Bigg[ -\pi
m^{i\alpha} (\frac{1}{\Omega_2})_{\alpha \beta} G_{ij} m^{j\beta} -\pi 
n^{i\alpha} (\Omega_2 + \Omega_1 \frac{1}{\Omega_2} \Omega_1)_{\alpha \beta} G_{ij} n^{j\beta} \cr
&&+2\pi n^{i\alpha} (\Omega_1 \frac{1}{\Omega_2} )_{\alpha \beta} G_{ij} m^{j\beta} + 2i \pi m^{i\alpha} B_{ij} n^{j\alpha} \Bigg].
\eea
This can be Poisson resummed just as in the torus case. The result is known and the classical zero modes' part can be cast into the familiar form
\be
\label{Gpartitionfunction}
\textrm{Det}(\Omega_2)^{d/2} \sum_{ P^i_{L\alpha}, P^i_{R\alpha}  }
q^{\frac{1}{4} G_{ij} P^i_{L\alpha} \Omega_{\alpha \beta} P^j_{L\beta}  }\bar{q}^{\frac{1}{4} G_{ij} P^i_{R\alpha} \bar{\Omega}_{\alpha \beta} P^j_{R\beta} }
\ee
where $P^i_{L\alpha}, P^i_{R\alpha}$ are the higher-genera left and right momentum zero modes.
Just like in the case of genus one, the stringy instanton sums can be expressed in terms of theta functions
associated with Riemann surfaces of higher genus endowed with the period matrix (see Appendix~\ref{subsec:Theta}). The notion of modular covariance involves the transformation property of the higher-genus partition traces under the symplectic modular group $Sp(2g, \mathbb{Z})$ defined as all integral $2g \times 2g$ matrices $M$ satisfying the following property
\be
\label{sp2g}
M =  \left( \begin{array}{cc} A & B \\ C & D \end{array} \right) \in Sp(2g, \mathbb{Z}),\qquad
AC^T= CA^T,\,\,\, DB^T = BD^T,\,\,\,A^TD - B^T C = \mathds{1},
\ee
which act on the period matrices as
\be
\label{actionOnperiodmatrices}
\Omega' = \left( A\Omega +B\right)\left( C \Omega + D \right)^{-1}.
\ee
As we have seen in the genus one case, the partition traces of an orbifold are labelled by twisted boundary conditions along each canonical cycle and they must transform onto one another under elements of the symplectic modular group.  

In this section, we shall consider the case of $\mathbb{Z}_2$ for definiteness. As explained in the seminal work \cite{DVV}, it is useful to understand double-valued fields on a Riemann surface $\mathcal{S}$ of genus $g$ in terms of single-valued ones living on the double cover $\hat{\mathcal{S}}$ of $\mathcal{S}$. Suppose the field is anti-periodic along a cycle, say $b_g$, then there is a branch cut running along the cycle $a_g$. We can use this branch cut to define $\hat{\mathcal{S}}$ by taking two copies of $\mathcal{S}$, slicing each apart along $a_g$ and finally pasting the them together, yieiding a surface of genus $2g-1$. We can adopt a choice for the canonical cycles of $\hat{\mathcal{S}}$ by one that projects onto the corresponding cycle on $\mathcal{S}$. Let $\pi$ be the projection that takes $H_1 \left( \hat{\mathcal{S}}, \mathbb{Z} \right)$ to $H_1 \left( \mathcal{S}, \mathbb{Z} \right)$ such that $\pi (\hat{a}_k) = a_k$, $\pi (\hat{b}_k) = b_k$. Such a choice is unique up to modular transformations on $\hat{\mathcal{S}}$. The other $g-1$ pairs of canonical cycles can be formally defined by taking images of $\hat{a}_k, \hat{b}_k$ under an involution ($\iota$) which exchanges the two copies of $\mathcal{S}$, with $a_g, b_g$ being mapped back to themselves. We now introduce differential one-forms $\nu_i = \nu_i (\hat{z} ) d\hat{z} $ which are odd under the involution. Just like the holomorphic one-forms of $\mathcal{S}$ that define the period matrix, one can normalize them and construct a symmetric $(g-1) \times (g-1)$ period matrix $\Pi_{ij}$ corresponding to these differentials, i.e.
\be
\label{Prymdifferentials}
\oint_{\hat{a}_i} \nu_j = - \oint_{\iota (\hat{a}_i)} \nu_j = \delta_{ij}, \qquad \oint_{\hat{b}_i} \nu_j = - \oint_{\iota (\hat{b}_i)} \nu_j = \Pi_{ij},\,\,\, (i,j = 1, \ldots, g-1).
\ee
The one-forms $\nu$ are the Prym differentials and $\Pi$ the Prym period which is fixed by the original period matrix of $\mathcal{S}$ up to the action of the Torelli subgroup. An implicit relation between these periods can be obtained by using theta functions defined on Riemann surfaces of genus $g$ and $g-1$.  

Henceforth, we will restrict ourselves to genus two for an explicit account (see \cite{Aoki:TwoLoops} for a related analysis), though higher-genera generalizations should be straightforward.
The classical string instantons fall into $2^{2\times2}$ distinct sectors corresponding to $H_1 (\mathcal{S}, \mathbb{Z}_2 )$ and like in characterizing the genus two theta functions, we can introduce a two-by-two matrix valued characteristic $\epsilon$ where the vectors in the top and bottom indicate periodic/anti-periodic boundary conditions along the $a$ and $b$ cycles respectively. For definiteness, let's take the following twist to construct the unramified double cover described previously.
\be
\label{standardtwist}
\epsilon = \left(  \begin{array}{cc} 0 & 0 \\ 0 & \frac{1}{2} \end{array}   \right).
\ee
And we shall denote the Prym period in \eqref{Prymdifferentials} by $\tau_{\epsilon}$.
It is a beautiful fact that the Prym period
can be related to the period matrix of $\mathcal{S}$ by the implicit relations
\bea
\label{Schottky}
&&\frac{\theta \left[ \delta^+_i \right] (0,\Omega)^2 \theta \left[ \delta^-_i \right] (0,\Omega)^2}{\theta_i (0, \tau_\epsilon)^4} = \frac{\theta [ \delta^+_j ] (0,\Omega)^2 \theta [ \delta^-_j ] (0,\Omega)^2}{\theta_j (0, \tau_\epsilon)^4},\qquad \textrm{for}\,\,\, i,j = 1,2,3, i\neq j  \cr
&& 
\delta^+_2 = \left( \begin{array}{cc} \frac{1}{2} & 0 \\ 0 & 0 \end{array} \right),
\delta^+_3 = \left( \begin{array}{cc} 0 & 0 \\ 0 & 0 \end{array} \right),
\delta^+_4 = \left( \begin{array}{cc} 0 & 0 \\ \frac{1}{2} & 0 \end{array} \right),
\delta^-_k \equiv \delta^+_k + \epsilon,
\eea
where the index on the twist of the genus two theta functions is chosen to be compatible with the characteristic labeling of the genus one theta functions. This `Schottky' relation can be proved by describing the genus two surface as a hyperelliptic curve and using Thomae identitites to relate branch points and the theta functions ( see for example \cite{Gunning:Theta} ). Given some period matrix $\Omega$, this fixes the Prym period up to a translations of integral multiples of eight. 
We note that the twist vectors in the first column of $\delta^{\pm}_i$ in \eqref{Schottky}
can be identified with the three even spin structures in the genus one case, and that $\theta_j (0,\tau_\epsilon)$ are genus one theta functions.

For the purpose of tracking the modular covariance of asymmetric orbifold CFTs on higher-genus curves, recall that our starting step involves finding an appropriate $\mathbb{Z}_2$-valued phase factor that accompanies the T-duality twist by seeking mutual locality consistency conditions for the string vertex operators. As we mentioned earlier, such a condition shouldn't depend on the global properties of the string worldsheet, so roughly speaking, we should expect our considerations to generalize straightforwardly for worldsheet of higher genera. Nonetheless, recall that the $\mathbb{Z}_2$-valued phase factors involve winding modes that make sense with reference to the topology of the worldsheet. Below, we wish to explore this fact explicitly for the simplest T-fold, leaving generalizations to other asymmetric orbifolds for future work. 

We begin with the case of the geometric $\mathbb{Z}_2$ orbifold of the self-dual boson - the symmetric counterpart of the simplest T-fold. To any genus expansion, it was explained beautifully in \cite{DVV} that this CFT has the simple equivalent description in terms of the CFT of a boson of twice or half the self-dual radius (in our units, it's the unity). We can begin with the partition function corresponding to the orbifold twist specified in \eqref{standardtwist}  
\be
\label{twistedChoice}
Z_{\epsilon} \left(  \Omega \right) = Z^{quant.}_{\epsilon} \sum_{P_{L,R}} q^{\frac{1}{4}P_L^2 }\bar{q}^{\frac{1}{4} P_R^2}, \qquad q \equiv e^{2\pi i \tau_\epsilon}, \bar{q} \equiv e^{-2\pi i \bar{\tau}_\epsilon},  
\ee
where the residual left and right momenta modes descend from the string winding around those cycles that project onto the untwisted cycles in the surface $\mathcal{S}$, and we have temporarily denoted the excited stringy states' contributions to be $Z^{quant.}_{\epsilon}$. The other partition traces can be obtained from \eqref{twistedChoice} by acting on the latter with elements of $Sp(4,\mathbb{Z})$. 

Now, $Z^{quant.}_\epsilon$ depends solely on the determinant of the Laplace equation on the curve, with global boundary conditions specified by the twist $\epsilon$. In the genus one case, the Laplace equation can be solved explicitly with the oscillators as the Fourier modes, and $Z^{quant.}_\epsilon$ is the inverse of the absolute square of the Dedekind eta function. As we reviewed in Section two, in some twisted sector, one can compute the twisted determinants easily to yield the chiral blocks expressible in terms of the genus one theta functions and the eta function. For higher-genera, it is not so obviously clear how does one go about computing these twisted determinants, but as a start, we shall first review the trick in \cite{DVV} where the $\mathbb{Z}_2$ twisted determinant can be expressed in terms of the untwisted one. Of course, once this is obtained, it is valid for any orbifold of any compact boson of any radius.

The trick employed delicately in \cite{DVV} is to use the fact that this symmetric orbifold background is equivalent to that of another compact boson of either half or two. For the purely toroidal CFT, there is no notion of `twisted sectors', but it turns out that one can decompose the partition function in terms of a sum of partition traces equivalent to the decomposition in a symmetric $\mathbb{Z}_2$
orbifold. Explicitly as explained in \cite{DVV},
\be
\label{equivalence}
\sum_{\epsilon, \gamma \in \{0,\frac{1}{2} \}} Z^{quant.}_\epsilon \left| \theta \left[ \begin{array}{c} \gamma \\ 0 \end{array} \right] (0|2 \tau_\epsilon) \right|^2 = \sum_{\epsilon, \gamma \in \{0,\frac{1}{2} \}} Z^{quant.}_0\left| \theta \left[ \begin{array}{c} \frac{1}{2}\epsilon_a +  \gamma \\ \epsilon_b \end{array} \right] (0|2 \Omega) \right|^2 
\ee
where we have displayed the toroidal partition function on RHS and $\epsilon_{a,b}$ are the rows of $\epsilon$ in \eqref{standardtwist}.
We note the appearance of another $\mathbb{Z}_2$-valued index $\gamma$. On the LHS, this simply arises from the fact that the partition function of the self-dual boson can be expressed as the sum of the absolute square of two genus one theta functions. On the RHS, this index can be interpreted as a projection on even momentum states. Such an equivalence (sector by sector in $\gamma$) allows us to express the twisted determinant in terms of the untwisted one, and implies the modular covariance of the simplest T-fold to all orders in string perturbation theory. Now, the ratio should be independent of the index $\gamma$, and indeed this is nothing but the Schottky relation we encountered earlier in \eqref{Schottky}. Thus, 
\be
\label{quantumequiv}
Z^{quant.}_\epsilon (\Omega) = \left| \frac{      \theta \left[ \begin{array}{c} \frac{1}{2}\epsilon_a +  \gamma \\ \epsilon_b \end{array} \right] (0|2 \Omega)   }{  \theta \left[ \begin{array}{c} \gamma \\ 0 \end{array} \right] (0|2 \tau_\epsilon)       } \right|^2 
\,\, Z^{quant.}_0 (\Omega)
\ee
for any choice of $\gamma= \{0,\frac{1}{2}\}$. 

In the case of the simplest T-fold, we have observed earlier that by virtue of some identities among the theta functions and eta function, the partition traces are equivalent to that of the symmetric $\mathbb{Z}_2$ orbifold. This equivalence allows us to write down the higher-genus partition traces where the $\mathbb{Z}_2$-valued phase factors are manifest. We begin by intuitively assembling the holomorphic and anti-holomorphic pieces for each partition trace. (This is sometimes referred to as `chiral-splitting' or `holomorphic factorization', as in for example \cite{Aoki}.) 
Taking the twist to act only on the right-movers, the excited stringy states should assemble to yield 
\be
\label{quantumhigherSimplest}
Z^{quant.}_\epsilon (\Omega)= \frac{     \overline{ \theta \left[ \begin{array}{c} \frac{1}{2}\epsilon_a +  \gamma \\ \epsilon_b \end{array} \right] (0|2 \Omega)  } }{ \overline{ \theta \left[ \begin{array}{c} \gamma \\ 0 \end{array} \right] (0|2 \tau_\epsilon)}}\,\,Z^{quant.}_0 (\Omega),
\ee
where now $Z^{quant.}_\epsilon$ pertains to the asymmetric $\mathbb{Z}_2$ orbifold. 
To be more careful, for one to deduce the form of \eqref{quantumhigherSimplest}, one needs the untwisted determinant to be holomorphically factorizable too, i.e. it can be written as the absolute square of some complex function of the period matrix. It turns out that this can be done by a description of the Riemann surface via Schottky uniformization, i.e. representing the curve as the quotient of the Riemann sphere by discrete subgroups of $SL(2,\mathbb{C})$. 
In such a description (see \cite{Zograf, McIntyre}), there are appropriate higher-genera analogues of the variables $q,\bar{q}$ carrying with them information about the twist and length of each handle.
and the untwisted determinant can be expressed as the absolute square of some complex function of $q$.\footnote{Of course, the full string path-integral involves integrating over the moduli space of Riemann surfaces, and it is an open question of how to perform this integration in these variables for generic genus.} 
Since we do not need explicit details of such a construction in this paper, we leave it to the interested reader to refer to \cite{Zograf, McIntyre} for the explicit mathematical proof and also the Appendix of \cite{Tan} for a review.

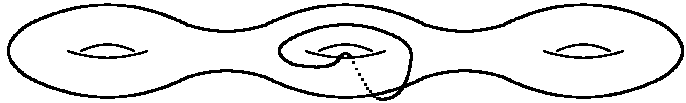
\begin{figure}[h]
\centering
\begin{picture}(250,80)
\linethickness{0.225mm}
      \qbezier(0,50)(30,65)(60,50)
      \qbezier(60,50)(75,45)(90,50)
      \qbezier(90,50)(120,65)(150,50)
      \qbezier(150,50)(165,45)(180,50)
      \qbezier(180,50)(210,65)(240,50)
    \qbezier(0,50)(-15,40)(0,30)
     \qbezier(0,30)(30,15)(60,30)
      \qbezier(60,30)(75,35)(90,30)
      \qbezier(90,30)(120,15)(150,30)
      \qbezier(150,30)(165,35)(180,30)
      \qbezier(180,30)(210,15)(240,30)
    \qbezier(240,50)(255,40)(240,30) 
\linethickness{0.15mm}
\qbezier(15,40)(30,35)(45,40)
  \qbezier(20,40)(30,45)(40,40)
   \qbezier(105,40)(120,35)(135,40)
  \qbezier(110,40)(120,45)(130,40)
 \qbezier(195,40)(210,35)(225,40)
  \qbezier(200,40)(210,45)(220,40)
\linethickness{0.23mm}
\qbezier(100,45)(120,55)(140,45)
\qbezier(100,45)(90,40)(100,35)
\qbezier(100,35)(116,32)(118,38)  
\qbezier(118,38)(120,40)(122,38) 
\qbezier[10](122,38)(128,25)(132,22) 
\qbezier(132,22)(143,20)(145,38)  
\qbezier(145,38)(145,42)(140,45) 
\end{picture}
\caption{The unramified double covering corresponding to a $\mathbb{Z}_2$ twist inserted in a $b$-cycle. We have put in a classical string configuration with an unit winding number (i.e. $n_2 = 1$ in \eqref{rightMom}) along each cycle of the twisted handle. The twist phase factor reads $(-1)^{n_2}$.}
\label{fig:covertwist1}
\end{figure}

What about the instanton part? Let us denote by $f(p)$ the phase factors that accompany the orbifold twists. We can then write it as 
\be
\label{classicalHigher}
\sum_{P_{L,R}} f(P_{L,R}) e^{ \frac{i\pi}{2} P_L \cdot \Omega \cdot P_L} e^{-\frac{i\pi}{2} P_R \cdot \bar{\tau_\epsilon} \cdot P_R}
\ee 
based on the fact that we can create the momentum states in each sector independently by the appropriate vertex operators. Formally, we need the proper machinery of an operator formalism of CFT on higher-genus Riemann surfaces, such as the one proposed in \cite{DVV}, but for the simplest T-fold we take good advantage of its equivalence to the ordinary circle theory to justify the form of \eqref{classicalHigher}. The final ingredient is the derivation of the phase factors $f(P_{L,R})$. This we can do easily by equating it to the partition function of the symmetric $\mathbb{Z}_2$ orbifold which yields
\be
\label{equivalence2}
\sum_{P_{L,R}} f(P_{L,R}) e^{ \frac{i\pi}{2} P_L \cdot \Omega \cdot P_L} e^{-\frac{i\pi}{2} P_R \cdot \bar{\tau_\epsilon} \cdot P_R}
= 
\frac{\theta \left[ \begin{array}{c} \frac{1}{2}\epsilon_a +  \gamma \\ \epsilon_b \end{array} \right] (0|2 \Omega)}{\theta \left[ \begin{array}{c} \gamma \\ 0 \end{array} \right] (0|2 \tau_\epsilon)} \sum_{P_{L,R}} e^{\frac{i\pi}{2} (p^2_L \tau_\epsilon - p^2_R \bar{\tau}_\epsilon)}.
\ee
Since the zero mode summation in the RHS of \eqref{equivalence2} is identical to the ordinary circle theory on a torus worldsheet with complex structure $\tau_\epsilon$, we can write it as a sum of theta functions as 
\be
\label{instantonsumTheta}
\sum_{P_{L,R}} e^{\frac{i\pi}{2} (p^2_L \tau_\epsilon - p^2_R \bar{\tau}_\epsilon)}=\sum_{\gamma' = \{0,\frac{1}{2}\}} \Bigg| \theta 
\left[ \begin{array}{c} \gamma' \\ 0 \end{array} \right](0|2\tau_\epsilon ) \Bigg|^2.
\ee
Since the RHS of \eqref{equivalence2} is also independent of $\gamma$, we can choose it to be identical to $\gamma'$ appearing in the RHS of \eqref{instantonsumTheta} in each term in the summation, and obtain 
\be
\sum_{P_{L,R}} f(P_{L,R}) e^{i\pi P_L \cdot \Omega \cdot P_L} e^{-i\pi P_R \cdot \bar{\tau_\epsilon} \cdot P_R}
= 
\theta \left[ \begin{array}{c} 0\,\,0 \\ 0\,\, \frac{1}{2} \end{array} \right](0|2\Omega) \sum_m e^{\frac{-i\pi \bar{\tau}_\epsilon }{2}(2m)^2 } + \theta \left[ \begin{array}{c} \frac{1}{2}\,\,0 \\ 0\,\, \frac{1}{2} \end{array} \right](0|2\Omega) \sum_m e^{\frac{-i\pi \bar{\tau}_\epsilon }{2}(2m+1)^2 }.
\ee
The final step involves reading off the phase factors $f(P_{L,R})$. This turns out to be very easy since
\bea
\theta \left[ \begin{array}{c} 0\,\,0 \\ 0\,\, \frac{1}{2} \end{array} \right](0|2\Omega)
&=&\sum_{n_1,n_2} e^{2\pi i (n^2_1 \tau_1 +4\beta n_2 n_1 + n^2_2 \tau_2)} e^{\pi i n_2} \\
\theta \left[ \begin{array}{c} \frac{1}{2}\,\,0 \\ 0\,\, \frac{1}{2} \end{array} \right](0|2\Omega)
&=&\sum_{n_1,n_2} e^{2\pi i ((n_1+\frac{1}{2})^2 \tau_1 +4\beta n_2 (n_1+\frac{1}{2}) + n^2_2 \tau_2)} e^{\pi i n_2}
\eea
from which we see that the phase factors are nothing but $e^{\pi i n_2}$ after identifying 
\be
\label{rightMom}
P_R = (n_1-m,0), P_L = (n_1+m, 2n_2).
\ee
This also tells us that the quantum numbers running along the twisted handle and the untwisted ones in the unramified double cover are related in the usual way for the left and right momenta in the Narain lattice, such that we can take them to be two sets of integers with identical parity. In Figure ~\ref{fig:covertwist1}, we sketched a string instanton configuration on the unramified cover which will receive a non-trivial twist phase factor. Now, one can also check that this is consistent with the separating and pinching limits of the genus-two worldsheet. Thus, this seems to suggest that the twist phase factors simply depend on the residual winding numbers which are defined on the twisted handle.

\begin{figure}[h]
\centering
\begin{picture}(250,120)
\linethickness{0.2mm}
   \qbezier(90,30)(120,45)(150,30)
   \qbezier(90,10)(120,-5)(150,10)
 \qbezier(90,80)(120,95)(150,80)
   \qbezier(90,60)(120,45)(150,60)
\qbezier(90,80)(40,45)(90,10) 
\qbezier(92,62)(72,45)(92,28) 
\qbezier(148,62)(168,45)(148,28)
\qbezier(150,80)(200,45)(150,10)
\linethickness{0.15mm}
   \qbezier(105,20)(120,15)(135,20)
  \qbezier(110,20)(120,25)(130,20)
\linethickness{0.15mm}
   \qbezier(105,70)(120,65)(135,70)
  \qbezier(110,70)(120,75)(130,70)
\linethickness{0.23mm}
\qbezier(100,75)(120,85)(140,75)
\qbezier(100,75)(90,70)(100,65)
\qbezier(100,65)(116,62)(118,68)  
\qbezier(118,68)(120,70)(122,68) 
\qbezier[10](122,68)(128,55)(132,52) 
\qbezier(132,52)(143,50)(145,68)  
\qbezier(145,68)(145,72)(140,75) 

\linethickness{0.23mm}
\qbezier(100,25)(120,35)(140,25)
\qbezier(100,25)(90,20)(100,15)
\qbezier(100,15)(116,12)(118,18)  
\qbezier(118,18)(120,20)(122,18) 
\qbezier[10](122,18)(128,5)(132,2) 
\qbezier(132,2)(143,0)(145,18)  
\qbezier(145,18)(145,22)(140,25) 
\end{picture}
\caption{The unramified double covering corresponding to a $\mathbb{Z}_2$ twist (see \eqref{standardtwist2}) inserted in both $b$-cycles. Like in Fig.~\ref{fig:covertwist1}, we have depicted a disconnected classical string configuration with an unit winding number along each cycle of the twisted handle (i.e. $n_1 = n_2 = 1$ in \eqref{standardtwist2}). For this string instanton, the twist phase factor is trivial since $(-1)^{n_1+n_2}=1$.}
\label{fig:covertwist2}
\end{figure}
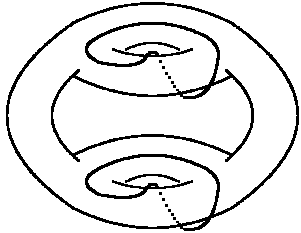

Our preceding discussion pertains to the partition trace $Z^{(0,1)}_{(0,0)}$. It is useful to study how these phase factors appear in the partition trace $Z^{(1,1)}_{(0,0)}$, by performing a suitable modular transformation 
on $Z^{(0,1)}_{(0,0)}$. The $Sp(4,\mathbb{Z})$ element we need is 
\be
\left( \begin{array}{cccc}1 & -1 & -1 & 1 \\ 0 & 1 & 1 & 0\\0 & 0 & 1 & 0 \\0 & 0 & 1 & 1 \end{array} \right)
\ee
and the corresponding twist reads 
\be
\label{standardtwist2}
\epsilon = \left(  \begin{array}{cc} 0 & 0 \\ \frac{1}{2} & \frac{1}{2} \end{array}   \right),\qquad
P_R = (0,0), P_L = (2n_1, 2n_2).
\ee
From the modular property of the theta functions, we found the partition trace to read
\be
\label{twistedpt}
Z^{(1,1)}_{(0,0)} = 
\left( 
\left| 
\theta \left[ \begin{array}{cc}0&0\\ \frac{1}{2}&\frac{1}{2}\end{array} \right](2\Omega)
\right|^2 +
\left| 
\theta \left[ \begin{array}{cc} \frac{1}{2} & \frac{1}{2} \\ \frac{1}{2}&\frac{1}{2}\end{array} \right](2\Omega)
\right|^2
 \right) Z^q_0 (\Omega).
\ee
Expanding the theta functions in \eqref{twistedpt}, we find the twist phase factors of the form $e^{i\pi (n_1 + n_2)}$, where $n_1, n_2$ are integer quantum numbers that can be interpreted as the residual winding numbers along each handle (see Figure ~\ref{fig:covertwist2} ). In the separation limit (see \eqref{sep}), this trace turns into a product of the one-loop partition traces $Z^1_0 (\tau_1) Z^1_0 (\tau_2)$ of the simplest T-fold, as expected.

\section{Discussion}
\label{sec:Dis}

In this paper, we have presented a study of asymmetric orbifolds of tori, with the orbifold group being some $\mathbb{Z}_N$ subgroup of the T-duality group and, in particular, provide a concrete understanding of certain phase factors that may accompany the T-duality operation on the stringy Hilbert space in toroidal compactification. We have explicitly explained how this phase factor is related to the symmetry and locality properties of the closed string vertex operator algebra, and clarified the role that it plays in the modular covariance of the orbifold theory. The mutual locality of vertex operators requires the presence of 2-cocycle maps which help to realize a simple set of constraint equations for the T-duality twist phase factor. These equations descend from preserving the corresponding symmetry of the operator algebra after it has been decomposed into eigenspaces of the twist. They can be interpreted as solving for a certain ratio of the two-cocycles -  $\epsilon \left(\alpha,\beta \right) / \epsilon \left(g(\alpha),g(\beta) \right)$ to be a trivial class of $H^2 (\Lambda, U(1))$, subject to certain orbifold group action-dependent constraints for the one-cochains or twist phase factors. Evaluated upon the invariant sublattices, the twist phase factors are trivial elements of $H^1 (\Lambda, U(1))$ and they should also furnish a representation of $\mathbb{Z}_N$.

As a start, we have focussed on those orbifolds of which twist is trivial in one chiral sector. When the toroidal lattice is the root lattice of some simple Lie algebra, the allowed twists belong to its inner automorphism, and we have computed the T-duality twist phase factors by solving the triviality condition for two-dimensional and six-dimensional examples, the latter being motivated by thinking about asymmetric orbifold points of $CY_3$ compactification of the heterotic string. Upon evaluation on the residual sublattices in the partition traces $Z^g_0$, the twist phase factors ensure that 
\be
\label{levelMatching2}
Z^0_h (\tau + N_h) = e^{i\delta} Z^0_h (\tau),
\ee
where $N_h$ is the order of the twist $h$ and $\delta$ is some real constant. These twist phase factors arise as necessary conditions for T-duality to be an automorphism of the operator algebra, and the constant phases $\delta$ in \eqref{levelMatching2} will appear in level-matching conditions together with other phases that appear after tensoring the bosonic orbifold CFT with other CFTs like that of twisted fermions, shift orbifolds, etc., in the larger string theory. It should be interesting to furnish an equivariant geometric understanding of the modular covariance of asymmetric orbifolds by studying how the methods of \cite{Freed:1987qk} and \cite{Dong} extend to twist phase factor-refined lattice sums. In the seminal work \cite{Moore:1988}, it was shown that modular covariance and fusion rule algebras are related via imposing certain cohomological conditions on the fusing matrices, and thus, it would be nice to study if the twist phase factors can be understood more so in such a manner. If so, it would enable us to study their appearances in other types of orbifolds in particular those which can be described in the language of defect lines (\cite{Frohlich:2009,Fuchs:1999zi,Fuchs:1999xn}).

For the worldsheet theory at higher genus, we have also taken some preliminary steps towards understanding the twist phase factors. Of course, at least in principle, what is required is an appropriate Hamiltonian formalism for CFT at higher genus (such as that proposed in \cite{DVV}) that is within our grasp such that we can generalize our derivation of the T-duality twist phase factor. Nonetheless, as shown in Section 6, we manage to do this for the simplest T-fold - basically by virtue of its equivalence to a geometric orbifold (\cite{DVV}), and in this case, we saw that the twist phase factor can be simply described in terms of the residual winding numbers defined on the handle cut by the twist. It would be interesting to develop this further for generic asymmetric orbifolds (see also Section 2.1 of \cite{Narain2} in this aspect).

The other natural generalizations of this work include a more systematic classification of asymmetric orbifolds (and the corresponding twist phase factors) along the lines of that done in \cite{Kreuzer:1993tf,Fischer}, uncovering their M and F theory origins in the spirit of \cite{Dasgupta:1999ss},
and extending our study of twist phase factors to asymmetric toroidal orientifolds (see for example \cite{Angelantonj:2000xf,Blumenhagen:1998uf}). We have focussed entirely on the closed string sector, and it would be worthwhile to study the role of these phase factors in the boundary states of D-branes that couple to the asymmetric orbifolds \cite{Work}. There has been a number of interesting papers on this issue in the past (see \cite{Gaberdiel:2002jr,Brunner,Craps:2002rw,Bianchi:2008cj,Bianchi:2009mu}), and a more systematic understanding would possibly yield some new non-geometric brane backgrounds via orbifold construction apart from those studied in \cite{deBoer:2010ud,deBoer:2012ma}.

Finally, for those keen in studying modern duality-covariant frameworks like `Double Field Theory', it would be interesting to see how these phase factors arise in those settings where the non-geometric twists may at least naively appear as geometric ones. Already in the seminal papers \cite{Narain1, Narain2}, the twist phase factors are motivated right from the outset by requiring consistent holomorphic factorization of a larger non-chiral bosonic theory, and they are indispensable for one being able to take the square root of stringy instanton sum in the latter as well as modular covariance properties. Our work clearly supports this philosophy. Asymmetric shift orbifolds presented in Section~\ref{subsec:Shift} were first explored in the context of a T-duality covariant sigma model \cite{Hull}
in \cite{Tan}. It would be interesting to see how these twist phase factors appear in the path-integral of `doubled' string sigma models.

\section*{Acknowledgments}
It is a pleasure to thank Miranda Cheng, Simeon Hellerman, Dan Israel and Daniel Robbins for helpful discussions. In particular, I am indebted to Jan de Boer and Ori Ganor for memorable conversations, comments on previous drafts, references and many words of wisdom and encouragement without which this paper would not be completed. I dedicate this work to both of them for being great inspirations to me. Also, I would like to acknowledge support from the Foundation for Fundamental Research on Matter (FOM) which is part of the Netherlands Organization for Scientific Research during the course of completion of this work.

\appendix

\section{On modular covariance}
\label{sec:OnMod}

In this section, we will derive the modular covariance of genus-one characters. Let $\Phi$ denote some target space field, and introduce the periodic worldsheet coordinates $\sigma_{1,2} \sim \sigma_{1,2} + 1$ with worldsheet metric 
\be
ds^2 = \frac{1}{\textrm{Im}(\tau)} \left| d\sigma_1 + \tau d\sigma_2 \right|^2
\ee
Under $SL(2,\mathbb{Z})$, the coordinates and complex structure transform as 
\be
\tau = \frac{a\tilde{\tau} + b}{c\tilde{\tau}+d},\,\,\, \tilde{\sigma}_1 = d\sigma_1 + b\sigma_2,\,\,
\tilde{\sigma}_2 = c\sigma_1 + a\sigma_2.\,\,
\ee
Taking $\sigma_{1,2}$ to be the space and time directions respectively, by definition,
\be 
\Phi (\sigma_1 +1, \sigma_2) = h \circ \Phi (\sigma_1,\sigma_2),\,\,\Phi (\sigma_1, \sigma_2+1) = g \circ \Phi (\sigma_1,\sigma_2),
\ee
and we then have 
\bea
\Phi (\tilde{\sigma}_1 \rightarrow \tilde{\sigma}_1 +1, \tilde{\sigma}_2) = \Phi(\sigma_1 \rightarrow \sigma_1 + a, \sigma_2 \rightarrow \sigma_2 -c) &=& 
h^a g^{-c} \circ \Phi (\tilde{\sigma}_1, \tilde{\sigma}_2), \\
\Phi (\tilde{\sigma}_1, \tilde{\sigma}_2 \rightarrow \tilde{\sigma}_2 +1) = \Phi(\sigma_1 \rightarrow \sigma_1 -b, \sigma_2 \rightarrow \sigma_2 +d) &=& 
h^{-b} g^{d} \circ \Phi (\tilde{\sigma}_1,\tilde{\sigma}_2),
\eea
which is the modular covariance relation in \eqref{Ztrace}. We can pick $\tau \rightarrow \tau + 1$ and $\tau \rightarrow -1/\tau$ to be the two generators of $SL(2,\mathbb{Z})$ and study if \eqref{traceSL2} is satisfied for consistent asymmetric orbifolds. In general, the partition traces do not mix entirely among one another, and there is a $U(1)$ phase degree of freedom $\epsilon (g,h)$ that we can assign to each twisted sector when we compute the complete partition sum. Let $N$ be the order of the finite abelian\footnote{For non-abelian groups, the sum over $g$ involves summing over the maximal subgroup that commutes with $h$.} orbifold group $G$, then
\be
\label{paritionfunction}
Z(\tau) = \frac{1}{N} \sum_{g,h} \epsilon (g,h) Z^g_h (\tau)
\ee
where we have also inserted the discrete torsion $\epsilon (g,h)$ that is related to the two-cocycles $\xi(g,h)$ of the cohomology of the orbifold group valued in $U(1)$, i.e. $H^2\left( G, U(1) \right)$, via the relation
$\epsilon (g,h) = \xi (g,h)/ \xi (h,g).$\footnote{
The two-cocycles satisfy the defining relation 
\be
\xi(g,hf) \xi(h,f) = \xi(g,h) \xi(gh,f)
\ee
and the phases $\epsilon(g,h)$ can be interpreted as measuring the discrete torsion of the cohomology (see for example \cite{Vafa,Sharpe}). It represents the extra ambiguity that one can associate to $Z^g_h (\tau)$ in preserving the relation \eqref{Ztrace}.} For our purpose, we will be dealing with $\mathbb{Z}_N$ orbifolds in our explicit examples, in which case the discrete torsion can be set to unity. They may however arise when there are multiple $\mathbb{Z}_N$ actions.

\subsection{Modular Covariance of Chiral Blocks}
\label{subsec:ChiralBlocks}

Let $g_R$ denote a $\mathbb{Z}_N$ orbifold generator acting on just the right-movers of a closed bosonic string, where the $\mathbb{Z}_N$ acts on a flat $T^2$ with eigenvalues $e^{\pm 2\pi i /N}$.
We begin by considering the following chiral block in some twisted sector where states are twisted by the element $\theta^k$, and in which we insert a chiral orbifold generator $g^l_R$.
\be
\chi^l_k (\tau) \equiv \textrm{Tr}_k \left(  g^l_R q^{L_0}  \right),
\ee
where $l=1,2,\ldots,N-1$. Similarly, we can consider an anti-chiral block in the left sector and write
\be
\overline{\chi}^l_k (\bar{\tau}) \equiv \textrm{Tr}_k \left(  g^l_L \bar{q}^{\bar{L}_0} \right)
\ee
Multiplying these blocks together give us the partition traces of the orbifold theory where the orbifold generator $g = g_R \otimes g_L$,
\be
Z^l_k  (\tau ) = \chi^l_k (\tau) \bar{\chi}^{l}_{k} (\bar{\tau})
\ee
For the moment, we shall exclude all zero mode contributions to $L_0$ (which is suitably normal-ordered, i.e. it contains the casimir energy relevant to periodic boundary conditions ). This counts the oscillators' modes for the right-movers, with the insertion of the operator $g^l_R$. For a generic flat $T^2$ compactification, this character can be expressed in terms of theta functions with characteristics. For example, in the untwisted sector,
\bea
\chi^l_0 (\tau) &=& q^{-\frac{1}{12}} \prod_{m=1}^\infty \left( 1 - q^m e^{\frac{2\pi i l}{N}} \right)^{-1} \left( 1 - q^m e^{-\frac{2\pi i l}{N}} \right)^{-1} \cr
&=&2 \, \sin \left(\frac{l\pi}{N} \right) \frac{\eta(\tau)}{\theta \left[ \frac{1}{2} | \frac{l}{N} -\frac{1}{2} \right] (\tau)}
\eea
where the Jacobi theta function is defined as
\bea
\theta \left[\alpha|\beta \right] (\tau) &\equiv& \eta(\tau) e^{2 \pi i \alpha \beta}q^{\frac{\alpha^2}{2} - \frac{1}{24}} \prod_{m=1}^\infty \left( 1 + q^{m+ \alpha -\frac{1}{2}} e^{2\pi i \beta} \right)
\prod_{m=1}^\infty \left( 1 + q^{m - \alpha - \frac{1}{2}} e^{-2\pi i \beta} \right) \cr
&=& \sum_{m=-\infty}^{\infty} e^{i\pi (n+\alpha)^2 \tau + 2\pi i (n+\alpha)\beta}.
\eea
Similarly, we can write down the chiral block in some twisted sector $k$, with some insertion of $g_R^l$. 
\bea
\chi^l_k (\tau) &=& ie^{i\pi \frac{l}{N} (\frac{k}{N}-1)} \sqrt{\chi (g_R^k, g_R^l)} q^{-\frac{k}{2N} (\frac{k}{N}-1) - \frac{1}{12}} \prod^\infty_{n=1} \left( 1 - q^{n-\frac{k}{N}}e^{2\pi i \frac{l}{N}} \right)^{-1}
\left( 1 - q^{n-1+\frac{k}{N}}e^{-2\pi i \frac{l}{N}} \right)^{-1} \cr
&=& e^{-i\pi \frac{k}{N} (\frac{l}{N}-1)} \sqrt{\chi (g_R^k, g_R^l)} \frac{\eta(\tau)}{\theta \left[ \frac{1}{2}-\frac{k}{N} | \frac{l}{N} -\frac{1}{2} \right] (\tau)}
\eea
where $\chi (g_R^k, g_R^l)$ is the number of common fixed points of $g_R^k$ and $g_R^l$ where for the moment, $g_R$ is taken to be a geometric $\mathbb{Z}_N$ twist. We note that $\chi (\theta^l) = 4 \sin^2 \frac{\pi l}{N}$. The terms in the product count the excited states created by the oscillators (in the diagonal basis of the twist) $\hat{a}_{-(n-\frac{k}{N})}, \hat{a}^*_{-(n-1+\frac{k}{N})}, n>0$, while the vacuum energy in the twisted sector reads $-\frac{2}{24}+\frac{k}{2N}(1-\frac{k}{N})$. The factor $\sqrt{\chi (g_R^k, g_R^l)}$ is reminiscent of a similar factor (without the square root) in the corresponding symmetric orbifold, in which the fixed points label distinct Fock vacua. Last but not least, the factor $e^{i\pi \frac{l}{N} (\frac{k}{N}-1)}$ is inserted so that the chiral block transforms covariantly under the modular $SL(2,\mathbb{Z})$.\footnote{Under a $SL(2,\mathbb{Z})$ element $\gamma$ which takes $\tau$ to $ (a\tau +b)/ (c\tau + d)$, it transforms as
$
\theta \left[ \epsilon| \epsilon' \right] \left( 0 , \gamma (\tau) \right) = \kappa \left[\epsilon|\epsilon';\gamma \right]\sqrt{c\tau +d} \theta \left[ a\epsilon + c\epsilon' - ac| b\epsilon + d\epsilon' + bd \right] (0,\tau), 
\kappa \left[ \epsilon| \epsilon'; \gamma \right] \equiv e^{2\pi i \left( -\frac{1}{4}(a\epsilon + c\epsilon')bd - \frac{1}{8}(ab\epsilon^2 + cd \epsilon'^2 + 2bc \epsilon \epsilon')  \right)} \kappa\left( \gamma \right)
$
where $\kappa \left( \gamma \right)$ is a $\gamma$ dependent eighth root of unity. For our purpose, we only need the values $
\kappa (-\mathds{1}) = -i, \,\,\, \kappa\left(  \gamma(\tau) = -1/\tau \right) = e^{-\frac{\pi i }{4}} ,\,\,\,\kappa\left(  \gamma(\tau) =\tau+1 \right) =1.
$}
Using the relations
\bea
\theta \left[ \alpha |  \beta \right](\tau+1) &=& e^{-i\pi(\alpha^2-\alpha)} \theta \left[ \alpha | \alpha + \beta - \frac{1}{2} \right] (\tau),\,\,\, \theta \left[ \alpha |  \beta \right](-\frac{1}{\tau}) = \sqrt{-i\tau} e^{2\pi i \alpha \beta} \theta \left[ -\beta | \alpha  \right] (\tau)
\cr
\eta(\tau + 1) &=& e^{\frac{i\pi}{12}} \eta(\tau),\,\,\,\eta(-1/\tau) = \sqrt{-i\tau} \eta(\tau).
\eea
we can show that chiral blocks $\chi^l_k(\tau)$ transform as
\be
\chi^l_k (\tau + 1)  = e^{-\frac{i\pi}{6}} \chi^{l-k}_k (\tau) ,\,\,\, \chi^l_k (-1/\tau)  = e^{-\frac{i\pi }{2} } \chi^{-k}_l (\tau)
\ee
and with conjugate phase factors for the anti-chiral blocks.
Even without the extra phase factors 
$e^{i\pi \frac{l}{N} (\frac{k}{N}-1)}$ in the partition trace, any $U(1)$ valued modular anomaly in the chiral sector is cancelled away by the opposite factor in the anti-chiral half, so there is no concern for one-loop modular anomaly arising in this manner. This is of course provided that the insertion of these phases is justified from the operator point of view (or from other considerations like discrete torsion). For symmetric orbifolds, we can do away with these factors because regardless of the sector, the phase factors come in conjugate pairs and thus cancel away in the partition trace $Z^l_k$. 

There is another feature about these blocks that is important because it relates to level-matching in the string theory. This is the fact these blocks do not furnish a faithful representation of the $\mathbb{Z}_N$ group, a problem which is sometimes referred to as a `global anomaly'. To see this, one can easily verify that
\be
\chi^{l+N}_k (\tau) = e^{i\pi (\frac{k}{N}-1)} \chi^l_k (\tau), \qquad \chi^l_{k+N} (\tau) = -e^{-i\pi \frac{l}{N}} \chi^l_k (\tau)
\ee
We shall adopt, as a consistency principle of the asymmetric orbifold, the rule that when all the various chiral blocks are assembled together, the phase factors should sum up to be trivial. By definition, this leads to constraints on the allowed twists and thus the ground state energies of $L_0$ and $\bar{L}_0$. We should mention that without the additional phase factors $e^{i\pi \frac{l}{N} (\frac{k}{N}-1)}$, the blocks $\tilde{\chi}$ now transform as 
\be
\tilde{\chi}^{l+N}_k (\tau) =\tilde{\chi}^l_k (\tau), \qquad \tilde{\chi}^l_{k+N} (\tau) = -e^{-2i\pi \frac{l}{N}} \tilde{\chi}^l_k (\tau)
\ee
For a symmetric orbifold, there is thus no global anomaly of the partition traces.

For higher-genus worldsheets, the stringy instanton sums can be expressed in terms of theta functions
associated with Riemann surfaces endowed with the period matrix in \eqref{periodmatrix} which are defined as (see for example \cite{Alvarez,Gunning:Theta})
\be
\theta \left[  \begin{array}{c} \vec{a} \\ \vec{b}  \end{array}   \right] \left( \vec{z}, \Omega \right)
=\sum_{\vec{n} \in \mathbb{Z}^g} 
\textrm{exp}\left[  i\pi (\vec{n} + \vec{a} )\cdot \Omega (\vec{n} + \vec{a} ) + 2\pi i (\vec{n} + \vec{a})(\vec{z} + \vec{b}) \right]
\ee
where the $g$-dimensional vectors $\vec{a}, \vec{b} \in \mathbb{R}^g$ are known as its characteristics. 
On the theta functions characteristics, an element of $Sp(2g,\mathbb{Z})$ acts as
\be
\label{highergenusthetafunction}
\left[  \begin{array}{c} \vec{a'} \\ \vec{b'}  \end{array}   \right] 
= \left( \begin{array}{cc} D & -C\\ -B & A \end{array} \right) 
\left[  \begin{array}{c} \vec{a} \\ \vec{b}  \end{array}   \right] + 
\frac{1}{2} \left[  \begin{array}{c} (CD^T)_{diag.} \\ (AB^T)_{diag.}  \end{array}   \right] 
\ee
with the complete transformation law being 
\bea
\label{transformationII}
&&\theta \left[  \begin{array}{c} \vec{a'} \\ \vec{b'}  \end{array}   \right] (\Omega')
= \xi (M) e^{-i\pi \phi (\vec{a},\vec{b},\Omega )} \textrm{Det} \left( C\Omega + D \right)^{\frac{1}{2}} \theta \left[  \begin{array}{c} \vec{a} \\ \vec{b}  \end{array}   \right] (\Omega) \cr
&&\phi (\vec{a},\vec{b},\Omega) = \vec{a} \cdot D^TB\cdot \vec{a} + \vec{b} \cdot C^T A \cdot \vec{b} 
-2\vec{a} \cdot B^T C \cdot \vec{b} + (\vec{a} \cdot D^T - \vec{b} \cdot C^T ) \cdot (AB^T)_{diag.}\nonumber \\
\eea
where $\xi(M)$ is a constant eighth root of unity and is equal to $e^{\frac{i\pi}{4}\textrm{Tr} (D-1)}$ if $M$ is equivalent to the identity matrix modulo two. 
The symplectic group is isomorphic to the quotient of the mapping class group modulo the Torelli subgroup which consists of Dehn twists along homologically trivial cycles on the Riemann surface. We can represent the canonical cycles as a $2g$-dimensional vector on which the $Sp(2g,\mathbb{Z})$ matrices act on. Each symplectic element can be taken as some product of Dehn twists around the canonical cycles. 

It is useful to briefly discuss degeneration limits which we have used as consistency checks in Section \ref{sec:Genera}. To this end, we parametrize the period matrix $\Omega$ in terms of three independent parameters as follows 
\be
\label{GenustwoPeriod}
\Omega = \left( \begin{array}{cc} \tau_1 & 2\beta \\ 2\beta & \tau_2 \end{array} \right),\,\,\, \tau_{1,2},\beta \in \mathbb{C}
\ee
Recall that there are two classes of degeneration limits corresponding to whether one is squeezing a homologically trivial or non-trivial cycle. One can pinch any of the two handles by taking $\tau_1 \rightarrow i \infty$ or $\tau_2 \rightarrow \infty$, yielding a torus with a double point. The genus two theta functions then reduce to those defined on the torus in the following manner
\bea
\label{pinch}
\lim_{\tau_2 \rightarrow i\infty} \theta \left[ \begin{array}{cc} a_1 & a_2 \\ b_1 & b_2 \end{array} \right] \left( 0| \Omega \right) 
&= &\theta \left[ \begin{array}{c} a_1 \\ b_1 \end{array} \right] (0|\tau) \,\,\, \text{if}\,\,a_2\in \mathbb{Z}, \cr
\lim_{\tau_2 \rightarrow i\infty} \theta \left[ \begin{array}{cc} a_1 & a_2 \\ b_1 & b_2 \end{array} \right] \left( 0| \Omega \right) 
&=&
e^{\frac{i\pi \tau_2}{4}} \left(  e^{\frac{i\pi b_2}{2}} + e^{i\pi (a_1 b_1 - \frac{b_2}{2})}   \right)
\theta \left[ \begin{array}{c} a_1 \\ b_1 \end{array} \right] (\beta|\tau) \,\,\, \text{if}\,\,a_2\in \mathbb{Z}+1/2, \cr
\lim_{\tau_1 \rightarrow i\infty} \theta \left[ \begin{array}{cc} a_1 & a_2 \\ b_1 & b_2 \end{array} \right] \left( 0| \Omega \right) 
&= &\theta \left[ \begin{array}{c} a_2 \\ b_2 \end{array} \right] (0|\tau) \,\,\, \text{if}\,\,a_1\in \mathbb{Z}, \cr
\lim_{\tau_1 \rightarrow i\infty} \theta \left[ \begin{array}{cc} a_1 & a_2 \\ b_1 & b_2 \end{array} \right] \left( 0| \Omega \right) 
&=&
e^{\frac{i\pi \tau_1}{4}} \left(  e^{\frac{i\pi b_1}{2}} + e^{i\pi (a_2 b_2 - \frac{b_1}{2})}   \right)
\theta \left[ \begin{array}{c} a_2 \\ b_2 \end{array} \right] (\beta|\tau) \,\,\, \text{if}\,\,a_1\in \mathbb{Z}+1/2, \nonumber \\
\eea
whereas pinching a homologically trivial cycle leads to two tori linked by a long tube in the limit $\beta = 0$, and in this separation limit, the theta function factorizes because
\be
\label{sep}
\lim_{\beta = 0} \theta \left[ \begin{array}{c} a_1\,\,a_2 \\ b_1\,\,b_2 \end{array} \right] = 
\theta \left[ \begin{array}{c}  a_1 \\ b_1 \end{array} \right]\theta \left[ \begin{array}{c}  a_2 \\ b_2 \end{array} \right] 
-\frac{i \beta}{\pi} \partial_z \theta \left[ \begin{array}{c} a_1 \\ b_1 \end{array} \right] (z|\tau_1) 
\partial_z \theta \left[ \begin{array}{c} a_2 \\ b_2 \end{array} \right] (z|\tau_2)\vert_{z=0} + \ldots
\ee

\subsection{Twisted chiral fermions with GSO projections}
\label{subsec:Fermions}

We can perform a similar analysis for complex fermions which we briefly review below. Apart from the spin structures defined along the two homology cycles of the torus, one can compute the twisted genus-one characters. Just like for the complex bosons, let $i=1,2$ denote the toroidal directions in a basis where the orbifold action is diagonalized, and define the complex chiral fermionic field $\psi = \frac{1}{\sqrt{2}} \left(  \psi^1 + i\psi^2 \right)$ with the following boundary conditions\footnote{The negative sign arises as the path integral is performed with anti-periodic boundary conditions. It can of course be removed with the insertion of $(-1)^F$.}
\bea
\psi \left( \sigma_1 +1, \sigma_2 \right) &=& -e^{2\pi i \alpha} \psi \left( \sigma_1, \sigma_2 \right), \cr
\psi \left( \sigma_1 , \sigma_2 +1 \right) &=& -e^{2\pi i \beta} \psi \left( \sigma_1, \sigma_2 \right).
\eea
where $\alpha,\beta$ are twist parameters in the worldsheet space and time directions. We should note that in the absence of orbifold twists, they refer to the sector and the GSO projection respectively. The character can be computed easily after realizing the twist operator in terms of the expansion modes of $\psi$. Writing $\psi(z) = \sum_{n\in \mathbb{Z} } \hat{\psi}_{n+\alpha + \frac{1}{2}} e^{-i(n+\alpha + \frac{1}{2})z}$, the operator $g$ inserted in the trace is realized as $g = e^{2\pi i \beta \sum_{k>0} ( \psi_{-k}\bar{\psi}_k - \bar{\psi}_{-k} \psi_k )}$, where $\bar{\psi}$ refers to its complex conjugate. Then the fermionic partition trace can be expressed as 
\bea
\label{tracefermion}
\chi^\beta_\alpha (\tau) &=& q^{\frac{\alpha^2}{2} - \frac{1}{24}} \prod_{m=1}^\infty 
\left( 1 + q^{n+\alpha - \frac{1}{2}} e^{-2\pi i \beta} \right)
\left( 1 + q^{n-\alpha - \frac{1}{2}} e^{2\pi i \beta} \right) \cr
&=& e^{2\pi i \alpha \beta} \frac{\theta[\alpha|-\beta] (\tau)}{\eta(\tau)}
\eea
where as usual, the prefactor arises from a regularized one-point function of the fermion's energy momentum tensor. Now for $\mathbb{Z}_N$ orbifolds, in some fixed twisted sector labelled by $k$ and with the insertion of $g^l$, we can redefine the character \eqref{tracefermion} after summing up over the spin structures and appropriate GSO projections. Thus, letting $\alpha, \beta \in \{ 0,\frac{1}{2} \}$, we can write\footnote{We are taking $|\alpha + \frac{k}{N}| < \frac{1}{2}$, otherwise, we have to send $\alpha \rightarrow \alpha -1$.}
\be
\label{fermionblocks}
\chi^l_k  (\tau) = \sum_{\alpha, \beta} \mathfrak{C}_{\alpha \beta} (k,l) \frac{ \theta \left[\alpha + \frac{k}{N} | -\beta - \frac{1}{N} \right]}{\eta (\tau)} 
\ee
where $\mathfrak{C}_{\alpha \beta}$ are some constant spin-structure coefficients that can be possibly managed to preserve modular invariance, with $\alpha = 0, \frac{1}{2}$ labelling the NS and R sectors respectively. This has of course been well-understood since a long time ago. Recall that in \eqref{fermionblocks}, the GSO projection is manifest 
in the insertion of $(-1)^F$ in the partition trace but we have to specify the phase factors that accompany each such insertion. For a critical string theory in the light-cone gauge, we have four complex fermions. In the absence of the twists, the partition trace in the left-moving sector then reads
\be
\label{untwistedGSO}
Z_{\pm} (\tau) = \frac{1}{2\eta^4(\tau)} \left( \theta [ 0 | 0 ]^4(\tau) - \theta \left[\frac{1}{2} \Bigg| 0 \right]^4(\tau) - \theta \left[0 \Bigg| \frac{1}{2} \right]^4(\tau) \pm \theta \left[\frac{1}{2} \Bigg| \frac{1}{2} \right]^4(\tau) \right)
\ee
where the various signs are picked to preserve modular covariance. We should note that the last term is identically zero and so is the entire partition trace by a Jacobi's identity, indicating spacetime supersymmetry. Let us now insert the orbifold twists, and generalize the various signs with the coefficients $\mathcal{C}_{\alpha \beta} (k,l)$. Requiring $Z^l_k (\tau + 1) \sim Z^{l-k}_k (\tau)$ and $Z^l_k (-1/\tau) \sim Z^{-k}_l (\tau)$ up to phase factors yields the relations, after setting $\mathcal{C}_{00}=1$,
\be
\mathcal{C}_{0\frac{1}{2}} (k,l) = -e^{i\pi \sum k_i}, \mathcal{C}_{\frac{1}{2} 0} (k,l)= -1, \mathcal{C}_{\frac{1}{2} \frac{1}{2}} (k,l) = e^{i\pi \sum k_i} 
\ee
where we have adopted the positive sign in the last term of the untwisted sum of \eqref{untwistedGSO}, and importantly, we find
\bea
Z^l_k \left( -\frac{1}{\tau} \right) &=& e^{-2\pi i \sum_m l_m k_m } Z^{-k}_l (\tau) \\
Z^l_k (\tau +1) &=& e^{-\pi i \sum_m k_m^2}e^{\frac{2\pi i}{3}} Z^{l-k}_k (\tau).
\eea
For the orbifolds we considered in the previous sections, the $\mathbb{Z}_N$ twists sum to zero, 
so the spin-structures coefficients are identical as in \eqref{untwistedGSO} and we have the same GSO projection in each twisted sector $Z^l_k$, i.e.
\be
\label{GSOtwisted}
Z^l_k (\tau) =\frac{1}{2} \textrm{Tr}_{k,NS} \left[ \hat{g}^l q^{L_0} (1-(-1)^F) \right] + \frac{1}{2}  \textrm{Tr}_{k,R} \left[ \hat{g}^l q^{L_0} (1+(-1)^F) \right] 
\ee
The condition that the twists sum to zero yields a vanishing partition trace too, by virtue of a generalized Jacobi identity that reads
\be
\sum_{\alpha,\beta \in \{ 0,\frac{1}{2} \}} e^{2\pi i (\alpha + \beta)} \prod_{j=1}^4
e^{2\pi i \alpha k_j} \theta \left[ \alpha + l_j |\beta + k_j \right] (\tau) = 0,\qquad \textrm{if}\,\, \sum_m k_m = \sum_m l_m = 0.
\ee

\subsection{Theta functions and modular covariance of the simplest T-fold}
\label{subsec:Theta}

We begin with the partition trace $Z^1_0 (\tau)$.
For the right-movers, there is a residual instanton sum that counts the distinct configurations invariant under the twist. These are string geometries with winding number equal to momentum number. Taking into account the twist phase factor \eqref{simplestTfoldtwist} that refines the orbifold twist element, the chiral block reads 
\be
\chi^1_0 (\tau) = \frac{1}{\eta(\tau)} \sum_m (-1)^m q^{m^2} = \frac{1}{\eta(\tau)} \theta_4 (2\tau)
\ee
while the anti-chiral block reads 
\be
\bar{\chi}^1_0 (\bar{\tau}) = \bar{q}^{-\frac{1}{24}} \prod_{m=1}^\infty \left( 1+\bar{q}^m \right)^{-1}
= \sqrt{\frac{2\eta (\bar{\tau})}{\theta_2 (\bar{\tau})}}
\ee
Under $\tau \rightarrow \tau + 1$, each block develops a phase of $e^{\pm \frac{i\pi}{12}}$ which thus cancels away, with the instanton sum being invariant. Under $\tau \rightarrow -\frac{1}{\tau}$, we obtain
the partition trace in the twisted sector, with 
\be
\chi^0_1 (\tau) = \frac{\theta_2(\frac{\tau}{2})}{\sqrt{2} \eta(\tau)},\qquad 
\bar{\chi}^0_1 (\bar{\tau}) = \sqrt{\frac{2\eta(\bar{\tau})}{\theta_4(\bar{\tau})}}.
\ee
Further performing $\tau \rightarrow \tau + 1$, we arrive at 
\be
\chi^1_1 (\tau)= \frac{\theta_2 (\frac{\tau+1}{2})}{\sqrt{2} e^{\frac{i\pi}{12}} \eta(\tau)} = e^{\frac{i\pi}{24}} \frac{ \theta_2 \left( \frac{1}{4} ; \frac{\tau}{2} \right)}{\eta(\tau)},\qquad
\bar{\chi}^1_1 (\bar{\tau}) = e^{-\frac{i\pi}{24}} \sqrt{\frac{2\eta(\bar{\tau})}{\theta_3 (\bar{\tau})}}
\ee
The partition trace $Z^1_1$ should be invariant under $\tau \rightarrow -\frac{1}{\tau}$, and further performing $\tau \rightarrow \tau + 1$ should bring it back to $Z^0_1$. These can be straightforwardly verified using the properties of the theta and eta functions, with perhaps the only slighty trickier step being to show that
\bea
\label{thetaplus}
\theta_2 \left( \frac{1}{4};\frac{\tau}{2} \right) &=& \sum_n e^{\frac{i\pi}{2} (n+\frac{1}{2}) + \frac{i\pi \tau}{2} (n+\frac{1}{2} )^2} \cr 
&=& e^{\frac{i\pi \tau}{8} + \frac{i\pi}{4}}  \left(  \sum_m (-1)^m e^{i\pi \tau (2m^2 -m)} + e^{-\frac{\pi i}{2}(2m+1) + \pi i \tau (2m+1)m} \right) \cr
&=& \sqrt{2} \sum_m (-1)^m q^{(m-\frac{1}{4})^2}\cr
&=& \frac{1}{\sqrt{-i\tau}} \sum_m e^{-\frac{i\pi}{2\tau}(m-\frac{1}{2})^2 -\frac{i\pi}{2} (m-\frac{1}{2})}
\eea
where we have performed a Poisson resummation in the last step. Then it is clear that under $\mathcal{S}$,
\be
\theta_2 \left(\frac{1}{4} ; \frac{\tau}{2} \right) \rightarrow \sqrt{-i\tau}\, \theta_2 \left( \frac{1}{4}; \frac{\tau}{2} \right) 
\ee
Symbolically, we summarize the action of the mapping class group elements $\mathcal{S}$ and $\mathcal{T}$ on the partition traces as follows.
\begin{figure}[h]
\centering
\begin{tikzpicture}
\begin{scope}
[bend angle=65, post/.style={->, shorten >=1pt,>=stealth'}]
\node (z11) at (4,0) {$Z^1_1$}
edge [loop right] node[right,swap]{\footnotesize{$\mathcal{S}$}} (z11);
\node (z01) at (2,0) {$Z^0_1$}
edge [post] node[above,swap] { \footnotesize{$\mathcal{T}$}} (z11);
\node (z10) at (0,0) {$Z^1_0$}
edge [post] node[above,swap] { \footnotesize{$\mathcal{S}$}} (z01)
edge [loop left] node[left,swap]{\footnotesize{$\mathcal{T}$}} (z10);
\end{scope}
\end{tikzpicture}
\caption{The modular covariance of the genus-one characters of a $\mathbb{Z}_2$ orbifold.}
\label{fig:Z2Mod}
\end{figure}
We have started with a refined $\mathbb{Z}_2$ orbifold element, then generating the rest of the partition traces by the action of $\mathcal{S}$ and $\mathcal{T}$. For the twisted sector $Z^0_1$, in the case of the geometric $\mathbb{Z}_2$ orbifold, there is an overall factor of two which corresponds to the two sectors of Fock vacua labelled by the two zero modes $x_0= \{ -\pi, \pi \}$ which are fixed points under the geometric reflection twists. The reflection kills off the zero mode contributions, yet the stringy Hilbert space decomposes into two separate sectors each labelled by one value of $x_0$. When the twist is asymmetric, the right-moving sector has surviving zero modes, and the instanton sum replaces the factor of two that appears in the twisted sector of the geometric $\mathbb{Z}_2$ orbifold. Of course, in the generic case, the twisted sectors of an asymmetric orbifold can have degeneracies too, and as first mentioned in \cite{Narain1} and \cite{Narain2}, it is a non-trivial fact that the degeneracy factors are integers (as they should be) and can be expressed generally as 
\be
\label{degoftwistedsector}
D = \sqrt{\frac{ \textrm{Det}(1-\theta_L)  \textrm{Det}(1-\theta_R)}{| I^*/I|}}
\ee
where $I$ is the sublattice of $\Lambda$ invariant under the orbifold twist, and $I^*$ its dual. As mentioned in Section ~\ref{subsec:generalpoints}, for chiral asymmetric orbifolds, the degeneracy in \eqref{degoftwistedsector} reads $\sqrt{4\sin^2 \frac{\pi k}{N}}/\sqrt{\textrm{Det}(2G)}$
in our notations where $G$ is the torus metric and the twist eigenvalue is $e^{2\pi i k/N}$. The origin of this factor was explained in \cite{Narain1} and \cite{Narain2} to be equivalent to the dimension of the irreducible representation of the vertex operators corresponding to untwisted states provided we tensor the vertex operators with a matrix-valued cocycle that acts only on the fixed points of the twist.

What happens when we decide not to augment the chiral reflection with the $U(1)$ factor $(-1)^n$ in $\chi^1_0 (\tau)$?  We find that the relation in Fig.~\ref{fig:Z2Mod} is not satisfied because instead of $\theta_4(2\tau)$ in $\chi^1_0 (\tau)$, we have $\theta_3(2\tau)$. After performing $\tau \rightarrow -\frac{1}{\tau}$, we have $\theta_3 \left( \frac{\tau}{2} \right)$ instead of $\theta_2 \left( \frac{\tau}{2} \right)$ in $\chi^0_1 (\tau)$, yet $\theta_3 \left(  \frac{\tau + 1}{2} \right)$ does not have the same transformation property as $\theta_2 \left( \frac{\tau+1}{2} \right)$ in \eqref{thetaplus}. To see this explicitly, let's first Poisson resum to write
\bea
\theta_3 \left(  \frac{\tau + 1}{2} \right) &=& \frac{1}{\sqrt{-2i\tau}} \sum_n e^{-\frac{i\pi n^2}{2\tau}} + e^{\frac{i\pi \tau}{2} - \frac{i\pi}{2\tau}(n+\tau)^2} \cr
&=& \frac{1}{\sqrt{-2i\tau}} \sum_n \left(  1 + (-1)^n \right) e^{-\frac{i\pi n^2}{2\tau}} = \sqrt{\frac{2i}{\tau}}\theta_3\left( -\frac{2}{\tau} \right)
\eea
Under $\mathcal{S}$, we then have 
\be
\theta_3 \left( \frac{\tau+1}{2} \right) \overset{\mathcal{S}}{\rightarrow} \sqrt{-2i\tau} \, \theta_3 (2\tau)
\ee
Thus, up to a phase of $e^{-\frac{i\pi}{12}}$, we find that $\chi^1_1$ maps back to $\chi^1_0$ instead of being invariant under $\mathcal{S}$. Another way to see that it doesn't work is to see that $Z^0_1$ doesn't map back to itself under $\tau \rightarrow \tau +2$, i.e. no level-matching.

\section{Twist phase factors of some chiral asymmetric orbifolds of $T^6$}
\label{sec:asymHet}

\subsection{$E_6$ orbifolds}
\label{subsec:E6orbifolds}

We first consider chiral $\mathbb{Z}_3$ and $\mathbb{Z}_{12}$ orbifolds of the $E_6$ torus, and pick our moduli matrix $E=G+B$ to be
\be
E = \left( \begin{array}{cccccc} 1 & -1 & 0 & 0 & 0 & 0 \\ 0 & 1 & -1 & 0 & 0 & 0\\ 
0 & 0 & 1 & -1 & 0 & -1\\ 0 & 0 & 0 & 1 & -1 & 0\\ 0 & 0 & 0 & 0 & 1 & 0\\ 
0 & 0 & 0 & 0 & 0 & 1 \end{array} \right)
\ee
The $\mathbb{Z}_3, \mathbb{Z}_{12}$ twists which we shall discuss below are constructed by taking suitable products of the Weyl reflections. Let $\alpha_i,\, i=1,2,\ldots 6$ denote its six simple roots\footnote{Let $e_i$ denote the vector with unity as its $i$th component and zero for the rest, then $\alpha_i = e_i - e_{i+1}, \alpha_5 = e_4 + e_5, \alpha_6 = \frac{1}{2}(-e_1 -e_2 - e_3 -e_4 +e_5 +\sqrt{3} e_6)$.}, and let $r_i$ denote the Weyl reflection associated with the root $\alpha_i$. 
Realizing the twist as $\theta$ acting on the metric $G$ by $G \rightarrow \theta G \theta^T$, it is straightforward to compute them to be ($r_0 = -\alpha_1 -2\alpha_2 -3\alpha_3 -2\alpha_4 -\alpha_5-2\alpha_6$ is the lowest root which appears in the extended Dynkin diagram)
\be
\theta^{\mathbb{Z}_3} = r_1 r_2 r_4 r_5 r_6 r_0 =
\left( \begin{array}{cccccc} 
-1 & -1 & 0 & 0 & 0 & 0\\
1 & 0 & 0 & 0 & 0 & 0\\
-1 & -1 & -2 & -1 & 0 & -1\\
0 & 0 & 0 & -1 & -1 & 0\\
0 & 0 & 0 & 1 & 0 & 0\\
1 & 2 & 3 & 2 & 1 & 1
\end{array} \right)
\ee
\be
\theta^{\mathbb{Z}_{12}} = r_1 r_2 r_3 r_4 r_5 r_6  =
\left( \begin{array}{cccccc} 
-1 & -1 & -1 & -1 & -1 & -1\\
1 & 0 & 0 & 0 & 0 & 0\\
0 & 1 & 0 & 0 & 0 & 0\\
0 & 0 & 1 & 0 & 0 & 1\\
0 & 0 & 0 & 1 & 0 & 0\\
0 & 0 & 1 & 1 & 1 & 0
\end{array} \right)
\ee
When treated as geometric twists, they yield singular compact manifolds of Euler numbers 48 and 45 respectively, but as asymmetric twists, they are simply realized as symmetries on the stringy Hilbert space.

\subsubsection{$\mathbb{Z}_3$ orbifold}
\label{subsubsec:Z3}

Let us first consider the $\mathbb{Z}_3$ orbifold for which there is only one independent $SL(2,\mathbb{Z})$ orbit. 
We find the following twist phase factors characterized by the following $Q's$ (recall that $U = e^{i N^T Q N}, N=(n\,\, m)$).
\be
Q^{\mathbb{Z}_{3}}=
\pi
\left(
\begin{array}{cccccccccccc}
 \frac{a_1}{2} & \frac{1}{2} & \frac{3}{2} & \frac{3}{2} & \frac{3}{2} & 2 & 1 & \frac{1}{2} & \frac{3}{2} & \frac{1}{2} & \frac{1}{2} & \frac{1}{2} \\
 \frac{1}{2} & \frac{a_2}{2} & 3 & 3 & 3 & 4 & 1 & \frac{3}{2} & \frac{7}{2} & 1 & 1 & 1 \\
 \frac{3}{2} & 3 & \frac{a_3}{2} & \frac{3}{2} & 3 & 3 & 0 & 0 & \frac{9}{2} & 0 & 0 & 0 \\
 \frac{3}{2} & 3 & \frac{3}{2} & \frac{a_4}{2} & 2 & 1 & \frac{1}{2} & \frac{1}{2} & \frac{5}{2} & 1 & \frac{1}{2} & \frac{1}{2} \\
 \frac{3}{2} & 3 & 3 & 2 & \frac{a_5}{2} & 1 & 1 & 1 & \frac{1}{2} & 1 & \frac{3}{2} & 1 \\
 2 & 4 & 3 & 1 & 1 & \frac{a_6}{2} & 1 & 1 & 2 & 1 & 1 & \frac{3}{2} \\
 1 & 1 & 0 & \frac{1}{2} & 1 & 1 & \frac{a_7}{2} & 0 & \frac{3}{2} & 0 & 0 & 0 \\
 \frac{1}{2} & \frac{3}{2} & 0 & \frac{1}{2} & 1 & 1 & 0 & \frac{a_8}{2} & \frac{3}{2} & 0 & 0 & 0 \\
 \frac{3}{2} & \frac{7}{2} & \frac{9}{2} & \frac{5}{2} & \frac{1}{2} & 2 & \frac{3}{2} & \frac{3}{2} & \frac{a_9}{2} & \frac{3}{2} & \frac{3}{2} & \frac{3}{2} \\
 \frac{1}{2} & 1 & 0 & 1 & 1 & 1 & 0 & 0 & \frac{3}{2} & \frac{a_{10}}{2} & 0 & 0 \\
 \frac{1}{2} & 1 & 0 & \frac{1}{2} & \frac{3}{2} & 1 & 0 & 0 & \frac{3}{2} & 0 & \frac{a_{11}}{2} & 0 \\
 \frac{1}{2} & 1 & 0 & \frac{1}{2} & 1 & \frac{3}{2} & 0 & 0 & \frac{3}{2} & 0 & 0 & \frac{a_{12}}{2}
\end{array}
\right)
\ee
where the constants $a_i$ are partially fixed by \eqref{conditionGtwist4} to satisfy $( \text{mod}\,\,\,4 )$
\be
\label{Z3diagconstraint}
a_1 = a_7 = a_8 + a_9 + a_{10}, a_2 = 2 +a_9 + a_{10}, a_3 = 2 + a_{10}, a_4=0, a_5=a_{11}, a_6 = a_{10} + a_{12}.
\ee
In the partition trace $Z^1_0$, the twist phase factor reads, upon evaluated on the invariant sublattice,
\be
\label{Z3phasefactor}
U(\theta, \tilde{p} ) = e^{ i\frac{\pi}{2} \left(  (a_1 + a_7)m^2_1 + (2+a_1 + a_2 +a_8)m^2_2              
+ (a_2 + a_3 +a_9)m^2_3 + + (2+a_3 + a_4 +a_{10} )m^2_4 + + (a_4 + a_5 + a_{11})m^2_5 +
 (2+a_3 + a_6 +a_{12})m^2_6  \right) }
\ee
Imposing \eqref{Z3diagconstraint} in \eqref{Z3phasefactor} renders the latter trivial, in agreement with our earlier point that for chiral asymmetric orbifolds where there are no residual zero modes in the twisted half,  the twist phase factor as evaluated on the invariant sublattice has to be trivial for twist of odd orders. Thus, in this case, we simply have to consider the (unweighted) $E_6$ lattice sum which reads 
\be
\label{E6lattice}
\Theta_{E_6} (\tau) = \frac{1}{2} \left[ \theta_3 (3\tau) \theta^5_3 (\tau) + \theta_4(3\tau) \theta^5_4(\tau) + \theta_2(3\tau) \theta^5_2 (\tau) \right]
\ee
The dual lattice sum can be easily obtained in this case by invoking Jacobi's inversion formula which yields
\be
\label{E6duallattice}
\Theta_{E^*_6} (\tau) = \frac{1}{2} \left[  \theta_3 (\frac{\tau}{3} ) \theta^5_3 (\tau) + \theta_2 (\frac{\tau}{3} ) \theta^5_2 (\tau) + \theta_4 (\frac{\tau}{3} ) \theta^5_4 (\tau) \right]
\ee
and it can be checked that it is invariant under $\mathcal{T}^3$, and thus this asymmetric orbifold is perfectly modular covariant. 

\subsubsection{$\mathbb{Z}_{12}$ orbifold}
\label{subsubsec:Z12}

On the other hand, for the $\mathbb{Z}_{12}$ orbifold, there are five independent $SL(2,\mathbb{Z})$ orbits of which trace representatives we can take to be $\{ Z^0_2, Z^0_3, Z^0_4, Z^0_6, Z^0_1 \}$. Since $\theta, \theta^2$ and $\theta^4$ have no eigenvalue equal to unity, the invariant sublattice in these partition traces is nothing but the $E_6$ lattice. The twist phase factor reads 

\be
\label{Z12phase}
Q=\frac{\pi}{2}\left(
\begin{array}{cccccccccccc}
 a_1 & 1 & 1 & 1 & 1 & 1 & 1 & 0 & 0 & 0 & 0 & 0 \\
 1 & a_2 & 1 & 1 & 1 & 1 & 0 & 1 & 0 & 0 & 0 & 0 \\
 1 & 1 & a_3 & 1 & 1 & 1 & 0 & 0 & 1 & 0 & 0 & 0 \\
 1 & 1 & 1 & a_4 & 1 & 0 & 0 & 0 & 0 & 1 & 0 & 0 \\
 1 & 1 & 1 & 1 & a_5 & 0 & 0 & 0 & 0 & 0 & 1 & 0 \\
 1 & 1 & 1 & 0 & 0 & a_6 & 0 & 0 & 0 & 0 & 0 & 1 \\
 1 & 0 & 0 & 0 & 0 & 0 & a_7 & 0 & 0 & 0 & 0 & 0 \\
 0 & 1 & 0 & 0 & 0 & 0 & 0 & a_8 & 0 & 0 & 0 & 0 \\
 0 & 0 & 1 & 0 & 0 & 0 & 0 & 0 & a_9 & 0 & 0 & 0 \\
 0 & 0 & 0 & 1 & 0 & 0 & 0 & 0 & 0 & a_{10} & 0 & 0 \\
 0 & 0 & 0 & 0 & 1 & 0 & 0 & 0 & 0 & 0 & a_{11} & 0 \\
 0 & 0 & 0 & 0 & 0 & 1 & 0 & 0 & 0 & 0 & 0 & a_{12}
\end{array}
\right)
\ee
On the invariant sublattice, the phase factor in \eqref{Z12phase} reads 
\be
e^{i\frac{\pi}{2} \left( (2+a_1 +a_7)m_1 + (a_1 + a_8 + a_2)m_2 + (a_2 + a_3 + a_9) m_3 +  (a_3+a_4+a_{10})m_4 + (a_4 +a_5 +a_{11})m_5 + (a_3+a_6+a_{12})m_6                 \right)   }
\ee
There are no constraints on the parameters $a_i$ but we can choose all of them to vanish except for $a_1 = a_8 =2$ to get \eqref{Z12phase} to be trivial in $Z^1_0, Z^2_0$ and $Z^4_0$. Then one would find that the appropriate level-matching conditions below are satisfied.
\be
Z^0_1 (\tau + 12) = Z^0_1 (\tau), Z^0_2 (\tau + 6) = Z^0_2 (\tau), Z^0_4 (\tau + 3 ) = Z^0_4 (\tau)
\ee
by virtue of invariance of \label{E6duallattice} under $\tau \rightarrow \tau +3$. For the other two traces, one has find the invariant sublattices first. They are turn out to be the same eight-dimensional lattice in $Z^3_0$ and $Z^6_0$, and can be conveniently described by projecting the residual left and right Narain momenta onto the eight-dimensional integral vector $\vec{v}=\{n_1, n_2, m_1, m_2, m_3, m_4, m_5, m_6 \}$ with the projection matrices $\tilde{P}_L = \mathcal{P}_L \cdot \vec{v}, \tilde{P}_R = \mathcal{P}_R \cdot \vec{v}$. Explicitly, the projection matrices read 
\be
\mathcal{P}_L = \left(
\begin{array}{cccccccc}
 1 & 0 & 1 & 0 & 0 & 0 & 0 & 0 \\
 0 & 1 & -1 & 1 & 0 & 0 & 0 & 0 \\
 -1 & -1 & 1 & -1 & 1 & -1 & 0 & -1 \\
 1 & 0 & -1 & 1 & -1 & 2 & -1 & 0 \\
 0 & 1 & 0 & -1 & 1 & -1 & 2 & 0 \\
 0 & 0 & 0 & 0 & -1 & 0 & 0 & 2
\end{array}
\right),\,\,\,
\mathcal{P}_R =\left(
\begin{array}{cccccccc}
 1 & 0 & -1 & 1 & 0 & 0 & 0 & 0 \\
 0 & 1 & 0 & -1 & 1 & 0 & 0 & 0 \\
 -1 & -1 & 1 & 0 & -1 & 0 & 0 & 0 \\
 1 & 0 & -1 & 1 & 0 & 0 & 0 & 0 \\
 0 & 1 & 0 & -1 & 1 & 0 & 0 & 0 \\
 0 & 0 & 0 & 0 & 0 & 0 & 0 & 0
\end{array}
\right) 
\ee
from which one can compute the lattice matrix $\Upsilon$ straightforwardly, and check that with our choice of the parameters $a_i$, the twist phase factor becomes trivial, and also we have the level-matching conditions
\be
Z^0_6 (\tau+2) = Z^0_6 (\tau),\qquad Z^0_3 (\tau+4) = Z^0_3 (\tau).
\ee
Since we have taken into account the representatives of the five $SL(2,\mathbb{Z})$ orbits, we thus conclude that this orbifold theory is modular covariant.

\subsection{Asymmetric $SU(7)$ Orbifold}
\label{subsec:SU7}

We now consider chiral $\mathbb{Z}_7$ orbifolds of the $SU(7)$ torus, and pick our moduli matrix $E=G+B$ to be
\be
E = \left( \begin{array}{cccccc} 1 & -1 & 0 & 0 & 0 & 0 \\ 0 & 1 & -1 & 0 & 0 & 0\\ 
0 & 0 & 1 & -1 & 0 & 0\\ 0 & 0 & 0 & 1 & -1 & 0\\ 0 & 0 & 0 & 0 & 1 & -1\\ 
0 & 0 & 0 & 0 & 0 & 1 \end{array} \right)
\ee
The $\mathbb{Z}_{7}$ twist which we shall discuss below is constructed by the Coxeter element of the Weyl group. Let $\alpha_i,\, i=1,2,\ldots 6$ denote its six simple roots, then the twist is defined by the product of each Weyl reflection associated with $\alpha_i$, i.e. 
\be
\theta^{\mathbb{Z}_7} = r_1 r_2 r_3 r_4 r_5 r_6 =
\left( \begin{array}{cccccc} -1 & -1 & -1 & -1 & -1 & -1\\
1 & 0 & 0 & 0 & 0 & 0\\
0 & 1 & 0 & 0 & 0 & 0\\
0 & 0 & 1 & 0 & 0 & 0\\
0 & 0 & 0 & 1 & 0 & 0\\
0 & 0 & 0 & 0 & 1 & 0
\end{array} \right)
\ee
We find the following twist phase factor characterized by the following $Q$.
\be
Q=\frac{\pi}{2}
\left(
\begin{array}{cccccccccccc}
 a_1 & 1 & 1 & 1 & 1 & 1 & 1 & 0 & 0 & 0 & 0 & 0 \\
 1 & a_2 & 1 & 1 & 1 & 1 & 0 & 1 & 0 & 0 & 0 & 0 \\
 1 & 1 & a_3 & 1 & 1 & 1 & 0 & 0 & 1 & 0 & 0 & 0 \\
 1 & 1 & 1 & a_4 & 1 & 1 & 0 & 0 & 0 & 1 & 0 & 0 \\
 1 & 1 & 1 & 1 & a_5 & 1 & 0 & 0 & 0 & 0 & 1 & 0 \\
 1 & 1 & 1 & 1 & 1 & a_6 & 0 & 0 & 0 & 0 & 0 & 1 \\
 1 & 0 & 0 & 0 & 0 & 0 & a_7 & 0 & 0 & 0 & 0 & 0 \\
 0 & 1 & 0 & 0 & 0 & 0 & 0 & a_8 & 0 & 0 & 0 & 0 \\
 0 & 0 & 1 & 0 & 0 & 0 & 0 & 0 & a_9 & 0 & 0 & 0 \\
 0 & 0 & 0 & 1 & 0 & 0 & 0 & 0 & 0 & a_{10} & 0 & 0 \\
 0 & 0 & 0 & 0 & 1 & 0 & 0 & 0 & 0 & 0 & a_{11} & 0 \\
 0 & 0 & 0 & 0 & 0 & 1 & 0 & 0 & 0 & 0 & 0 & a_{12}
\end{array}
\right)
\ee
where the diagonal constants are partially fixed by \eqref{conditionGtwist4} to satisfy (mod 4)
\bea
\label{Z7diagconstraint}
a_1 &=& 2+a_7,\,\, a_2=2+a_7 + a_8, a_3 = 2 + a_7 + a_8 + a_9,\,\,a_4 = 2 + a_7 + a_8 + a_9 + a_{10},\cr
a_5 &=& 2 + a_7 + a_8 + a_9 + a_{10} + a_{11},\,\,
a_6 = 2 + a_7 + a_8 + a_9 + a_{10} + a_{11} + a_{12}
\eea
In the partition trace $Z^1_0$,  the twist phase factor reads, upon evaluated on the invariant sublattice,
\be
\label{Z7phasefactor}
U(\theta, \tilde{p} ) = e^{ i\frac{\pi}{2} \left(  (2 +a_1 + a_7)m^2_1 + (a_1 + a_2 +a_8)m^2_2              
+ (a_2 + a_3 +a_9)m^2_3 + + (a_3 + a_4 +a_{10} )m^2_4 + (a_4 + a_5 + a_{11})m^2_5 +
 (a_5 + a_6 +a_{12})m^2_6  \right) }
\ee
Imposing \eqref{Z7diagconstraint} on \eqref{Z7phasefactor} renders it trivial,
in agreement with our general observation for orbifold elements of odd order. This implies for the instanton 
sum, the phase factor $\delta_g$ is the same. Since the order is a prime number, all we need to compute is the phase $\delta_1$ that the dual $A_6$ lattice sum picks up under the Dehn twist $\mathcal{T}^7$. For a general $N$, the $A_{N-1}$ lattice sum reads 
\be
\Theta_{A_{N-1}} (\tau)= \frac{\sum_{k=0}^{N-1} \theta_3 \left( \frac{k\pi}{N} | z \right)^N }{N\theta_3(Nz)}.
\ee
By the Jacobi inversion formula, 
\be
\Theta_{A_{N-1}} (-1/\tau) = \frac{1}{\sqrt{N}} \left( -i\tau\right)^{\frac{N-1}{2}}
\Theta_{A^*_{N-1}} (\tau)
\ee
The factor $(-i\tau)$ is cancelled away by an identical term that arises from performing $\mathcal{S}$ on $\eta (\tau)$. Thus, we only need to consider the dual $A_6$ lattice sum. To check the monodromy under $\mathcal{T}^7$, it is slightly more convenient to scale the lattice and consider $\Theta_{A^*_{N-1}} (N\tau)$ that is associated with $\sqrt{N} A^*_{N-1}$. The Gram matrix can be chosen such that the quadratic form reads \cite{Conway} (sometimes called the Voronoi's principal form of the first type)
$
(N-1) \sum_{j=1}^{N-1} x_j^2 - \sum^{N-1}_{i\neq j} x_i x_j
$
for integers $x_i$, and thus 
\be
\Theta_{A^*_{N-1}} (N\tau) = \sum_{x} q^{(N-1) \sum_{j=1}^{N-1} x_j^2 - \sum^{N-1}_{i\neq j} x_i x_j
}
\ee
from which it is easy to see that the phase $\delta_1 = 0$ since $\Theta_{A^*_{7-1}} (\tau+ 7) = \Theta_{A^*_{7-1}} (\tau)$. Alternatively, it turns out that the dual $A_6$ theta function was presented in a beautiful form by Ramanujan in his `lost' notebook \cite{Ramanujan}. Following Ramanujan, let's first define the function 
$$f(-q^2) \equiv \sum_{k=-\infty}^{\infty} (-1)^{\frac{k^2}{2}} q^{k(3k-1)}.$$
Ramanujan found that 
\be
\Theta_{A^*_6} (\tau) = \frac{f^7(-q^{\frac{2}{7}}) }{f(-q^2)} + 7 q^{\frac{4}{7}} \frac{f^{-1}(-q^{\frac{2}{7}}) }{f^{-7}(-q^2)}+ 7 q^{\frac{2}{7}} f^3(-q^{\frac{2}{7}}) f^3(-q^2)
\ee
from which it is elementary to see that $\Theta_{A^*_6} (\tau +7) = \Theta_{A^*_6} (\tau)$.

\subsection{Asymmetric $SU(4) \times SU(4)$ Orbifold}
\label{subsec:SU4}

We consider the $SU(4)$ root lattice with the following moduli and $\mathbb{Z}_4$ coxeter twist
\be
E= \left( \begin{array}{ccc}
1 & -1 & 0\\
0 & 1 & -1\\
0 & 0 & 1 
\end{array} \right),
\qquad
\theta^{\mathbb{Z}_4} = r_1 r_2 r_3  =  \left( \begin{array}{ccc}
-1 & -1 & -1\\
1 & 0 & 0\\
0 & 1 & 0 
\end{array} \right).
\ee
We find the following twist phase factor characterized by the following $Q$.
\be
Q=\frac{\pi}{2} \left(
\begin{array}{cccccc}
a_1 & 1 & 1 & 1 & 0 & 0 \\
1 & a_2 & 1 & 0 & 1 & 0\\
1 & 1 & a_3 & 0 & 0 & 1\\
1 & 0 & 0 & a_4 & 0 & 0\\
0 & 1 & 0 & 0 & a_5 & 0 \\
0 & 0 & 1 & 0 & 0 & a_6
\end{array}
\right).
\ee
where the diagonal constants are partially fixed by \eqref{conditionGtwist4} to satisfy
\be
\label{SU4diagconstraints}
a_1 = a_2 + a_3 +a_4 + a_6,\qquad \textrm{mod}\,\,4
\ee
with an arbitrary $a_5$. In the partition trace $Z^1_0$, the twist phase factor reads, upon evaluated on the invariant sublattice,
\be
\label{refineSU4}
U(\theta, \tilde{p}) = e^{i\frac{\pi}{2} \left( (a_1 +a_4 -2)m^2_1 + (a_1 + a_2 +a_5 )m^2_2 +(a_2 + a_3 + a_6 )m^2_3   \right) } \equiv e^{2\pi i \left( \delta_1 m_1 + \delta_2 m_2 + \delta_3 m_3  \right)}
\ee
where $m_i$ are the residual winding numbers and $\delta_i$ are valued in $\left\{ 0, \frac{1}{2} \right\}$.
The instanton sum is the theta function of the $A_3$ lattice which, in the absence of possible weights in \eqref{refineSU4}, reads 
\be
\label{A3lattice}
\Theta_{A_3} (\bar{\tau}) = \theta (4\bar{\tau})^3 + 3 \theta_3 (4\bar{\tau}) \theta_2 (4\bar{\tau})^2 = \frac{1}{2} \left[
\theta^3_3 (\bar{\tau}) + \theta^3_4 (\bar{\tau}) \right]
\ee
Appearing in the twisted sector is the dual lattice sum which derives from a Poisson resummation \eqref{latticeinversion}, and reads 
\be
\label{inA3lattice}
\Theta_{A^*_3} (\bar{\tau}) = \frac{1}{4} \left[ \theta^3_3 (\frac{\bar{\tau}}{4}) + 3 \theta_3 (\frac{\bar{\tau}}{4} ) \theta^2_4 (\frac{\bar{\tau}}{4})    \right] = \theta^3_2 (\bar{\tau}) + \theta^3_3 (\bar{\tau}) 
\ee
Let us now insert in $\mathbb{Z}_2$-valued periodic weights following \eqref{refineSU4}. 
The constraint \eqref{SU4diagconstraints} tell us while $\delta_2$ is arbitrary, 
\be
\delta_1 + \delta_3 = \frac{1}{2}.
\ee
which nicely agrees with \eqref{SU4diagconstraints}.
With such a shift, the dual lattice sum $(\Theta^{(\delta_1,\delta_2,\delta_3)}(\bar{\tau}))$ now reads 
\be
\Theta^{(\frac{1}{2},\delta_2,0)}(\bar{\tau})=\Theta^{(0,\delta_2,\frac{1}{2})}(\bar{\tau})
= 2 \bar{q}^{\frac{3}{32}} \left( 1 + 3\bar{q}^{1/4} + 3 \bar{q}^{1/2} + 4 \bar{q}^{3/4} + 6 \bar{q} + \ldots \right) = \frac{1}{4} \theta^3_2 (\frac{\tau}{4})
\ee
Since this appears in $Z^0_1$ (and thus $Z^0_3$), we should check its transformation under $\mathcal{T}^4$. 
\be
\Theta^\delta_{A^*_3} (\bar{\tau} + 4) = e^{-\frac{3\pi i}{4}} \Theta^\delta (\bar{\tau} ),
\ee
whereas without the shift, we have 
\be
\Theta_{A^*_3} (\bar{\tau} + 4) = -\theta^3_2 (\bar{\tau}) + \theta^3_3 (\bar{\tau}).
\ee
Thus, up to a constant phase anomaly of $e^{-\frac{3\pi i}{4}}$, the presence of the twist phase factor preserves the modular covariance of the theory. We should also look at the other independent $SL(2,\mathbb{Z})$ orbit containing the partition trace $Z^2_0$. The invariant sublattice turns out to be four dimensional, and similar to the $\mathbb{Z}_{12}$ orbifold of $E_6$, it can be conveniently described by projecting the residual left and right Narain momenta onto the four-dimensional integral vector $\vec{v}=\{n_1, m_1, m_2, m_3 \}$ with the projection matrices $\tilde{P}_L = \mathcal{P}_L \cdot \vec{v}, \tilde{P}_R = \mathcal{P}_R \cdot \vec{v}$. Explicitly, the projection matrices read 
\be
\mathcal{P}_L =\left(
\begin{array}{cccc}
 1 & 1 & 0 & 0 \\
 -1 & 0 & 1 & -1 \\
 1 & -1 & 0 & 2
\end{array}
\right),\,\,\,
\mathcal{P}_R = \left(
\begin{array}{cccc}
 1 & -1 & 1 & 0 \\
 -1 & 1 & -1 & 0 \\
 1 & -1 & 1 & 0
\end{array}
\right)
\ee
From the group composition law, the twist phase factor 
$$
U(\theta^2, p) = U(\theta,p)U(\theta,\theta (p))
$$
and on the invariant sublattice, it reads $e^{i\pi (n_1 + m_1 + m_2)}$ after \eqref{SU4diagconstraints} is
taken into account, again preserving invariance of $Z^0_2$ under $\tau \rightarrow \tau + 2$ up to a constant phase anomaly of $e^{-\frac{3\pi i}{2}}$. When the other $A_3$ lattice is taken into account, we are left with constant $\mathbb{Z}_4$ phase factors which can be cancelled by appropriate shifts in the internal lattice.

\section{On the operator algebra in the eigenbasis of the $\mathbb{Z}_N$ twist}
\label{sec:eigenbasisKM}

In this Section, we will present the Kac-Moody algebra at level one discussed in Section~\ref{subsec:Constructing} in the basis which we have used to compute the constraints for the 2 cocycle $\epsilon (\alpha, \beta)$ by relating them to the vertex operators' cocycles. It is the one induced by an orbifold twist, where the stringy Hilbert space decomposes into $N$ eigenspaces of the $\mathbb{Z}_N$ twist. First, we rewrite equation \eqref{RHSOPEtwist} to read
\be
\label{orbVV}
V(\alpha,z)_{[a]} \times V(\beta,w)_{[b]} \sim \frac{1}{N}
\sum_{\delta = 0}^{N-1} \epsilon \left( \theta^\delta (\alpha), \beta \right) e^{-\frac{2\pi i \delta a}{N}}
\frac{U(\theta^\delta, \alpha) V\left(  \theta^\delta (\alpha) + \beta, w \right)_{[a+b]}}  
{ (z-w)^{-\frac{1}{2} \alpha_R \theta_R^\delta \beta_R} (\bar{z}-\bar{w})^{-\frac{1}{2} \alpha_L \theta_L^\delta \beta_L}          } 
\ee
For the chiral asymmetric orbifolds considered earlier, either of the chiral sectors has no surviving momenta zero modes, and our choice of the lattice metric leads to a Kac-Moody algebra with level one. In the basis above, the singular terms appear whenever the conditions $\alpha.\theta^\delta.\beta = -2$ or $\beta+\theta^\delta.\alpha=0$ are satisfied in \eqref{orbVV}. Thus we can write \eqref{orbVV} in the following form. 
\bea
V(\alpha,z)_{[a]} \times V(\beta,w)_{[b]} &\sim& \frac{1}{N}
\sum_{\delta = 0}^{N-1} \sum_{\alpha_R \theta^\delta_R \beta_R = -2} 
\epsilon \left( \theta^\delta (\alpha), \beta \right) e^{-\frac{2\pi i \delta a}{N}}
\frac{U(\theta^\delta, \alpha) V\left(  \theta^\delta (\alpha) + \beta, w \right)_{[a+b]}}  
{ (z-w)} \cr
&&+\frac{1}{N}
\sum_{\delta = 0}^{N-1} \sum_{\beta_R + \theta^\delta_R \alpha_R = 0} 
e^{-\frac{2\pi i \delta a}{N}}
U(\theta^\delta, \alpha) 
\left(
\frac{\delta_{a+b,0}}{(z-w)^2} 
+ \frac{i \alpha_{[a+b]} \partial X(w)  }{z-w}
\right) \nonumber \\
\eea
where the projected momenta $\alpha_{[a]}$ are defined as 
$$
\alpha_{[a]} = \frac{1}{N} \sum_s e^{-\frac{2\pi i sa}{N}} \theta^{s} \cdot \alpha
$$
Defining a set of projected vector $\epsilon_{[a]}$ to contract with the primaries $\partial X$, the other relevant OPEs read
\bea
\epsilon_{[a]k} \partial X^k (z) \times V(\alpha, w)_{[b]}
&=& \frac{1}{N^2} \sum_{r,s=0}^{N-1} 
e^{-\frac{2\pi i (sa+rb)}{N}}
U(\theta^r,\alpha)  \left( \theta^s \right)_k^m \epsilon_m \partial X^k e^{i\theta^r(\alpha) \cdot X(w)} \cr
&=& \frac{1}{N^2} \left(
\sum_{s-r} e^{\frac{-2\pi i (s-r)a}{N}} \alpha \cdot \theta^{s-r} \cdot \epsilon \right)
\left(
\sum_r e^{-\frac{2\pi i r(a+b)}{N}} U(\theta^r, \alpha) \frac{V(\theta^r(\alpha),w)}{z-w}
\right) \cr
&=& \alpha \cdot \epsilon_{[a]} \frac{V(\alpha,w)_{[a+b]}}{z-w}
\eea
and finally, between the oscillaors, we have
\bea
\epsilon_{[a]k} \partial X^k (z) \times \eta_{[b]l} \partial X^l (w)
&=& -\frac{1}{2N^2} \sum_{r,s=0}^{N-1} e^{-\frac{2\pi i (s-r)a}{N}} \epsilon \cdot \left(\theta^T \right)^{s-r} \eta e^{-\frac{2\pi i r(a+b)}{N}} \frac{1}{(z-w)^2} \cr
&=& -\frac{1}{2} \frac{\epsilon_{[a]} \eta }{(z-w)^2} \delta_{a+b,0}.
\eea
There is an analogous construction for the twisted sectors. The operator algebra is generated by twisted vertex operators acting on a vacuum that has a non-zero conformal weight that depends on the twist. Such a vacuum can be constructed by including orbifold twist fields acting on the untwisted vacuum. The twist fields modify the integral modding to be fractional for the oscillators, whereas the momenta zero modes should be generated by untwisted vertex operators invariant under the twist. The enhanced affine symmetries that arise correspond to the subalgebra associated with the automorphism of the original operator algebra. This is the notion of `twisted affine algebras' \cite{Kac}. For example, for the class of chiral asymmetric orbifolds discussed in the previous section, the twisted affine algebra is isomorphic to the original algebra because the orbifold twist originates from an inner automorphism of the finite Lie algebra of which roots generate the toroidal lattice. 
The equivalence of the unorbifolded toroidal theories (ADE) to WZW theories at level one prompts the question of whether there exists a corresponding map between asymmetric orbifolds of tori that enjoys enhanced affine symmetries, and WZW orbifolds \cite{deBoer,Birke:1999ik}.

\end{document}